\documentclass[%
reprint,
amsmath,
amssymb,
nolongbibliography,
aip,
jcp,
]{revtex4-2}
\usepackage{graphicx}
\usepackage[hidelinks]{hyperref}
\usepackage{mathtools}
\usepackage{amsthm} 
\usepackage{booktabs} 

\usepackage[linesnumbered,ruled,longend]{algorithm2e}

\SetCommentSty{mycommfont}
\SetAlgoNlRelativeSize{0}
\SetNlSty{texttt}{\footnotesize}{}
\SetNlSkip{1ex}
\SetKwComment{cc}{$//$\ }{}
 
\SetAlgoCaptionSeparator{.} 

\newcommand{\kB}{{k_\mathrm{B}}}
\newcommand{\iu}{{\mathrm{i}}}
\newcommand{\invcm}{{\mathrm{cm}^{-1}}}

\newcommand{\set}[1]{{\left\{{#1}\right\}}}
\newcommand{\ket}[1]{\left| {#1} \right\rangle}

\newcommand{\bra}[1]{\left\langle {#1} \right|}

\newcommand{\ip}[2]{\left\langle {#1} \middle| {#2} \right\rangle}
\newcommand{\op}[2]{{\ket{#1}\bra{#2}}} 
\newcommand{\mel}[3]{\bra{#1}{#2}\ket{#3}}
\newcommand{\comm}[2]{\left[ {#1} , {#2}\right]}
\newcommand{\Tr}{\mathrm{Tr}}
\newcommand{\order}[1]{\mathcal{O}(#1)}

\newcommand{\dv}[1]{\frac{\mathrm{d}}{\mathrm{d}{#1}}}
\newcommand{\pdv}[1]{\frac{\partial}{\partial{#1}}}

\newcommand{\abs}[1]{\left\lvert {#1} \right\rvert} 
\newcommand{\ie}{\emph{i.e.}}
\newcommand{\cf}{\emph{cf.}}  

\newcommand{\dd}[1]{\mathrm{d}{#1}}

\newcommand{\PyName}{{\texttt{TENSO}}}

\newcommand{\fig}[1]{Fig.~\ref{#1}} 

\newcommand{\Fig}[1]{Figure~\ref{#1}}

\newcommand{\eq}[1]{Eq.~\eqref{#1}}

\newcommand{\Eq}[1]{Equation~\eqref{#1}}

\newcommand{\stn}[1]{Sec.~\ref{#1}}

\begin{document}


%

\title{Tree tensor network hierarchical equations of motion based on time-dependent variational principle for efficient open quantum dynamics in structured thermal environments}%

\author{Xinxian Chen}
\affiliation{Department of Chemistry, University of Rochester, Rochester, New York 14627, United States}

\author{Ignacio Franco}
\email{ignacio.franco@rochester.edu}
\affiliation{Department of Chemistry, University of Rochester, Rochester, New York 14627, United States}
\affiliation{Department of Physics, University of Rochester, Rochester, New York 14627, United States}
\affiliation{The Institute of Optics, University of Rochester, Rochester, New York 14627, United States}

\date{\today}

\begin{abstract}
We introduce an efficient method TTN-HEOM for exactly calculating the open quantum dynamics for driven quantum systems interacting with highly structured bosonic baths  by combining the tree tensor network (TTN) decomposition scheme to the bexcitonic generalization of the numerically-exact hierarchical equations of motion (HEOM).  The method yields a series of quantum master equations for all core tensors in the TTN that efficiently and accurately capture the open quantum dynamics for non-Markovian environments to all orders in the system-bath interaction.
These master equations are constructed based on the time-dependent Dirac--Frenkel variational principle which isolates the optimal dynamics for the core tensors given the TTN ansatz. The dynamics converges to the HEOM when increasing the rank of the core tensors, a limit in which the TTN ansatz becomes exact.  We introduce \PyName, Tensor Equations for Non-Markovian Structured Open systems, as a general-purpose Python code to propagate the TTN-HEOM dynamics.  We implement three general propagators for the coupled master equations: Two fixed-rank methods that require a constant memory footprint during the dynamics, and one adaptive-rank method with variable memory footprint controlled by the target level of computational error. 
We exemplify the utility of these methods by simulating a two-level system coupled to a structured bath containing one Drude--Lorentz component and eight Brownian oscillators, which is beyond what can presently be computed using the standard HEOM.  Our results show that the TTN-HEOM is capable to simulate both dephasing and relaxation dynamics of driven quantum system interacting with structured baths, even those of chemical complexity,  with affordable computational cost.
\end{abstract}

\maketitle
\section{Introduction}\label{sec:intro}
Computation of the dynamics of open quantum systems with high precision is a central challenge in physics, chemistry, and quantum information science.\cite{Breuer2007, Vega2017,Weimer2021,Mulvihill2021,Landi2022}
It is necessary to capture the decoherence and eventual thermalization of quantum systems due to interactions with a quantum thermal environment.  From a molecular science perspective, such open quantum dynamics is central in our elementary description of photophysics, photochemistry, multidimensional optical spectroscopies,\cite{Mukamel1995, Biswas2022} coherent control,\cite{Shapiro2012, Rice2000,  Buerle2018, Heide2024} and charge and energy transfer,\cite{May2011, Cheng2009, Ortmann2009}  From a quantum information science perspective, correctly capturing such open quantum dynamics is needed to understand the evolution and decay of coherence and entanglement in qubits, simulate the operation of digital and analog quantum processors, design quantum control strategies, and develop strategies to minimize decoherence effects in next-generation quantum devices.\cite{Schlosshauer2007, Nielsen2010, Gu2018, Hu2018, Kim2022, Kim2024a, Korol2025, Cao2020, Chiesa2024} 

In open quantum dynamics,\cite{Breuer2007, Schlosshauer2007, Joos2003} it is customary to divide the Hamiltonian of the quantum universe $H= H_\mathrm{S} + H_\mathrm{B} + H_\mathrm{SB}$ into a system $H_\mathrm{S}$ which correspond to the degrees of freedom (DoFs) of interest,  an environment or bath $H_\mathrm{B}$ and  their interaction $H_\mathrm{SB}$.
The state of the system is completely described by its reduced density matrix $\rho_{\text{S}}(t) = \Tr_\mathrm{B}[{\rho}(t)]$ obtained by tracing out the bath DoFs from the density matrix of the composite quantum system ${\rho(t)}$.
The dynamics of the system $\rho_{\text{S}}(t)$ can be obtained by either following the dynamics of both system and bath and then tracing over the bath, or by solving so-called quantum master equations satisfied by $\rho_{\text{S}}(t)$ that implicitly capture the influence of the environment on the system DoFs. 
The former approach is preferred\cite{Lyu2022, Wang2021, Popp2021, Ren2018, Hu2018, Meyer1990} but is often computationally impractical as the bath can be macroscopic.
For this reason, there has been impressive progress in formulating increasingly accurate quantum master equations and developing computational techniques to propagate them.\cite{Redfield1957, Scully1967, Mollow1969, Tanimura1989, Ishizaki2009, Ikeda2020, AntoSztrikacs2023,Mulvihill2021}

Of particular interest are numerically ``exact'' master equations, such as the hierarchical equations of motion (HEOM)\cite{Tanimura2020, Yan2021, Ikeda2020, Chen2024}, as they can be used to model a large class of problems of interest in chemistry and quantum information science with an accuracy that can be assumed. This contrasts with common strategies based on the Born-Markov approximation, such as the Lindblad\cite{Lindblad1976} and the Redfield master equation\cite{Redfield1957,Redfield1965}, that are only valid for quantum systems weakly coupled to an environment that has a short memory time with respect to the system's dynamics; conditions that are often violated by chemically and physically relevant systems. The HEOM approach is complemented by other numerically exact methods to capture the open quantum dynamics that do not have a master equation form such as the quasi-adiabatic propagator path integral (QuAPI).\cite{Topaler1993, Makri1995, Makri1995a, Kundu2023, Bose2023}

The HEOM is based on decomposing the dynamics of the  bath correlation function (BCF) into a series of $K$ complex exponential functions or \emph{features}.  The influence of the thermal environment on the system is captured by introducing an infinite hierarchy of auxiliary density matrices (ADMs) that evolve as the system decoheres and reaches thermal equilibrium.  The dimensionality $M\times M$ of each ADM is the same as the reduced density matrix of the $M$-level system.  This HEOM dynamics was recently shown to be identical to the system in interaction with $K$ fictitious bosonic quasiparticles called bexcitons that are born, oscillate and decay during the dynamics in such a way that the system has the correct dynamics.\cite{Chen2024} 

This hierarchy can be truncated to a given order $N$ for each feature, which is referred to as the depth of the HEOM. The number of required ADMs is $N^K$, resulting in the overall space complexity of HEOM $\order{M^2N^K}$.   Thus, the computational cost of the HEOM increases exponentially with the number of features $K$. This number of features grows with the complexity of the chemical environment or when low-temperature corrections are needed in the dynamics. For this reason, to date, using the HEOM we are able to investigate illustrative model problems with simple environment models with only a few features in the BCF decomposition such as single Drude--Lorentz or Brownian oscillator model\cite{Tanimura2020,Ikeda2020,Lindoy2023}.
However, the HEOM becomes intractable  for realistic highly structured chemical environments. 

To reduce this curse of dimensionality, one strategy is to employ the filtering technique that removes ADMs that are almost zero\cite{Shi2009}, thus enabling HEOM simulations with more complex environments with larger $K$.  While helpful, the strategy is still insufficient to model chemically realistic problems and, further, it is not applicable to environments at low temperature.\cite{Dan2023} A second strategy is to more efficiently capture the time-dynamics of the BCF thus reducing the number of required features.\cite{LeDe2024, Xu2022, Nakatsukasa2018, Takahashi2024, Lednev2024, Lambert2019, Potts2013, Chen2022} More general strategies to curb this curse of dimensionality are needed to apply this numerically-exact HEOM method to chemically complex systems and environments.

In this paper, we introduce a tree tensor network (TTN) decomposition of the HEOM, TTN-HEOM, that enables efficient simulation of open quantum dynamics in structured thermal environment, even those of chemical complexity.  Our approach is based on the recent bexcitonic generalization of the HEOM which recovers all HEOM variants and, thus, is of general applicability to the HEOM family of quantum master equations.  The method further admits arbitrary time-dependence in the system Hamiltonian as needed to investigate driven dynamics of qubits and molecular systems in the presence of a thermal environment. In addition, we develop a general-purpose Python-based computational implementation of the TTN-HEOM that we name \PyName.  For computational efficiency, our implementation takes advantage of NumPy\cite{Harris2020} and PyTorch\cite{Ansel2024} which contain a series of libraries specifically designed and optimized to deal with tensor manipulation on CPUs and GPUs.  

Tensor network decomposition are the basis of highly successful simulation strategies\cite{Wang2003,Schollwoeck2011} in many-body science. In unitary quantum dynamics, tensor trains (or matrix product states) have been successfully used to enable wavefunction propagation in high-dimensions \cite{Cazalilla2002, Luo2003, Feiguin2005a, Haegeman2011, Keller2015, Greene2017, Lyu2022}. In turn, TTN decompositions of the multi-DoF wavefunction propagated by the time-dependent Schr\"odinger equation is the basis of widely employed methods such as the multi-layer multi-configurational time-dependent Hartree method (ML-MCTDH)\cite{Wang2003, Dorfner2024} and related strategies.\cite{Legeza2014, Murg2015} For open quantum system, tensor network techniques have also been proposed to accelerate simulations. This includes efforts in approximate methods such as Lindblad master equations\cite{Werner2016, Somoza2019} and strategies where thermal effects are included just at initial time through purification\cite{Ren2018, Li2020b}.  They have also been used in numerically exact approaches,  such as the thermalized time-evolving density operator with orthogonal polynomials algorithm\cite{Tamascelli2019}, path-integral process tensor methods\cite{Strathearn2018, Bose2022, Kundu2020}, and also the HEOM.\cite{Shi2018, Borrelli2019, Borrelli2021, Yan2021, Li2022, Ke2022, Mangaud2023, Ke2023}  Overall, the current view that has emerged from these efforts is that tensor network strategies can be successful in curbing the curse of dimensionality in quantum dynamics by efficiently encoding the entanglement among DoFs.

In this paper, we advance a rigorous and practical TTN decomposition of the HEOM by taking inspiration from ideas and methods in the MCTDH literature. Specifically, 
we arrange the collection of all ADMs in HEOM into an extended density operator (EDO) containing both the physical DoFs and the DoFs arising from the decomposition of the BCF into features. 
We then use a TTN decomposition to express this high-order EDO tensor into a contraction of low-order \emph{core} tensors. We provide a rigorous derivation of the quantum master equation satisfied by each core tensor in the TTN  by invoking the Dirac--Frenkel time-dependent variational principle (TDVP), which is also used in the MCTDH\cite{Meyer1990,Raab2000b}.  These coupled master equations guarantee that the TTN decomposition of the EDO remains as accurate as possible during the dynamics. To propagate the dynamics, we use strategies used in MCTDH and adapt them and generalize them to the EDO and non-unitary dynamics of the HEOM. Specifically, we implement and test the projector-splitting \cite{Haegeman2016,Kloss2017,Lubich2015,Lubich2018} and direct-integration with regularization \cite{Wang2018a,Meyer2018,Wang2021} and demonstrate that they provide stable propagation of TTN-HEOM. 

We exemplify our method and computational implementation in a two-level molecule coupled to a highly structured thermal environment with a spectral density extracted from experiments,\cite{Gustin2023} a system that is challenging to model using standard HEOM due to the large memory requirements. To this end, we decompose the spectral density in terms of Drude-Lorentz features to represent the solvent and underdamped Brownian oscillators to represent intramolecular vibrations.\cite{Gustin2023, Lorenzoni2024} This approach is more efficient to invoke in HEOM compared to the discretization approach where finite-many discrete vibrational modes are used for modeling the bath as they do not capture the dissipative nature in open quantum dynamics \cite{Tanimura2020}.

Compared to recent advances in combining tensor network techniques into HEOM, our proposed method and computational implementation admits both tensor trees and tensor trains and thus builds up but generalizes initial efforts using tensor train strategies \cite{Shi2018, Borrelli2019, Borrelli2021, Ke2022, Mangaud2023}.

Compared to recent efforts to introduce TTN into the HEOM,\cite{Yan2021, Ke2023} we have been able to develop and implement three numerically stable propagation strategies based on projector-splitting (PS) and direct integration. 
PS is a form of Trotterization where a single-step in the overall propagation is split into step-wise propagation of each core tensor in the TTN.  We implemented two versions of PS, a constant rank PS1 method where memory requirements are fixed, and an adaptive rank PS2 method controlled by the target level of computational error.  

In addition, we introduce a direct integration method which arises from our derivation of the quantum master equation of the core tensors through TDVP.
The method simultaneously integrates the master equations for all core tensors. As such, it admits any numerical integration scheme such as high-order Runge-Kutta schemes\cite{Dormand1980} and thus can be parallelized and have better scaling with the integration time step $\Delta t$ with respect to PS. However, the scheme requires  regularization\cite{Wang2018a,Meyer2018,Wang2021} which can add a small error to the dynamics.

We implemented the TTN-HEOM with these numerically propagation strategies into a Python package, called Tensor Equations for Non-Markovian Structured Open systems (\PyName). We discuss the merits and limitations of these numerical propagation strategies in TTN-HEOM using \PyName. Overall, our developments provide a TTN-HEOM method and computational implementation of full functionality that enables investigates dissipative dynamics of quantum systems immersed in highly structured thermal bosonic environments.

The paper is organized as follows. 
We first summarize the bexcitonic generalization of the HEOM (\stn{sec:heom}).
Then, we introduce its TTN decomposition (\stn{sec:ttn-main}) and isolate the equations of motion satisfied by the core tensors (\stn{sec:ttn-me}). 
Next, we introduce the three propagation methods (\stn{sec:propagation}) and discuss the computational implementation of the TTN-HEOM (\stn{sec:code}).
In Sec.~\ref{sec:results}, we exemplify the utility of the method simulating dissipative quantum dynamics due to interactions of quantum systems with highly-structured thermal baths using the three propagation methods and different TTN topologies. 
We summarize our main findings in Sec.~\ref{sec:con}.
\section{Theory}\label{sec:theory}
\subsection{Hierarchical equations of motion and bexcitonic picture}\label{sec:heom}

HEOM is capable of following the dissipative dynamics of general driven quantum systems coupled to multiple independent thermal baths through system operators that do not need to commute.\cite{Ikeda2020} For clarity in presentation, and without loss of generality, we consider coupling to one thermal harmonic bath with Hamiltonian
\begin{equation}\label{eq:bath-h}
  {H}_\text{B} = \sum_{j}\left(\frac{{p}^2_j}{2m_j} +\frac{m_j\omega_j^2{x}_j^2}{2}\right),
\end{equation}  
where ${x}_j$ and ${p}_j$ are the position and momentum operators of the $j$-th harmonic mode of effective mass $m_j$ and frequency $\omega_j$.
The system--bath coupling $H_{\text{SB}} = Q_{\text{S}} \otimes X_{\text{B}}$ is linear to a system operator $Q_{\text{S}}$ and a collective bath coordinate 
\begin{equation}\label{eq:bath-x}
  X_{\text{B}} = \sum_j c_{j} {x}_j,
\end{equation}
where $c_{j} $ quantifies the coupling strength between the $j$-th bath mode and the system operator.

While the dynamics of the density matrix of the composite system $\rho(t)$ is unitary, the dynamics of the system's density matrix $\rho_{\text{S}}(t) = \Tr_{\text{B}}(\rho(t))$ is non-unitary and satisfies\cite{Ishizaki2009}
\begin{equation}\label{eq:prop}
    {\tilde{\rho}_{\text{S}}(t)}= \mathcal{T} \tilde{\mathcal{F}}(t,0) \rho_{\text{S}} (0),
\end{equation}
where $\mathcal{T}$ is the time-ordering operator,
\begin{equation}\label{eq:if}
    \tilde{\mathcal{F}}(t,0) = e^{- \int_0^t\dd{s} \tilde{Q}^{\times}_{\text{S}}(s) \int_0^s \dd{u} \left(C(s-u) \tilde{Q}_{\text{S}}(u) \right)^{\times}},
\end{equation}
and $C(t) = \Tr\left(\tilde{X}_\text{B}(t) \tilde{X}_\text{B}(0) \rho_\text{B}^{\text{eq}}\right)$ is the BCF.
Throughout we use atomic units where $\hbar=1$ and the notation $A^>B = AB$ and $A^<B = BA^\dagger$ for the ordering of matrix multiplications, and $A^{\times} = A^> - A^<$ for the commutator super-operator generated from $A$.\cite{Takagahara1977}
In writing \eq{eq:prop} we have adopted the interaction picture of $H_0(t) = H_{\text{S}}(t) + H_{\text{B}}$,  where $\tilde{O}(t) = {\left(\mathcal{T} e^{-\iu\int_0^t H_0(t') \dd{t'}}\right)^\dagger} O(t) \mathcal{T} e^{-\iu\int_0^t H_0(t') \dd{t'}}$. 
\Eq{eq:if} provides a formal solution to the open quantum dynamics at all temperatures and to all orders in the system-bath interaction. As seen, $C(t)$ contains \emph{all} the information needed to capture the influence of the bath on $\rho_{\text{S}}(t)$. 

The BCF is related to the bath spectral density  $J(\omega) = \sum_j  \abs{c_{j}}^2 \delta(\omega-\omega_j)/ (2m_j\omega_j)$ through\cite{Callen1951,May2011}
\begin{equation}\label{eq:bcf}
    C(t) = \int_{0}^{\infty} J(\omega) (\coth(\omega/(2k_{\text{B}}T))\cos(\omega t) - \iu \sin(\omega t)) \dd{\omega}.
\end{equation}
The integral can be resolved by using the residue theorem through analytical continuation of $J(\omega)$ and Matsubara\cite{Ishizaki2005} 
or Pad{\'{e}}\cite{Hu2010} expansion of the thermal $\coth(\omega/(2k_\text{B}T))$ component. 
That analysis shows that $C(t)$ and its complex conjugate $C^\star(t)$ can always be decomposed in terms of a series of complex exponential functions as
\begin{equation}\label{eq:bcfdec}
    C(t) =  \sum_{k=1}^K c_k e^{\gamma_k t} \quad\text{and}\quad C^\star(t) = \sum_{k=1}^K \bar{c}_k e^{\gamma_k t},
\end{equation}
where $c_k$, $\bar{c}_k$, $\gamma_k$ are complex numbers. 
Other numerical methods can also be used to fit the BCF into the form in \eq{eq:bcfdec}.\cite{LeDe2024, Xu2022, Nakatsukasa2018, Takahashi2024, Lednev2024, Lambert2019, Potts2013, Chen2022} 
Each $k$ in the series \eq{eq:bcfdec} defines a \emph{feature} of the bath. 
This decomposition of $C(t)$ into $K$ features can capture any physical dynamics including exponential decay, oscillations and their combination.\cite{Chen2024}


The HEOM results from introducing this decomposition of the BCF into the exact dynamical map in \eq{eq:prop} and calculating the time-derivatives.  
What this shows is that the influence of the thermal environment on the dynamics of the system is exactly captured through a collection of auxiliary $M\times M$ density matrices (ADMs) $\{{\varrho}_{\vec{n}}(t) \}$ with the same dimensionality of $\rho_\text{S}(t)$.
Here,  $\vec{n}$ is a $K$-dimensional index $\vec{n} = (n_1, \ldots, n_k, \ldots, n_K)$ with $n_k=0,\ 1,\ 2,\ \ldots,$ and the series runs \emph{ad infinitum}.
We arrange these ADMs as a vector of matrices that we call the \emph{extended density operator} 
\begin{equation}\label{eq:edo}
\ket{{\Omega}(t)} =\sum_{\vec{n}} {\varrho}_{\vec{n}}(t) \ket{\vec{n}} 
\end{equation} 
in a basis $\{\ket{\vec{n}}\equiv \ket{n_1} \otimes \cdots \otimes \ket{n_k} \otimes \cdots \otimes \ket{n_K} \}$
such that ${\varrho}_{\vec{n}}(t) = \ip{\vec{n}}{{\Omega}(t)}$.  
The physical system's density matrix ${\rho}_{\text{S}}(t)= {\varrho}_{\vec{0}}(t)$ is located at $\vec{n} = \vec{0} \equiv (0, \ldots, 0)$.

In this context, we find that the exact quantum dynamics for this extended density operator $\ket{\Omega(t)}$ is \cite{Chen2024} 
\begin{equation}\label{eq:eom}
     \dv{t} \ket{{\Omega}(t)} =\left( - {\iu} {H}^{\times}_{\text{S}}(t) + \sum_{k=1}^{K}\mathcal{D}_k \right) \ket{{\Omega}(t)},
\end{equation}
with
\begin{equation}\label{eq:dso}
        \mathcal{D}_k 
       = \gamma_{k} {\hat{\alpha}}^\dagger_k {\hat{\alpha}}_{k}
       +  \left(
        c_k {Q}_\text{S}^> - \bar{c}_k {Q}_\text{S}^< \right)
        \hat{z}_{k}^{-1} \hat{\alpha}^\dagger_k 
       - {Q}^{\times}_\text{S}
        \hat{\alpha}_k\hat{z}_{k} 
\end{equation}  
and initial conditions 
\begin{equation}\label{eq:heom-ic}
\ket{\Omega(0)} = \rho_{\text{S}}(0)|\vec{0}\rangle,
\end{equation}
where $\rho_{\text{S}}(0)$ is the initial state of the system.
The first term in Eq.~\eqref{eq:eom} is the unitary dynamics, while $\mathcal{D}_k$ captures the dissipation due to the $k$-th feature of the bath.
Here the bosonic creation  $\hat{\alpha}^\dagger_k$ and annihilation $\hat{\alpha}_k$ operators ($[\hat\alpha_k, \hat\alpha_{k'}^\dagger] = \delta_{k,k'}$) associated to the $k$-th bath feature connect the different ADMs as 
\begin{equation}
    \hat{\alpha}^\dagger_k \ket{{n_k}} = \sqrt{n_k + 1} \ket{{n_k+1}},\quad\hat{\alpha}_k \ket{{n_k}} = \sqrt{n_k} \ket{{n_k-1}}. 
\end{equation}
In turn, the $\hat{z}_k$ is any invertible operator that satisfies $\comm{\hat{z}_k}{\hat{\alpha}^\dagger_k\hat{\alpha}_k} = 0$, that we refer to as the metric.
Eq.~\eqref{eq:eom} defines a class of exact quantum master equations as there is choice in the representation of $\ket{\vec{n}}$ (position $\ket{\vec{x}}$, momentum $\ket{\vec{p}}$ or number $\ket{\vec{n}}$) and the metric $\hat{z}_k$. The standard HEOM \cite{Shi2009, Liu2014,Tanimura2020} are a specific case of Eq.~\eqref{eq:eom} obtained when  the number representation and $\hat{z}_k = \iu (\hat{\alpha}^\dagger_k\hat{\alpha}_k)^{-1/2}$  is chosen.

As discussed in Ref.~\onlinecite{Chen2024}, based on \eq{eq:eom}, the open quantum dynamics can be interpreted as the system interacting with a collection of fictitious bosonic quasiparticles that we call \emph{bexcitons}. For this, we associate $\ket{\vec{n}}$ with the creation of bexcitons with respect to vacuum $|\vec{0}\rangle$.
Specifically, we associate a bexciton of label $k$, a $k$-bexciton, for each feature of the bath $k$. The state $\ket{\vec{n}}$ corresponds to a situation in which $n_k$ $k$-bexcitons have been created for each $k$. In this picture, $\hat{\alpha}_k^\dagger$ creates and  $\hat{\alpha}_k$ destroys a $k$-bexciton. 
The commutation relation between $\hat{\alpha}_k$ and $\hat{\alpha}_k^\dagger$ dictates that bexcitons are bosons.  
While the bath can be macroscopic, only $K$ effective bexcitons are needed to capture the relevant component that influences the system. 
Thus, the bexcitons offer a coarse-grained, but still exact, view of the correlated non-Markovian system-bath dynamics to all orders in $H_{\text{SB}}$. The dissipators $\{\mathcal{D}_k\}$ in Eq.~\eqref{eq:eom} describe the bexcitonic dynamics and their interaction with the system. 
 As the composite system evolves toward a stationary state, bexcitons are created and destroyed. 
Each version of Eq.~\eqref{eq:eom} constitutes an exact map of the open quantum dynamics to the system-bexciton dynamics.   While the system's dynamics is common to all maps, the bexcitonic one is not. For this reason, the bexcitons are unphysical quasiparticles and bexcitonic properties should  be seen as a way to monitor the open quantum dynamics and its numerical convergence. 

\subsection{Tree tensor network decomposition}\label{sec:ttn-main}
The computational challenge of the HEOM is that the number of bexcitons $K$ needed to accurately describe the dynamics increases as the complexity of the spectral density grows and with decreasing temperature as needed to appropriately decompose $C(t)$, see Eq.~\eqref{eq:bcf}. 
Further, the ladder of states for each $n_k$ needs to be truncated at a given $(N_k-1)$ that defines the \emph{depth} of the $k$-bexciton, a quantity that needs to be increased until convergence. 
The overall space complexity of Eq.~\eqref{eq:eom} for a $M$-state system and $K$ bath features all truncated at a depth of $N_k=\order{N}$ is $\order{M^2N^K}$, and thus shows exponential growth with the number of bath features $K$. 
This is the reason why the HEOM computations have been limited to relatively simple models of the bath.\cite{Lindoy2023} Our hypothesis is that the HEOM has a lot of redundancy in state-space that can be efficiently compressed through a tensor network strategy, and used to curb this curse of dimensionality.

In the same way that the density matrix of the system $\rho_\text{S}(t)$ has matrix elements $[\rho_\text{S}]_{ij}= \bra{i}\rho_S(t)\ket{j}$, where $\{\ket{i}\}$ is a basis that spans the Hilbert space of the system,  the extended density operator has tensor elements 
\begin{equation}
[\Omega(t)]_{i j n_1 \cdots n_K} = \mel{i}{ \ip{n_1 \cdots n_K}{\Omega(t)} }{j}
\end{equation} 
where $\{\ket{n_k}\}_{n_k=0}^{N_k-1}$ is the number basis that spans the space of  the $k$-bexciton truncated at the level of $(N_k -1)$. 
The bexcitonic dynamics \eq{eq:eom} for this extended tensor can be written as
 $\dv{t} \Omega (t) = \mathcal{L}(t) \Omega(t)$
where 
$\mathcal{L}(t)$ is the tensor representation of the super-operator that generates the dynamics ($- {\iu} {H}^{\times}_{\text{S}}(t) + \sum_{k=1}^{K}\mathcal{D}_k$) in the given basis. Because the basis is a tensor product of individual elements $\ket{i}\otimes \bra{j}\otimes \ket{n_1}\otimes \cdots \otimes \ket{n_K}$, then from \eq{eq:eom} we can define local operators ${h_m^{\kappa}}$ ($\kappa= >,\ <,\ 1,\ \ldots,\ K$) such that
\begin{equation}\label{eq:sop}
\sum_{m=1}^{5K+2}  h_m^{>}(t)\otimes h_m^{<}(t) \otimes h_{m}^{(1)} \otimes \cdots \otimes h_{m}^{(K)} \equiv \mathcal{L}(t).
\end{equation}
The label $m$ runs over the individual terms in Eq.~\eqref{eq:eom}. 
Each $\mathcal{D}_{k}$ in Eq.~\eqref{eq:eom} gives five terms and the system Liouvillian $-\iu {H}_{\text{S}}^{\times}(t)$ two more. 
Each term consists of a component $h_m^{>}(t)$ that acts on basis $\{\ket{i}\}$, a component $h_m^{<}(t)$ that acts on $\{\bra{j}\}$, and components  $h_m^{(k)}$ that act on $\{\ket{n_k}\}$. 

\begin{figure}[tb]
  \centering
  \includegraphics[width=\linewidth]{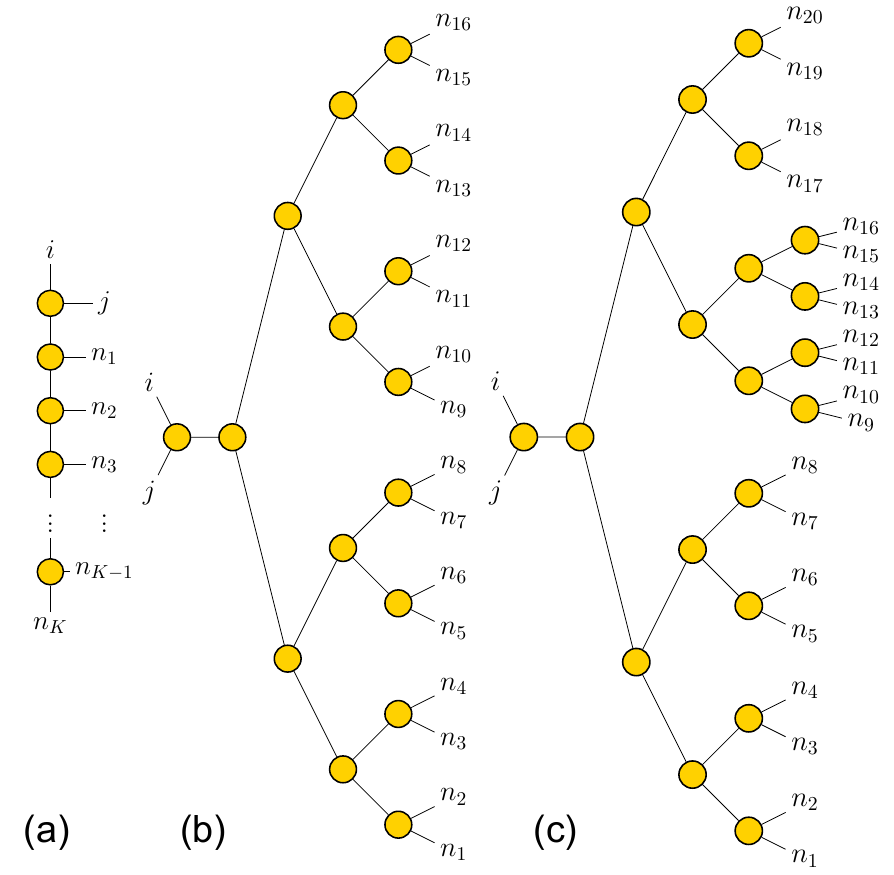}
  \caption{The topological structure of 
  (a) a tensor train with $K$ bexcitons for the EDO, (b) a perfectly balanced tensor tree with $16$ bexcitons, and (c) a balanced tensor tree with $20$ bexcitons. TTN-HEOM admits all these topologies with core tensors of variable order;  the figure focuses on order-3. }
  \label{fig:tn}
\end{figure}

The extended density tensor is high dimensional and can be compressed through a TTN which contains a collection of many low-order \emph{core tensors} with a given contraction-ordering that can be topologically described by a tree graph,  see Fig.~\ref{fig:tn}. 
The TTN may contain core tensors with different tensor orders.
We want to decompose the high-order tensor into a series of low-order tensors, as the operations between high-order tensor are computationally expensive. 
For instance, the space complexity of a $D$-order tensor $A_{a_1\cdots a_D}$ with $R$ as the range of all indexes $a_d$ is $\order{R^D}$. Since the  space-complexity of a tensor grows as a power of its order $D$, naturally we want the order of each core tensor in a TTN to be as small as possible. However, one cannot use only order-2 tensors in TTN for such decomposition in the presence of a thermal environment as it cannot give a tree, including a train for $K>0$. Therefore, the minimal non-trivial order for the core tensors is $3$.  For this reason, below we focus on the TTN-HEOM where all core tensors are of order-$3$. The generalization of the TTN to arbitrary order for each of the core tensors is included in the Supplementary Material.

No matter what the topology of the TTN is, the number of order-$3$ core tensors in the decomposition will be $K$, and the number of indexes for contractions in the TTN will be $K-1$. The simplest example is a tensor train (Fig.~\ref{fig:tn}(a)), which can be formulated as
\begin{equation}\label{eq:tt}
  \begin{multlined}
    \Omega_{ijn_1 \cdots n_K}   = \\\sum_{a_1 a_2 \cdots a_{K-1}}^{R_1 R_2 \cdots R_{K-1}}  A^{(0)}_{ija_1}   
    U^{(1)}_{a_1 n_1 a_2}
    U^{(2)}_{a_2 n_2 a_3}
    \cdots U^{(K-1)}_{a_{K-1} n_{K-1} n_K}.
  \end{multlined}
\end{equation} 
Here $\{a_s\}_{s=1}^{K-1}$ are the introduced indexes for contractions with $a_s = 0,\ \ldots,\ R_{s}-1$, with $R_s$ being the \emph{rank} of index $a_s$. 
In turn, Fig.~\ref{fig:tn}(b) and Fig.~\ref{fig:tn}(c) show tensor trees for a $16$-bexciton and $20$-bexciton EDO, respectively. 
In Fig.~\ref{fig:tn} each node represents a core tensor while each bond represents an index in 
$\{i,\ j\} \cup \{a_s\}_{s=1}^{K-1} \cup \{n_k\}_{k=1}^{K}$. The bonds $\{\alpha, \beta, \gamma\}$ attached to a node indicate that the tensor $U^{(s)}$ represented by that node will have indexes $\alpha,\ \beta,\ \gamma$ as $U^{(s)}_{\alpha \beta \gamma}$. 
Notice that the arrangement of indexes $\alpha,\ \beta,\ \gamma$ in tensor $U^{(s)}$ is not reflected in Fig.~\ref{fig:tn}.
As a convention (and without loss of generality), for $U^{(s)}$ we always place the index $a_s$ in the first index position $\alpha$.

Generally, the core tensors are isolated through hierarchical Tucker decomposition\cite{Grasedyck2010,Grasedyck2011} which repeatedly applies singular value decompositions (SVDs).
The final result of such a decomposition on a high-order EDO gives a TTN that can be formally represented as 
\begin{equation}\label{eq:ttn-main}
\begin{multlined} 
     [\Omega(t)]_{ijn_1\cdots n_K} = \\
    \sum_{a_1\cdots a_{K-1}}^{R_1\cdots R_{K-1}} 
    A^{(0)}_{ija_1}
    U^{(1)}_{a_1\beta_{1}\gamma_{1}}
    \cdots
    U^{(K-1)}_{a_{K-1}\beta_{K-1}\gamma_{K-1}}
    \\
    \equiv
    [\mathsf{Con}(A^{(0)}(t),U^{(1)}(t),\ldots,U^{(K-1)}(t))]_{ijn_1\cdots n_K}
\end{multlined}
\end{equation} 
where $\mathsf{Con}$ represents contractions among all core time-dependent tensors $ A^{(0)}(t),\ U^{(1)}(t),\ \ldots,\ U^{(K-1)}(t)$ in the TTN.
For each $U^{(s)}(t)$, its second and third indexes $\beta_s,\ \gamma_s \in \{a_s\}_{s=1}^{K-1} \cup \{n_k\}_{k=1}^{K}$ ($s = 1,\ \ldots,\ K-1$).  
From the SVD, the $U^{(s)}(t)$ are semi-unitary core tensors in the sense that the {matrix} $[U^{(s)}(t)]_{a_s, \beta \gamma}$ reshaped from the tensor $[U^{(s)}(t)]_{a_s \beta \gamma}$ satisfies
\begin{equation}\label{eq:onb}
  \sum_{\beta \gamma} [U^{(s)}(t)]_{a'_s, \beta \gamma}^\star [U^{(s)}(t)]_{a_s, \beta \gamma} = \delta_{a'_s a_s}.
\end{equation}
By contrast, there is only one  \emph{root} core tensor $A^{(0)}$ in the TTN which does not need to be semi-unitary. 
As a design principle, we choose to include the system's indexes $i$, $j$, and the index $a_1$ in the root tensor, such that the influence of all bexcitons is captured through compressed index $a_1$. 
In this way, the unitary component of the system's is exact and not compressed, while the influence of the bexcitons is compactly captured by the TTN. 

For a given set of ranks $\{R_s\}$ such that $R_s = \order{R}$ and a bexciton depth $N$ such that $N_k = \order{N}$, the space complexity of the TTN is $\order{M^2R + KNR (N+R)}$\cite{Grasedyck2011}, which no longer grows exponentially with the number of bath features $K$. The smaller the rank that can be used, the more efficient the compression of the TTN.

It is useful to define the \emph{height} $L$ of a core tensor $U^{(s)}$ as the number of bonds on the path between $U^{(s)}$ and the root.
For instance, the root $A^{(0)}$ is of height $0$, and in our TTN ansatz Eq.~\eqref{eq:ttn-main}, the $U^{(1)}$ is always of height $1$.
We sort the label $s$ of the core tensor $U^{(s)}$ according to its height $L(U^{(s)})$ without loss of generality.
That is, for two core tensors $U^{(r)}$ and $U^{(s)}$ if  $L(U^{(r)}) < L(U^{(s)})$ then $r < s$.

\subsection{Master equations for a tree tensor network}\label{sec:ttn-me}

To develop the master equations for the TTN, we invoke the Dirac--Frenkel TDVP\cite{Meyer1990, Raab2000b} and adapt it to $\Omega(t)$ as
\begin{equation}\label{eq:tdvp}
    \sum_{i j n_1 \cdots n_K} [\delta{\Omega (t)}]^\star_{ijn_1\cdots n_K} \left[\left(\mathcal{L}(t) - \dv{t}\right) \Omega(t)\right]_{ijn_1\cdots n_K} = 0,
\end{equation}
where $\delta{\Omega (t)}$ denotes a small variation of $\Omega(t)$. 
By doing so, it yields optimal dynamics for the core tensors that capture the dynamics of $\Omega(t)$ in a space with reduced dimensionality that changes dynamically during the quantum evolution.  
To guarantee that the TTN decomposition remains, we further require that Eq.~\eqref{eq:onb} holds during the propagation by demanding  
\begin{equation}\label{eq:gauge}
    \sum_{\beta \gamma} [U^{(s)} (t)]^\star_{a'_k \beta \gamma} [\dv{t} U^{(s)}(t)]_{a_k \beta \gamma} = 0,
\end{equation}
for all $t$, a condition that is referred to as the {gauge condition}.\cite{Haegeman2011} 
From Eq.~\eqref{eq:tdvp} and Eq.~\eqref{eq:gauge} we can systematically develop equations of motion for the semi-unitary and root core tensors for an arbitrary tensor tree. Appendix~\ref{sec:eom} derives the equations of motion for order-3 tensors, and the Supplementary Material offers their generalization using graph notation to a general TTN containing tensors with arbitrary order.

With the TTN decomposition in \eq{eq:ttn-main} and the gauge condition in \eq{eq:gauge},
the master equation of the root tensor $A^{(0)}(t)$ depends on $f_{m}^{(1)}(t)$ (defined later) as
\begin{equation}\label{eq:root-eom}
  \dv{t} A^{(0)}_{i' j' a'_1} = \sum_{m} \sum_{ij a_1} [h_m^{>}]_{i'i}[h_m^{<}]_{j'j} [f_{m}^{(1)}]_{a'_1 a_1} A^{(0)}_{i j a_1}.
\end{equation}
This is the simplest of the equations because the semi-unitary properties of the $U^{(s)}(t)$ and the gauge condition cancels out the terms involving direct contraction between the $U^{(s)}(t)$ and its time-derivatives. 
In turn, the master equations of the semi-unitary tensors $U^{(s)}(t)$ are:  
\begin{equation}\label{eq:su-eom}
  \begin{multlined}
    \sum_{a_{s}'} D^{(s)}_{a_{s}'a_{s}''}  \dv{t} U^{(s)}_{a_{s}'\beta'\gamma'} = 
    \sum_{m} \sum_{a_{s}'a_{s}\beta\gamma} [D^{(s)}_m]_{a_{s}'a_{s}''}\\
     \left([F_m^{(s2)}]_{\beta'\beta} [F_m^{(s3)}]_{\gamma'\gamma}U^{(s)}_{a'_{s}\beta\gamma} - 
    U^{(s)}_{a_{s}\beta'\gamma'}[f_m^{(s)}]_{a_{s}a_{s}'}\right).
  \end{multlined}
\end{equation}
Note that in our notation $F_m^{(s2)}(t)$ always contracts with the second index of $U^{(s)}(t)$ while $F_m^{(s3)}(t)$ with the third one. 
The definition of matrix $F_m^{(s\kappa)}(t)$ for $\kappa=2$ and $3$ depends on the location of the tensor $U^{(s)}(t)$ in the TTN.
$F_m^{(s\kappa)}(t) \equiv f_{m}^{(u)}(t)$  ($u > s$) if the $\kappa$-th index corresponds to a contracted index or bond in the TTN. That is, when the $\kappa$-th index in $U^{(s)}(t)$ is an $a_u$ in Eq.~\eqref{eq:ttn-main}. 
In turn, $F_m^{(s\kappa)}(t) \equiv h_{m}^{(k)}$ [\cf\ Eq.~\eqref{eq:sop}] when the $\kappa$-th index also occurs in the original EDO tensor $\Omega(t)$, and thus, corresponds to an open bond in the TTN.
That is, when the $\kappa$-th index in $U^{(s)}(t)$ is an $n_k$ in Eq.~\eqref{eq:ttn-main}.

The matrices $f_{m}^{(s)}(t)$ ($s=1,\ \ldots,\ K-1)$ are defined as 
  \begin{equation}\label{eq:f-recursive}
    [f_{m}^{(s)}]_{a_{s}'a_{s}} \equiv \sum_{\beta'\beta\gamma'\gamma}
    U^{(s)\star}_{a_{s}'\beta'\gamma'} 
    [F_{m}^{(s2)}]_{\beta'\beta}
    [F_{m}^{(s3)}]_{\gamma'\gamma}
    U^{(s)}_{a_{s}\beta\gamma}.
 \end{equation}
Notice that this definition is recursive as $f_{m}^{(s)}(t)$ depends on $F_{m}^{(s\kappa)}(t)=f_{m}^{(u)}(t)$ ($u>s$) if the $\kappa$-th index corresponds to a contracted bond. 
The matrices ${D^{(s)}}(t)$ and matrices ${D^{(s)}_m}(t)$ for given label $m$ in Eq.~\eqref{eq:sop} are also defined recursively.
For $s=1$,
 \begin{equation}\label{eq:d-recursive-1}  
 \begin{aligned}
    D^{(1)}_{a'_1 a_1} &\equiv \sum_{ij}
    A^{(0)}_{i j a'_1} 
    A^{(0)\star}_{i j a_1},\\
    [D^{(1)}_m]_{a'_1 a_1} &\equiv \sum_{i' i  j' j} 
    [h_m^{>}]_{ii'}[h_m^{<}]_{jj'} 
    A^{(0)}_{i' j' a'_1}
    A^{(0)\star}_{i j a_1}.
 \end{aligned}
\end{equation} 
For $s > 1$, there is a bond in the TTN corresponding to $a_s$ in Eq.~\eqref{eq:ttn-main} that contracts tensors $U^{(s)}(t)$ and $U^{(r)}(t)$ ($r < s$). If $U^{(r)}(t)$ has $a_s$ as its third index in Eq.~\eqref{eq:ttn-main} then
\begin{equation}\label{eq:d-recursive-2}  
\begin{aligned}
D^{(s)}_{a_{s}'a_{s}} &\equiv \sum_{a_{r}'a_{r}\varepsilon}
U^{(r)}_{a_{r}'\varepsilon a_{s}'} 
D^{(r)}_{a_{r}'a_{r}}
U^{(r)\star}_{a_{r}\varepsilon a_{s}},\\
[D^{(s)}_m]_{a_{s}'a_{s}} &\equiv \sum_{a_{r}'a_{r}\varepsilon'\varepsilon}
[F_{m}^{(r2)}]_{\varepsilon\varepsilon'}  
U^{(r)}_{a_{r}'\varepsilon'a_{s}'} 
[D^{(r)}_m]_{a_{r}'a_{r}}
U^{(r)\star}_{a_{r}\varepsilon a_{s}}
\end{aligned} 
\end{equation}
In turn, if $U^{(r)}$ has $a_s$ as the second index in Eq.~\eqref{eq:ttn-main}, then
\begin{equation}\label{eq:d-recursive-3}  
\begin{aligned}
D^{(s)}_{a_{s}'a_{s}} &\equiv \sum_{a_{r}'a_{r}\varepsilon}
U^{(r)}_{a_{r}' a_{s}' \varepsilon} 
D^{(r)}_{a_{r}'a_{r}}
U^{(r)\star}_{a_{r} a_{s} \varepsilon},
\\
[D^{(s)}_m]_{a_{s}'a_{s}} &\equiv \sum_{a_{r}'a_{r}\varepsilon'\varepsilon}
[F_{m}^{(r3)}]_{\varepsilon\varepsilon'}  
U^{(r)}_{a_{r}' a_{s}' \varepsilon'} 
[D^{(r)}_m]_{a_{r}'a_{r}}
U^{(r)\star}_{a_{r} a_{s} \varepsilon}.
\end{aligned} 
\end{equation}
In \stn{sec:code} we discuss the order in which these terms ($f_{m}^{(s)}(t)$,  $D^{(s)}(t)$ and $D_m^{(s)}(t)$) need to be evaluated.  As an explicit example, Appendix~\ref{sec:tree-example} details the TTN-HEOM scheme with $K=4$.

From \eq{eq:heom-ic}, the initial condition for the EDO is
$\Omega_{ij n_1 \cdots n_K }(0) = [\rho_{\text{S}}(0)]_{ij} \delta_{0n_1}\cdots\delta_{0n_K}$, where $\rho_{\text{S}}(0)$ is the initial state of the system.  In addition, we need to determine the initial conditions for the core tensors, $A^{(0)}(t=0)$ and $U^{(s)}(t=0)$, in the TTN. Except for minimal rank case when all $R_s=1$, this choice is not unique. 
We choose 
\begin{equation}\label{eq:root-init}
    A^{(0)}_{ija_1}(0) = [\rho_{\text{S}}]_{ij} \delta_{0a_1}
\end{equation}
for the root tensor. In turn,  $U^{(s)}_{a_s \beta \gamma}(0)$ for given $a_s=0,\ 1,\ \ldots,\ R_{s}-1$ is filled as:
$U^{(s)}_{0 \beta \gamma} = \delta_{0 \beta} \delta_{0 \gamma}$, 
$U^{(s)}_{1 \beta \gamma} = \delta_{1\beta}\delta_{0\gamma}$, 
$U^{(s)}_{2 \beta \gamma} = \delta_{0\beta}\delta_{1\gamma}$, 
$U^{(s)}_{3 \beta \gamma} = \delta_{2\beta}\delta_{0\gamma}$, 
$U^{(s)}_{4 \beta \gamma} = \delta_{1\beta}\delta_{1\gamma}$, 
$U^{(s)}_{5 \beta \gamma} = \delta_{0\beta}\delta_{2\gamma}$, 
$\ldots$
More explicitly, in each page of tensor $U^{(s)}_{a_s\beta\gamma}(0)$, the matrix $\mathbf{U}^{(s)}_{a_s}(0)$ for $a_s = 0,\ 1,\ \ldots,\ R_s-1$ is chosen as
\begin{align*}
    \mathbf{U}^{(s)}_{0} &= \left(\begin{matrix}
    1 & 0 & 0 & \cdots \\
    0 & 0 & 0 & \cdots \\
    0 & 0 & 0 & \cdots \\
    \vdots & \vdots & \vdots & \ddots 
\end{matrix}\right), \quad
\mathbf{U}^{(s)}_{1} &= \left(\begin{matrix}
    0 & 0 & 0 & \cdots \\
    1 & 0 & 0 & \cdots \\
    0 & 0 & 0 & \cdots \\
    \vdots & \vdots & \vdots & \ddots 
\end{matrix}\right), \\
\mathbf{U}^{(s)}_{2} &= \left(\begin{matrix}
    0 & 1 & 0 & \cdots \\
    0 & 0 & 0 & \cdots \\
    0 & 0 & 0 & \cdots \\
    \vdots & \vdots & \vdots & \ddots 
\end{matrix}\right), \quad 
\mathbf{U}^{(s)}_{3} &= \left(\begin{matrix}
    0 & 0 & 0 & \cdots \\
    0 & 0 & 0 & \cdots\\
    1 & 0 & 0 & \cdots\\
    \vdots & \vdots & \vdots & \ddots 
\end{matrix}\right), \\
\mathbf{U}^{(s)}_{4} &= \left(\begin{matrix}
    0 & 0 & 0 & \cdots \\
    0 & 1 & 0 & \cdots \\
    0 & 0 & 0 & \cdots \\
    \vdots & \vdots & \vdots & \ddots 
\end{matrix}\right), \quad 
\mathbf{U}^{(s)}_{5} &= \left(\begin{matrix}
    0 & 0 & 1 & \cdots \\
    0 & 0 & 0 & \cdots \\
    0 & 0 & 0 & \cdots \\
    \vdots & \vdots & \vdots & \ddots 
\end{matrix}\right),
\end{align*}
and so on. 
The choice of $U^{(s)}_{0 \beta \gamma}$ satisfies the correct initial condition for the EDO. In addition, the choice of $U^{(s)}_{a_s \beta \gamma}$ further guarantees the semi-unitary property of $U^{(s)}(0)$ at initial time. This choice is sufficient and is one of the simplest possible as all non-zero elements are $1$ and there is only one non-zero element in each  $\mathbf{U}^{(s)}_{a_s}(0)$.  Further, it locally balances all bonds associated with a node across the TTN.

\subsection{Propagation methods} \label{sec:propagation}

\subsubsection{Direct integration}

The main idea of direct integration is to simultaneously integrate the non-linear coupled series of ordinary differential equations (ODEs) in Eqs.~\eqref{eq:root-eom} and \eqref{eq:su-eom} using standard integration techniques. The advantage of this strategy is that it enables coupling the TTN-HEOM to well-developed ODEs solvers based on Runge-Kutta\cite{Dormand1980} and other schemes that allow for large integration time step, adaptive time steps, and even parallelization. 

To isolate the exact derivatives for the semi-unitary tensors, we need to multiply both sides of Eq.~\eqref{eq:su-eom} by $\left(D^{(s)}(t)\right)^{-1}$.  The challenge of this direct integration strategy is that this inverse does not always exist. In particular, for initially separable states, as required by the HEOM, $D^{(s)}(t)$ is singular. To see this, consider $[D^{(1)}(0)]_{a_1 a'_1} = \sum_{ij}A_{ija_1}(0)A^{\star}_{ija_1}(0) = (\Tr\rho^2_{\text{S}}(0))\delta_{0a_1}\delta_{0a'_1} $ from Eq.~\eqref{eq:root-init}. This leads to a singular matrix $D^{(1)}(0)$.
The remaining $D^{(s)}(0)$ are also singular according to the recursive relation [Eq.~\eqref{eq:d-recursive-2}] and the semi-unitary properties of $U^{(s)}$ [Eq.~\eqref{eq:onb}].

To make progress, we introduce the pseudo-inverse\cite{Penrose1955} $\left(D^{(s)}\right)^{+}$ of a matrix $D^{(s)}$.
The pseudo-inverse is a well-known generalization of the inverse of a matrix.
It can be constructed from the SVD of $D^{(s)} = U \sigma V^{\dagger}$ where $U$, $V$ are unitary matrix which are readily invertible as $U^\dagger = U^{-1}$.  The pseudo-inverse $\left(D^{(s)}\right)^{+} = V \sigma^{+} U^{\dagger}$ where $\sigma^+$ is the pseudo-inverse of the diagonal matrix $\sigma$ obtained by replacing the nonzero singular values $\sigma_b $ with their multiplicative inverses $\sigma_b^{-1}$.  Since the SVD always exists, then $(D^{(s)})^{+}$ also can always be defined.

Multiplying both sides of \eq{eq:su-eom} by $(D^{(s)})^{+}$ yields,
\begin{equation}\label{eq:su-eom2}
  \begin{multlined}
  \sum_{a'_s} \mathcal{P}^{(s)}_{a'_s a''_s} \dv{t} U^{(s)}_{a'_s \beta' \gamma'} = 
  \sum_{m}  \sum_{a'_s a_s  \beta \gamma} 
  [\mathcal{C}^{(s)}_{m}]_{a'_s a''_s} \ \\ 
   \times\left([F_m^{(s2)}]_{\beta'\beta} [F_m^{(s3)}]_{\gamma'\gamma} U^{(s)}_{a'_{s}\beta\gamma}  - U^{(s)}_{a_{s}\beta'\gamma'} [f_m^{(s)}]_{a_{s}a_{s}'}  \right) 
   .
  \end{multlined}
\end{equation}
Here   
\begin{equation}
   \mathcal{P}^{(s)}_{a'_k a''_k} \equiv \sum_{a_k} D^{(s)}_{a'_k a_k} (D^{(s)})^{+}_{a_k a''_k},
\end{equation}
and
\begin{equation}\label{eq:csadj}
  [\mathcal{C}^{(s)}_{m}]_{a'_k a''_k} \equiv \sum_{a_k} [D_m^{(s)}]_{a'_k a_k} \left(D^{(s)}\right)^{+}_{a_k a''_k}.
\end{equation}
If matrix $D^{(s)}$ is invertible, then $\left(D^{(s)}\right)^{+} = \left(D^{(s)}\right)^{-1}$ and $\mathcal{P}^{(s)}_{a'_k a''_k} = \delta_{a'_k a''_k}$ becomes an identity matrix. 
In turn, when $D^{(s)}$ is singular, then $\mathcal{P}^{(s)}$ is a projector to the column space of the matrix $D^{(s)}$
as 
\begin{equation}
\begin{multlined}
\mathcal{P}^{(s)}(t) = D^{(s)}(t) \left(D^{(s)}(t)\right)^+ 
= \sum_{\substack{b\ \text{s.t.}\\\sigma_b(t)>0}} u_b(t) u^\dagger_b(t),
\end{multlined}
\end{equation} 
where $u_b(t)$ is the $b$-th column of $U(t)$ and $\sigma_b(t)$ is the $b$-th singular value of matrix $D^{(s)}(t)$.
Hence, a singular $D^{(s)}(t)$ will introduce a loss of information if we let $\dv{t} U^{(s)}(t) \approx \mathcal{P}^{(s)}(t) \dv{t} U^{(s)}(t)$.

To handle this singularity issue we invoke the regularization technique developed in MCTDH and adapt it to this TTN-HEOM.\cite{Wang2018a,Meyer2018,Wang2021}  In one simple strategy, all singular values $\sigma_b$ in $D^{(s)}(t)$ that are smaller than a threshold $\epsilon$ are replaced by $\epsilon$. 
That is, if the SVD of $D^{(s)}_{a'_sa_s} = \sum_b U_{a'_s b} \sigma_{b}  V^*_{a_s b}$,
the regularization approximate $D^{(s)}_{a'_s a_s}\approx \sum_b U_{a'_s b} \max(\sigma_{b}, \epsilon) V^*_{a_s b}$.
This makes $D^{(s)}(t)$ invertible,  but introduces an error of $\order{\epsilon}$ in $D^{(s)}(t)$.

Our choice of regularization described below reduces the introduced error in $D^{(s)}(t)$ from $\order{\epsilon}$ to $\order{\epsilon^2}$, increasing the stability and overall accuracy of this propagation scheme. To do so, what is needed is to regularize both $D^{(s)}(t)$ and every $D^{(s)}_m(t)$. 
Starting from $s=1$, we perform the SVD for $A^{(0)}_{ija_1} = \sum_{b_1} {W}^{(1)}_{ij b_1}  {\sigma}^{(1)}_{b_1}  {V}^{(1)\star}_{a_1 b_1} $, and define
\begin{equation}\label{eq:csadj-recursive-1}  
 \begin{aligned} 
    [\bar{D}^{(1)}_m]_{a_1 b_1} &\equiv \sum_{i' i  j' j} 
    [h_m^{>}]_{ii'}[h_m^{<}]_{jj'} 
    A^{(0)}_{i' j' a_1}
    {W}^{(1)\star}_{i j b_1}.
 \end{aligned}
\end{equation} 
Here $\bar{D}^{(1)}_m(t)$, ${\sigma}^{(1)}(t)$ and ${V}^{(1)}(t)$ are all time-dependent quantities.
Further, together with the core tensor $U^{(1)}(t)$ we define a non-semi-unitary tensor $A^{(1)}(t)$ such that
\begin{equation}
A^{(1)}_{b_1\beta\gamma} \equiv \sum_{a_1} \sigma^{(1)}_{b_1} V^{(1)\star}_{a_1b_1}  U_{a_1\beta\gamma}^{(1)}.
\end{equation}
For $s>1$, the construction of $\bar{D}^{(s)}_m$, as well as $W^{(s)}$, $\sigma^{(s)}$, ${V}^{(s)}$ and $A^{(s)}$, is done by a recursive process over the TTN structure that is similar to the definition for ${D}^{(s)}$ and ${D}^{(s)}_m$ [\cf\ Eqs.~\eqref{eq:d-recursive-1}--\eqref{eq:d-recursive-3}]. 

For $s > 1$, there is a bond in the TTN corresponding to $a_s$ in \eq{eq:ttn-main}) that contracts tensors $U^{(s)}(t)$ and $U^{(r)}(t)$ ($r < s$). In this recursive argument, $\bar{D}^{(r)}_m(t)$ and $A^{(r)}(t)$ have already been determined from the previous step in the recursion.
If $U^{(r)}(t)$ has $a_s$ as its second index in \eq{eq:ttn-main} then
the SVD of $A^{(r)}$ is
\begin{equation}\label{eq:svd-reg1}
    A^{(r)}_{b_r a_s  \varepsilon} = \sum_{b_s}  {W}^{(s)}_{b_r b_s \varepsilon} {\sigma}^{(s)}_{b_s} {V}^{(s)\star}_{a_sb_s}.
\end{equation}
Then,
$\bar{D}^{(s)}_m(t)$ is defined as
\begin{equation}\label{eq:csadj-recursive-2}  
\begin{aligned}
[\bar{D}^{(s)}_m]_{a_s b_s} &\equiv \sum_{a_r b_r \varepsilon'\varepsilon}
[F_{m}^{(r3)}]_{\varepsilon\varepsilon'}  
U^{(r)}_{a_r a_s \varepsilon'} 
[\bar{D}^{(r)}_m]_{a_r b_r}
{W}^{(s)\star}_{b_r b_s \varepsilon}
\end{aligned} 
\end{equation}
In this case,
\begin{equation}\label{eq:dmat-reg1}
    D^{(s)}_{a_s a_s'} = \sum_{b_r \varepsilon} A^{(r)}_{b_r a_s  \varepsilon}  A^{(r)\star}_{b_r a_s'  \varepsilon}.
\end{equation}
In turn, if $U^{(r)}(t)$ has $a_s$ as its third index in Eq.~\eqref{eq:ttn-main},  the SVD of $A^{(r)}$ is
\begin{equation}\label{eq:svd-reg2}
A^{(r)}_{b_r \varepsilon a_s} = \sum_{b_s} {W}^{(s)}_{b_r \varepsilon b_s} {\sigma}^{(s)}_{b_s}  {V}^{(s)\star}_{a_s b_s}, 
\end{equation}
and
\begin{equation}\label{eq:csadj-recursive-3}  
\begin{aligned}
[\bar{D}^{(s)}_m]_{a_s b_s} &\equiv \sum_{a_r b_r \varepsilon'\varepsilon}
[F_{m}^{(r2)}]_{\varepsilon\varepsilon'}  
U^{(r)}_{a_r  \varepsilon' a_s} 
[\bar{D}^{(r)}_m]_{a_r b_r}
{W}^{(s)\star}_{b_r \varepsilon b_s }.
\end{aligned} 
\end{equation}
In this case, 
\begin{equation}\label{eq:dmat-reg2}
D^{(s)}_{a_sa_s'} = \sum_{b_r, \varepsilon} A^{(r)}_{b_r  \varepsilon  a_s}  A^{(r)\star}_{b_r   \varepsilon a_s'}
\end{equation}
For both cases, the definition of $A^{(s)}(t)$ continues as
\begin{equation}\label{eq:temp-root}
A^{(s)}_{b_s\beta\gamma} \equiv \sum_{a_s} \sigma^{(s)}_{b_s} {V}^{(s)\star}_{a_s b_s} U_{a_s\beta\gamma}^{(s)}.
\end{equation}
Substitute Eq.~\eqref{eq:svd-reg1} into Eq.~\eqref{eq:dmat-reg1}, and Eq.~\eqref{eq:svd-reg2} into Eq.~\eqref{eq:dmat-reg2}, $D^{(s)}$ become
\begin{equation}\label{eq:sqrt-d}
   D^{(s)}_{a_s a'_s} = \sum_{b_s} {V}_{a_s b_s}^{(s)\star}  \left({\sigma}^{(s)}_{b_s} \right)^{2} {V}_{a'_s b_s}^{(s)}.
\end{equation}
Further compare Eq.~\eqref{eq:d-recursive-2} with Eq.~\eqref{eq:csadj-recursive-2}, and Eq.~\eqref{eq:d-recursive-3} with Eq.~\eqref{eq:csadj-recursive-3}, we have
\begin{equation}\label{eq:d-reg}
   [D^{(s)}_m]_{a_s a'_s}= \sum_{b_s}[\bar{D}^{(s)}_m]_{a_s b_s} \sigma^{(s)}_{b_s}  V^{(s)}_{a'_s b_s}.
\end{equation}
From Eqs.~\eqref{eq:sqrt-d} and \eqref{eq:d-reg}, Eq.~\eqref{eq:csadj} becomes
\begin{equation}\label{eq:csadj-2}
  [\mathcal{C}^{(s)}_{m}]_{a_s a'_s} = \sum_{b_s} [\bar{D}^{(s)}_m]_{a_s b_s} \left(\sigma_{b_s}^{(s)}\right)^{-1}  V_{a'_sb_s}^{(s)}.
\end{equation}
The regularization is to replace
$\sigma_{b_s}^{(s)}$ that are less than $\epsilon$ by $\epsilon$. 
Hence, the equation of motion Eq.~\eqref{eq:su-eom2} becomes
\begin{equation}\label{eq:su-eom-main}
  \begin{multlined} 
  \dv{t} U^{(s)}_{a''_s \beta' \gamma'} \approx \\
  \sum_{m} \sum_{b_s a'_s a_s \beta \gamma}  [\bar{D}^{(s)}_m]_{a'_s b_s} \left(\max(\sigma_{b_s}^{(s)}, \epsilon)\right)^{-1}  V_{a''_sb_s}^{(s)} \\ 
   \times\left([F_m^{(s2)}]_{\beta'\beta} [F_m^{(s3)}]_{\gamma'\gamma} U^{(s)}_{a'_{s}\beta\gamma}  - U^{(s)}_{a_{s}\beta'\gamma'} [f_m^{(s)}]_{a_{s}a_{s}'}  \right) 
   ,
  \end{multlined}
\end{equation}
and the multiplicative inverse is now always achievable.
The key aspect of this regularization is that it introduces an error of $\order{\epsilon^{2}}$ in $D^{(s)}(t)$ [\cf\ Eq.~\eqref{eq:sqrt-d}].

\subsubsection{Projector-splitting propagator}
 
Another branch of propagation method is based on the so-called projector-splitting (PS) technique. These techniques avoid the errors introduced by the regularization, but introduce trotterization errors inherent to the approach. In a PS algorithm, instead of propagating all the core tensors at the same time as in the direct integration, the dynamics of each tensor is propagated individually and sequentially. 

The details of this algorithm, and proof of their validity, are discussed in the studies of tensor train and tensor tree\cite{Haegeman2016, Kloss2017, Lubich2015, Lubich2018, Lindoy2021a, Lindoy2021b} in the context of time-evolution of the matrix product state for a wavefunction.
The generalization of a one-site version of this algorithm (PS1) to HEOM with tree tensor network can be found in Ref.~\onlinecite{Ke2023}. 
Here we outline these algorithms and how they are used in TTN-HEOM, and generalize the two-site version (PS1) of this algorithm to the TTN-HEOM. PS1 is a static-rank method fixed memory algorithm where the rank is constant during propagation. In turn, the PS2 that we generalize is a dynamic-rank method that updates the rank in the TTN to achieve a target propagation accuracy.

The formal solution of the master equation $\dv{t} \Omega(t) = \mathcal{L}(t) \Omega(t)$ is $\Omega(t+\Delta) = e^{\Delta \mathcal{L}(t)}\Omega(t)$ for a small time step $\Delta$. In a Trotterization scheme in PS, $\mathcal{L}(t)$ is split into $\mathcal{L}(t) = \sum_{i=1}^{I_{\text{max}}}\mathcal{P}_i\mathcal{L}(t)$. 
The Trotter propagator is $\Omega(t+\Delta) \approx e^{\Delta\mathcal{P}_{I_{\text{max}}}\mathcal{L}(t)} \cdots e^{\Delta\mathcal{P}_1\mathcal{L}(t)}\Omega(t)$ to first order in $\Delta$, or $\Omega(t+\Delta) \approx e^{\frac{\Delta}{2}\mathcal{P}_1\mathcal{L}(t)} \cdots e^{\frac{\Delta}{2}\mathcal{P}_{I_{\text{max}}}\mathcal{L}(t)}e^{\frac{\Delta}{2}\mathcal{P}_{I_{\text{max}}}\mathcal{L}(t)}\cdots e^{\frac{\Delta}{2}\mathcal{P}_1\mathcal{L}(t)}\Omega(t)$ to second order in $\Delta$.
We employ the second Trotter where each time step is divided into a forward step  in the splitting of $\mathcal{L}$, $e^{\frac{\Delta}{2}\mathcal{P}_{I_{\text{max}}}\mathcal{L}(t)}\cdots e^{\frac{\Delta}{2}\mathcal{P}_1\mathcal{L}(t)}$, followed by a backward step in such splitting $e^{\frac{\Delta}{2}\mathcal{P}_1\mathcal{L}(t)} \cdots e^{\frac{\Delta}{2}\mathcal{P}_{I_{\text{max}}}\mathcal{L}(t)}$.  We denote each $e^{\tau\mathcal{P}_i\mathcal{L}(t)}$ as a split-step with a time $\tau$.

In the PS method, at each split-step, the TTN is transformed such that the propagation is always on the root tensor, for which the dynamics does not have the singularity issue. 
[\cf\ Eq.~\eqref{eq:root-eom}].
For this, one needs to construct $f_m^{(s)}$ [\cf\ Eq.~\eqref{eq:f-recursive}] during the propagation.
Suppose at a split-step that the root is now located at $A^{(r)}$, then the master equation is
\begin{equation}\label{eq:root-eom2}
  \begin{multlined}
    \dv{t} A^{(r)}_{\alpha' \beta' \gamma'} = 
  \sum_{m}  \sum_{\alpha \beta \gamma}  [F_m^{(r1)}]_{\alpha'\alpha}  [F_m^{(r2)}]_{\beta'\beta} [F_m^{(r3)}]_{\gamma'\gamma} A^{(r)}_{\alpha\beta\gamma},
  \end{multlined}
\end{equation}
where $F_m^{(r\kappa)}$ acts on the $\kappa$-th index of $A^{(r)}$.
The value of $F_m^{(r\kappa)}$ is $h^{>}_{m}$ if the $\kappa$-th index of $A^{(r)}$ is $i$,  $h^{<}_{m}$ if it is $j$, 
$h^{(k)}_{m}$ if it is $n_k$,  and $f^{(s)}_{m}$ if it is $a_s$.  
The transformation of the TTN moves the location of the root tensor from its current location to one adjacent semi-unitary tensor via additional SVD. 
In one step of PS propagator, we start from the original root and travel over every core tensor in the TTN. 
Once all core tensor in the TTN have been updated, the algorithm then returns the root tensor to its original position and proceeds to take another dynamical time step.

In the rest of this section, Einstein summation convention is assumed.

\paragraph{PS1 algorithm.}
The PS1 algorithm we implemented is in a second-order Trotter propagator form.
The key of the algorithm is to find a round-trip path over the whole tensor tree such that each contracted \emph{bond} in the tree is traveled exactly two times. 
This can be done by the depth-first-search algorithm\cite{Tarjan1972} from the root $A^{(0)}$.
We first travel over the tree: start from the root $A^{(0)}$, go pass every closed bond twice, and return to the origin $A^{(0)}$.
The forward path is a sequence $P = (A^{(0)},\ U^{(1)},\ \ldots,\ U^{(K-1)},\ \ldots,\ U^{(1)},\ A^{(0)} )$.
We propagate the whole TTN by $\Delta/2$ when we travel along the the forward path.
After that we use the reversed sequence of the forward path $P$ as the backward path to propagate another $\Delta/2$ to finish one step of propagation for the whole TTN. 
We use $P[i]$ to represent the core tensor at the $i$-th location of $P$, and range of $i = 1,\ \ldots,\ 2K-1$ for the order-$3$ TTN as in Eq.~\eqref{eq:ttn-main}.

Suppose that in the TTN the root tensor is $A^{(r)}$ and one of its neighbor $U^{(s)}$.
The one-site move function $\texttt{move1}(r, s, \tau)$ with a split-step time $\tau$ is showed in Algorithm~\ref{alg:move1}.

\begin{algorithm}[tb]
\DontPrintSemicolon
\cc{Without loss of generality, suppose that the indexes in $A^{(r)}$ and $U^{(s)}$ are $A^{(r)}_{\alpha\beta a_s}$ and $U^{(s)}_{a_s \gamma\varepsilon}$.}
\ShowLn
Perform the SVD of $A^{(r)}_{\alpha \beta a_s } = W_{\alpha \beta b} \sigma_{b} V_{a_s b}^{\star}$.
\label{step:svd1}
\;
Let $U_{\alpha \beta a_r}^{(r)} \leftarrow W_{\alpha \beta a_r}$.\;
Let $[f_{m}^{(r)}]_{a_{r}'a_{r}} \leftarrow 
    U^{(r)\star}_{\alpha\beta a'_{r}}
    [F_{m}^{(r1)}]_{\alpha'\alpha}
    [F_{m}^{(r2)}]_{\beta'\beta}
    U^{(r)}_{\alpha\beta a_{r}}$.\;
Let $M_{a_r a_s} \leftarrow\sigma_{a_r} V_{a_s a_r}^{\star}$.\;
\ShowLn Propagate $M$ by $\tau$ using the master equation $$\dv{t} M_{a'_r a'_s} = [f_{m}^{(r)}]_{a_{r}'a_{r}} [f_{m}^{(r)}]_{a_{r}'a_{r}} M_{a_r a_s}.$$\label{step:p1}\;
Let $A^{(s)}_{a_r \gamma \varepsilon} \leftarrow M_{a_r a_s} U^{(s)}_{a_s \gamma \varepsilon}$.
\;
Delete $A^{(r)}$, $U^{(s)}$ and $f_{m}^{(s)}$.
\;
\caption{One-site move function \texttt{move1}.}
\label{alg:move1}
\end{algorithm}

This algorithm generates   
$f_{m}^{(r)}$, a semi-unitary $U^{(r)}$ such that
\begin{equation}
    U^{(r)\star}_{\alpha \beta a'_r} 
    U^{(r)}_{\alpha \beta a_r} =\delta_{a'_r a_r}, 
\end{equation}
and the new root tensor $A^{(s)}$.
For the specific case when the time step $\tau=0$ (which is needed below) it is equivalent to skip the propagation at line~\ref{step:p1} in Algorithm~\ref{alg:move1}.

The iterative PS1 algorithm for the forward step in the splitting of $\mathcal{L}$ is showed in Algorithm~\ref{alg:ps1-fw}, and the backward one in Algorithm~\ref{alg:ps1-bw}.

\begin{algorithm}[tb]
\DontPrintSemicolon 
\For{$i \leftarrow 1,\ 2,\ \ldots,\ 2K-2$}{
    Suppose $P[i]$ is $A^{(r)}$, and $P[i+1]$ is $U^{(s)}$.
    \;
    \cc{Compare the heights of $A^{(r)}$ and $U^{(s)}$.}
    \uIf{${L}(A^{(r)}) < L(U^{(s)})$}{
        Call $\texttt{move1}(r, s, 0)$ to get $U^{(r)}$ and $A^{(s)}$.\;
    }
    \Else{
         Propagate $A^{(r)}$ by $\frac{\Delta}{2}$ using Eq.~\eqref{eq:root-eom2}.\;
         Call $\texttt{move1}(r, s, -\frac{\Delta}{2})$ to get $U^{(r)}$ and $A^{(s)}$.\;
    }
    Let $P[i] \leftarrow U^{(r)}$, and $P[i+1] \leftarrow A^{(s)}$.\;
}
Propagate $A^{(0)}$ by $\frac{\Delta}{2}$ using Eq.~\eqref{eq:root-eom2}.\;
\caption{Forward step of PS1.}
\label{alg:ps1-fw}
\end{algorithm}

\begin{algorithm}[tb]
\DontPrintSemicolon
Propagate $A^{(0)}$ by $\frac{\Delta}{2}$ using Eq.~\eqref{eq:root-eom2}.
\;
\For{ $i \leftarrow 2K-1,\ 2K-2,\ \ldots,\ 2$ }{
    Suppose $P[i]$ is $A^{(r)}$, and $P[i-1]$ is $U^{(s)}$.
    \;
     \uIf{${L}(A^{(r)}) < L(U^{(s)})$}{
        Propagate $A^{(r)}$ by $\frac{\Delta}{2}$ using Eq.~\eqref{eq:root-eom2}.
        \;
        Call $\texttt{move1}(r, s, -\frac{\Delta}{2})$ to get $U^{(r)}$ and $A^{(s)}$.
        \;
    }
    \Else{
         Call $\texttt{move1}(r, s, 0)$ to get $U^{(r)}$ and $A^{(s)}$.
         \;
    }
    Let $P[i] \leftarrow U^{(r)}$, and $P[i-1] \leftarrow A^{(s)}$.\;
}
\caption{Backward step of PS1.}
\label{alg:ps1-bw}
\end{algorithm}

\paragraph{PS2 algorithm.} In the two-site PS algorithm (PS2), it is possible to dynamically update the rank, \ie, the dimensionality of the dynamical space, in the TTN by constructing 4-order tensors, propagating them and then decomposing back to the 3-order tensor structure. 
The forward steps and backward steps for PS2 are similar to those in PS1, but PS2 implements a two-site move of the root tensor in the split steps in addition to the one-site move.

Suppose that in the TTN the root tensor is $A^{(r)}$ and one of its neighbor $U^{(s)}$.
The two-site move function $\texttt{move2}(r, s, \tau)$ with a split-step time $\tau$ is showed in Algorithm~\ref{alg:move2}.

\begin{algorithm}[tb]
\DontPrintSemicolon
\cc{Without loss of generality, suppose the indexes in $A^{(r)}$ and $U^{(s)}$ are $A^{(r)}_{\alpha\beta a_s}$ and $U^{(s)}_{a_s \gamma\varepsilon}$.}
Let $M_{\alpha \beta \gamma\varepsilon} \leftarrow A^{(r)}_{\alpha\beta a_s} U^{(s)}_{a_s \gamma\varepsilon}.$
\;
\ShowLn 
Propagate $M$ by $\tau$ using the master equation
\begin{equation*}
\begin{multlined}
  \dv{t} M_{\alpha \beta \gamma\varepsilon} = \\
  [F_{m}^{(r1)}]_{\alpha'\alpha} [F_{m}^{(r2)}]_{\beta'\beta}
  [F_{m}^{(s2)}]_{\gamma'\gamma} [F_{m}^{(s3)}]_{\varepsilon'\varepsilon}
  M_{\alpha \beta \gamma\varepsilon}.
\end{multlined}
\end{equation*}
\label{step:p2} 
\;
\ShowLn
Perform the SVD of $M_{\alpha \beta \gamma\varepsilon} = W_{\alpha \beta b} \sigma_{b} V_{\gamma\varepsilon b}^{\star}.$
\label{step:svd2}\; 
Let $U^{(r)}_{\alpha \beta a_r} \leftarrow W_{\alpha \beta a_r}$.\;
Let $[f_{m}^{(r)}]_{a_{r}'a_{r}} \leftarrow 
    U^{(r)\star}_{\alpha\beta a'_{r}}
    [F_{m}^{(r1)}]_{\alpha'\alpha}
    [F_{m}^{(r2)}]_{\beta'\beta}
    U^{(r)}_{\alpha\beta a_{r}}$.\; 
Let $A^{(s)}_{a_r \gamma \varepsilon} \leftarrow \sigma_{a_r} V_{\gamma\varepsilon a_r}^{\star}$.\; 
Delete $A^{(r)}$, $U^{(s)}$ and $f_{m}^{(s)}$.
\;
\caption{Two-site move function \texttt{move2}.}
\label{alg:move2}
\end{algorithm}
 
In the SVD at line~\ref{step:svd2} in Algorithm~\ref{alg:move2}, to control the rank of $a_r$ after the move, in practice we use a truncated version of SVD such that the range of $b$ is only for those $\sigma_{b}>\epsilon'$ as
\begin{equation}\label{eq:truncated-svd}
    M_{\alpha \beta \gamma\varepsilon} \approx 
    \sum_{\substack{b\ \text{s.t.}\\ \sigma_b\ge\epsilon'}}
    W_{\alpha \beta b} \sigma_{b} V_{\gamma\varepsilon b}^{\star}.
\end{equation}
where $\epsilon'$ is parameter that controls the error in the truncate SVD.
Similarly to $\mathtt{move1}$, for the specific case of time step $\tau=0$ it is equivalent to skip the propagation at line~\ref{step:p2} in Algorithm~\ref{alg:move2}.

The iterative PS2 algorithm for the forward step in the splitting of $\mathcal{L}$ is showed in Algorithm~\ref{alg:ps2-fw}, and the backward one in Algorithm~\ref{alg:ps2-bw}.

\begin{algorithm}[tb]
\DontPrintSemicolon 
\For{$i \leftarrow 1,\ 2,\ \ldots,\ 2K-2$}{
    Suppose $P[i]$ is $A^{(r)}$, and $P[i+1]$ is $U^{(s)}$.
    \;
    \uIf{${L}(A^{(r)}) < L(U^{(s)})$}{
         Call $\texttt{move1}(r, s, 0)$ to get $U^{(r)}$ and $A^{(s)}$.\;
    }
    \Else{
         Call $\texttt{move2}(r, s, \frac{\Delta}{2})$ to get $U^{(r)}$ and $A^{(s)}$.\;
         Propagate $A^{(s)}$ by $-\frac{\Delta}{2}$ using Eq.~\eqref{eq:root-eom2}.
    }
    Let $P[i] \leftarrow U^{(r)}$, and $P[i+1] \leftarrow A^{(s)}$.\;
}
Propagate $A^{(0)}$ by $\frac{\Delta}{2}$ use Eq.~\eqref{eq:root-eom2}.
\caption{Forward step of PS2.}
\label{alg:ps2-fw}
\end{algorithm}

\begin{algorithm}[tb]
\DontPrintSemicolon
Propagate $A^{(0)}$ by $\frac{\Delta}{2}$ use Eq.~\eqref{eq:root-eom2}.
\;
\For{ $i \leftarrow 2K-1,\ 2K-2,\ \ldots,\ 2$ }{
    Suppose $P[i]$ is $A^{(r)}$, and $P[i-1]$ is $U^{(s)}$.
    \;
     \uIf{${L}(A^{(r)}) < L(U^{(s)})$}{
        Propagate $A^{(r)}$ by $-\frac{\Delta}{2}$ using Eq.~\eqref{eq:root-eom2}.
        \;
        Call $\texttt{move2}(r, s, \frac{\Delta}{2})$ to get $U^{(r)}$ and $A^{(s)}$.
        \;
    }
    \Else{
         Call $\texttt{move1}(r, s, 0)$ to get $U^{(r)}$ and $A^{(s)}$.
         \;
    }
    Let $P[i] \leftarrow U^{(r)}$, and $P[i-1] \leftarrow A^{(s)}$.\;
}
\caption{Backward step of PS2.}
\label{alg:ps2-bw}
\end{algorithm}

The adaptive rank is dictated by our criterion in \eq{eq:truncated-svd} in PS2. This criterion is useful for the bulk of the dynamics. However, for trees that contain core tensors with all three bonds connected, the criterion needs modification at initial times.  This is because the tensor rank of $A^{(r)}$ is one for the initial HEOM state Eq.~\eqref{eq:root-init}. In Algorithm~\ref{alg:move2}, if  $A^{(r)}$ is of tensor rank one, then together with the semi-unitary property of $U^{(s)}$, the number of non-zero $\sigma_b$ in the SVD step \eq{eq:truncated-svd} is at most $1$, resulting the new rank $R_r$ to be fixed at $1$, no matter whether the propagation of time $\tau$ is done.

To address this challenge, for all times, in practice the range of $b$ is chosen to be twice the number that satisfies \eq{eq:truncated-svd} where the order of the SVD is chosen such that $\sigma_1> \sigma_2 > \cdots \ge 0$.

\subsubsection{Remarks}
In the same way that for wavefunction propagation there is no one propagation scheme that is better in all physical problems,\cite{GmezPueyo2018, Leforestier1991} we expect that for the TTN-HEOM the three proposed methods, direct integration, PS1 and PS2, will have specific regimes in which they have favorable properties. 
The advantage of PS propagators over direct integration is that it avoids the regularization error controlled by $\epsilon$. The disadvantage of the PS propagator is that it requires sequential SVDs during the dynamics which makes the algorithm more difficult to parallelize.\cite{Lindoy2021a, Lindoy2021b}
Further, since it individually propagates components of the tensor tree with a given time step $\Delta$, its propagation error is of $\order{\Delta^3}$ which is comparable to Trotter error.  
By contrast, for a Runge-Kutta method of order $n$ the direct integration can compute with the integration error $\order{\Delta^n}$.

Both direct integration and PS1 are limited by the assumption that the complete dynamics can be described by the TTN with a given rank.
By contrast, in PS2 the ranks are variable during the dynamics from a truncated SVD of 
controlled by an error of $\epsilon'$, which make it possible to change the size of TTN accordingly.

\subsection{Implementation considerations and capabilities}\label{sec:code}

We implemented the TTN-HEOM in a Python package, Tensor Equations for Non-Markovian Structured Open systems (\PyName), using the popular \texttt{NumPy}\cite{Harris2020} and \texttt{PyTorch}\cite{Ansel2024} libraries for the tensor data structure and tensor operations, as well as \texttt{torchdiffeq}\cite{torchdiffeq} for integrating the quantum master equations for tensors.\footnote{Available at: {\url{https://github.com/ifgroup/pytenso}}}
These packages offer high-level protocols for ease of programming that are compatible with various computational platforms such as CPUs and GPUs of different architectures. Details of the implementation will be provided in subsequent publication, but here we describe some of its key elements for TTN-HEOM.

\PyName\ admits system's Hamiltonians with any level structure and arbitrary time dependence, making it of utility to investigate driven open quantum systems.  The \PyName\ implementation admits arbitrary order for the core tensors and arbitrary tree structure. As such, it goes  beyond the order-3 tensor equations discussed in Secs.~\ref{sec:ttn-main}--\ref{sec:propagation} above, and beyond tensor-train approaches to the HEOM.  In Supplementary Material we detail this generalization using the language of graphs.

The system-bath coupling can include any number of terms $H_\text{SB} = \sum_{d} Q^{(d)}_\text{S} \otimes X^{(d)}_\text{B}$ and the $\{ Q^{(d)}_\text{S}\}$ do not need to commute. Thus, \PyName\ can be used to investigate a system that is coupled to two or more environments through non-commuting operators, something that is computationally challenging to adopt in path integral-based transfer tensor strategies.

This package currently implements the three propagation strategies discussed in Sec.~\ref{sec:propagation}: direct integration of the quantum master equations with fixed ranks, and the step-wise projector-splitting propagator including PS1 with fixed ranks and PS2 for variable ranks during the propagation.

To run a simulation using \PyName, in the input one needs to specify the parameters $c_k,\ \bar{c}_k,\ \gamma_k$ in the decomposition of the BCF \eq{eq:bcfdec}. Any decomposition compatible with \eq{eq:bcfdec} can be used. For common spectral density models including Drude-Lorentz and underdamped Brownian oscillator, we have implemented a helper function to obtain these parameters in \eq{eq:bcfdec} using either a  Pad{\'{e}}\cite{Hu2010}  or Matsubara\cite{Ishizaki2005} expansion for the thermal factor $\coth(\omega/2\kB T)$. 
The helper function gives both the high-temperature terms from the model spectral densities, and arbitrary order of low-temperature corrections terms from the expansion from the thermal factor.
Our code is also compatible with other BCF decomposition strategies that yield the form in \eq{eq:bcfdec}.\cite{LeDe2024, Xu2022, Nakatsukasa2018, Takahashi2024, Lednev2024, Lambert2019, Potts2013, Chen2022} 

To make use of the flexibility of the TTN, in the input one can also specify the topology of a TTN with the open bonds corresponding to all system and bexciton indexes.
The topology that is chosen for the TTN will automatically determine the quantum master equations for the core tensors. 
The code admits as input a list of the nodes in the TTN and their connectivity to either open bonds or to other nodes. We implemented templates for automatically generating  the train topology as exemplified in Fig.~\ref{fig:tn}(a) and the balanced tree topology as exemplified in Figs~\ref{fig:tn}(b) and \ref{fig:tn}(c). However, \PyName\ admits as input any type of TTN with open ends $i,\ j,\ n_1,\ \ldots,\ n_K$, with any one of the core tensors specified as the root initially.

Each tree structure has a unique version of the quantum master equations Eq.~\eqref{eq:root-eom}--\eqref{eq:d-recursive-3}.  These equations have common quantities that, for computational efficiency, must be evaluated in a specified order to avoid duplication of efforts. 
The order in which the $\{f_{m}^{(s)}(t)\}$  are computed is based on the structure of the tree. 
We compute the $\{f_{m}^{(s)}\}$ from $s = K-1$ to $1$. 
This can be seen in Eq.~\eqref{eq:f-recursive}, which shows that $f_m^{(s)}$ only depends on $f_m^{(u)}$ with $u > s$. 
That is, the computation of $f_{m}^{(s)}$ proceeds from the leaves of the TTN (the nodes with open bonds $n_1$, \ldots, $n_K$) to the root. 

Further, in the direct integration propagation method with regularization, we need to evaluate $\{\bar{D}_{m}^{(s)}(t)\}$, as well as $\{\sigma^{(s)} (t)\}$ and $\{V^{(s)} (t)\}$, for integrating Eq.~\eqref{eq:su-eom-main}. In this case, we proceed from the root to the leaves.
That is, we go from $s = 1$ to $K-1$ to evaluate all $\bar{D}_{m}^{(s)}(t)$, $\sigma^{(s)}(t)$ and $V^{(s)}(t)$ for Eq.~\eqref{eq:su-eom-main}.
This can be seen from Eqs.~\eqref{eq:csadj-recursive-1}--\eqref{eq:csadj-recursive-3} as $\bar{D}_m^{(s)}$, $\sigma^{(s)} (t)$ and $V^{(s)} (t)$ only depends on the $\bar{D}_m^{(r)}$, $\sigma^{(r)} (t)$ and $V^{(r)} (t)$ with ${r < s}$.
In this way, we provide a systematic sequential procedure to travel through the TTN and construct the needed $f_m^{(s)}$, $\bar{D}_m^{(s)}$. For the direct integration, this needs to be performed whenever the derivative of the core tensors are computed. For PS, this needs to be performed before each forward step in the algorithm.

Each propagation step runs over the whole TTN of $\Omega(t)$ that includes the dynamical information of the system and the collection of bexcitons. 
The reduced density operator of the system $\rho_{\text{S}}(t)$ is calculated from the $\Omega(t)$ for output times as
\begin{multline}\label{eq:get-rdo}
  [\rho_{\text{S}}(t)]_{ij} =  \sum_{n_1\cdots n_K} \\
  [\mathsf{Con}(A^{(0)}(t),U^{(1)}(t),\cdots,U^{(K-1)})(t)]_{ijn_1\cdots n_K} \\ \times \delta_{0n_1} \cdots \delta_{0n_K}
\end{multline}
In practice, the expression of TTN in Eq.~\ref{eq:ttn-main} is substituted in \eq{eq:get-rdo}.
To avoid reconstructing the full high-order EDO, the contractions is first done for the $n_k$ indexes, and then from $a_{K-1}$ to $a_1$.

Specifically, this is  done by constructing a series of vectors $\mathfrak{t}^{(s)}$ from $s=K-1$ to $1$.
The recursive definition of $\mathfrak{t}^{(s)}$ is
\begin{equation}\label{eq:t-recursive}
    \mathfrak{t}^{(s)}_{a_{s}} \equiv \sum_{\beta\gamma}
    [\mathfrak{T}_{m}^{(s2)}]_{\beta}
    [\mathfrak{T}_{m}^{(s2)}]_{\gamma}
    U^{(s)}_{a_{s}\beta\gamma}.
\end{equation}
Here the definition of vector $\mathfrak{T}^{(s\kappa)}(t)$ for $\kappa=2$ and $3$ depends on the location of the tensor $U^{(s)}(t)$ in the TTN.
$\mathfrak{T}^{(s\kappa)}(t) \equiv \mathfrak{t}^{(u)}(t)$  ($u > s$) if the $\kappa$-th index corresponds to a contracted index or bond in the TTN. That is, when the $\kappa$-th index in $U^{(s)}(t)$ is an $a_u$ in Eq.~\eqref{eq:ttn-main}.
In turn, $\mathfrak{T}^{(s\kappa)}_{n_k}(t) \equiv \delta_{0n_k}$ when the $\kappa$-th index is an $n_k$ in Eq.~\eqref{eq:ttn-main}. That is, when it corresponds to an open bond in the TTN.
Notice that this definition is recursive as $\mathfrak{t}_{m}^{(s)}(t)$ depends on $\mathfrak{T}^{(s\kappa)}(t)=\mathfrak{t}^{(u)}(t)$ ($u>s$) if the $\kappa$-th index corresponds to a contracted bond [\cf\ Eq.~\eqref{eq:f-recursive}].
After the construction of $\mathfrak{t}^{(s)}(t)$, the reduced density operator of the system $\rho_{\text{S}}(t)$ is calculated as
\begin{equation}\label{eq:get-rdo-main}
  [\rho_{\text{S}}(t)]_{ij} = \sum_{a_1}  [A^{(0)}(t)]_{ija_1} [\mathfrak{t}^{(1)}(t)]_{a_{1}}. 
\end{equation}
In this way, the explicit evaluation of the full high-order EDO tensor is avoided.

The HEOM is a numerically exact method for a given decomposition of the BCF Eq.~\eqref{eq:bcf}.  However, by construction it does not guarantee positivity of the reduced density operator of the system and, in fact, negativities can occur when employing inaccurate BCFs.  Since the TTN-HEOM is a decomposed version of HEOM, it can be numerically exact but its overall accuracy will also be limited by the quality of the spectral density that is employed in the model. 

In contrast to  HEOM, TTN-HEOM can be efficiently employed with highly structured spectral density and with low-temperature corrections, as needed to perform computations in chemically realistic systems.
{Our efforts complement a tensor train implementation of the HEOM \cite{Guan2024}, and a recent ML-MCTDH software package with HEOM capabilities \cite{Lindoy2025}.}

\section{Numerical Example}\label{sec:results}

\subsection{Model}
To illustrate the TTN-HEOM, we consider a two-electronic surface molecular system described by a two-level model coupled to a structured thermal bath. 
In the Hamiltonian, the electronic system is
\begin{equation} 
\label{eq:qubit}
    H_{\text{S}} = \frac{E}{2} \left(\op{1}{1} - \op{0}{0}\right) + V \left(\op{1}{0} + \op{0}{1}\right),
\end{equation}
where $\ket{0}$ and $\ket{1}$ denote two diabatic electronic states, $E$ is the energy level difference between them and $V$ their electronic coupling. In turn, 
the system is coupled to the bath via
\begin{equation}
 Q_{\text{S}} = \frac{1}{2}\left(\op{1}{1} - \op{0}{0}\right).
\end{equation}
That is, the bath is assumed to introduce energy fluctuations between $\ket{0}$ and $\ket{1}$.
As an initial state, we take the system to be in a pure superposition state of the form $\ket{\psi_\text{S}} = \left(\ket{0} + \ket{1}\right) / \sqrt{2}$. 

\begin{figure}[tb]
    \centering
    \includegraphics[width=\linewidth]{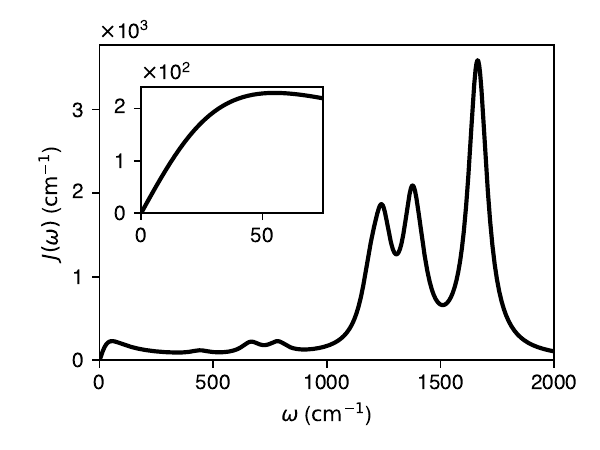}
    \caption{Bath spectral density $J(\omega)$ describing electron-nuclear interactions for thymine nucleotide in room temperature water\cite{Gustin2023} with broadenings $\gamma_b = 50~\invcm$ for each Brownian oscillator.}
    \label{fig:sd}
\end{figure} 

To characterize the system-bath interaction, in numerical exact simulation it is common to use simple model spectral densities, such as the Drude--Lorentz or the Brownian oscillator. As a computationally challenging example, 
in this paper we adopt the realistic spectral density\cite{Gustin2023} shown in Fig.~\ref{fig:sd} recently extracted from resonance Raman experiments for thymine nucleotide in room temperature water. This spectral density consists of one Drude--Lorentz component at low-frequencies that describes the solvent, and 8 Brownian oscillators at higher frequencies that represent the interaction of the electronic system with intramolecular vibrations.
That is, the bath spectral density is
\begin{equation}\label{eq:bath}
    J(\omega) = J_{\mathrm{DL}}(\omega) + \sum_{b=1}^{8} J_\mathrm{B}^{(b)}(\omega), 
\end{equation}
with $J_\mathrm{DL}(\omega) = \frac{2\lambda_0}{\pi} \frac{\gamma_0\omega}{\omega^2 + \gamma_0^2}$
and $J_\mathrm{B}^{(b)}(\omega)= \frac{4\lambda_{b}}{\pi} \frac{\gamma_{b} \omega_{b}^2 \omega}{(\omega^2 - \omega_{b}^2)^2 + 4\gamma_{b}^2 \omega^2}$.
Here, $\lambda_{0}$ is the reorganization energy of the solvent and $\gamma_{0}^{-1}$ its relaxation time. In turn, $\lambda_{b}$ is the reorganization energy of the $b$-th vibrational mode, $\omega_{b}$ its natural frequency, $\gamma_{b}^{-1}$ its lifetime,
and $\omega'_{b} = \sqrt{\omega_{b}^2-\gamma_{b}^2} > 0$ is its effective frequency under damping.  Spectral density parameters are listed in Table~\ref{tab:b}.

\begin{table}[tb]
\centering
\begin{tabular}{p{0.15\linewidth}p{0.25\linewidth}p{0.25\linewidth}p{0.25\linewidth}}
\toprule
$b$ & $\omega'_{b}$ ($\invcm$) & $\lambda_{b}$ ($\invcm$) & $\gamma_{b}$ ($\invcm$)\\
\midrule
0 & \phantom{0}{---} & 715.73 & 54.45 \\
1 & 1663 & 330.0\phantom{0} & 50\phantom{.00} \\
2 & 1416 & \phantom{0}25.6\phantom{0} & 50\phantom{.00} \\
3 & 1376 & 186.0\phantom{0} & 50\phantom{.00} \\
4 & 1243 & 161.7\phantom{0} & 50\phantom{.00} \\
5 & 1193 & \phantom{0}77.3\phantom{0} & 50\phantom{.00} \\
6 & \phantom{0}784 & \phantom{0}26.5\phantom{0} & 50\phantom{.00} \\
7 & \phantom{0}665 & \phantom{0}32.0\phantom{0} & 50\phantom{.00} \\
8 & \phantom{0}442 & \phantom{0}14.9\phantom{0} & 50\phantom{.00} \\
\bottomrule
\end{tabular}
\caption{Parameters in the spectral density \eq{eq:bath} for characterize the bath for thymine nucleotide in water at $300$~K. Parameters are taken from Ref.~\onlinecite{Gustin2023}.}
\label{tab:b}
\end{table}

\subsection{Tensor Tree and Bexcitonic Choices}

For the TTN, we use either a balanced binary tree or a tensor train, both of them containing order-3 core tensors only.  The balanced tree structure, in particular, minimizes the average distance between the index of each bexciton $n_k$ and the indexes of the system $i,\ j$.
To obtain the correct thermal state, we include $3$ low temperature correction terms from the Pad\'e expansion to evaluate Eq.~\eqref{eq:bcfdec}.  This results in overall $20$ bexcitons in the HEOM, and the resulting balanced tensor tree structure shown in Fig.~\ref{fig:tn}(c).  
Since the index $k$ for different terms in the BCF decomposition Eq.~\eqref{eq:bcfdec} is arbitrary, the correspondence of $n_k$ to different part of the spectral density is not unique. 
Here we choose $n_1$ to correspond to the high-temperature Drude-Lorentz, $n_2$--$n_{17}$ to the high-temperature Brownian oscillators, and $n_{18}$--$n_{20}$ to the overall low temperature corrections.  
The Brownian oscillators are sorted in descending order of their frequencies.
Each Brownian oscillator requires two bexcitons to be described, while the Drude-Lorentz feature requires just one.
As a metric in Eq.~\eqref{eq:dso}, we employ $\hat{z}_k = \iu \sqrt{\operatorname{Re}c_k}$.

\subsection{Open quantum dynamics of the model}

To test the performance of TTN-HEOM under different system settings, we set $V = 1000~\invcm$ and change the energy gap $E$ from $0$ to $5000~\invcm$. We monitor the dynamics through the population of state $\ket{0}$, $[\rho_{\text{S}}]_{00}$, and the purity $\Tr (\rho_{\text{S}}^2)$ which is a basis-independent measure of coherence (\ie, purity $= 1$ for pure system, $<1$ for mixed states, and $1/2$ for a maximally mixed two-level system).

Fig.~\ref{fig:ps2} shows the converged purity and $[\rho_{\text{S}}]_{00}(t)$ dynamics for the two-surface molecules with varying $E$. The system undergoes an initial decay of purity due to interaction with the bath until it reaches a minimum around 0.5. Subsequently, the purity recovers as the system relaxes to thermal equilibrium. For early-times, the decay of purity is Gaussian and independent of the details of the system Hamiltonian. In agreement with the theory of early-decoherence time scales,\cite{Gu2017,Gu2018b} this segment of the dynamics just depends on the initial-time quantum and thermal fluctuations of the operators coupling the system and bath. 
The subsequent  purity oscillations are due to the population transfer between $\ket{0}$ and $\ket{1}$, which are beyond the short-time limit.  For longer times $t > 100$~fs,  the purity and population oscillate as the system relaxes to thermal equilibrium. These deviations from exponential dynamics are clear signatures of non-Markovian open quantum dynamics that persist even for long times for this highly structured bath.

These results demonstrate that the TTN-HEOM can capture the numerically exact open quantum dynamics of systems interacting with highly structured thermal environments. {We further note that the TTN-HEOM and HEOM yield identical results. While HEOM computations for highly structured environments like those in \fig{fig:ps2} are not tractable, we numerically illustrate in Fig.~S4 in the Supplementary Material the coincidence between TTN-HEOM and HEOM using only the Drude--Lorentz component in the bath spectral density.}

\subsection{Propagator choice}

\begin{figure}[tb]
    \centering
    \includegraphics[width=\linewidth]{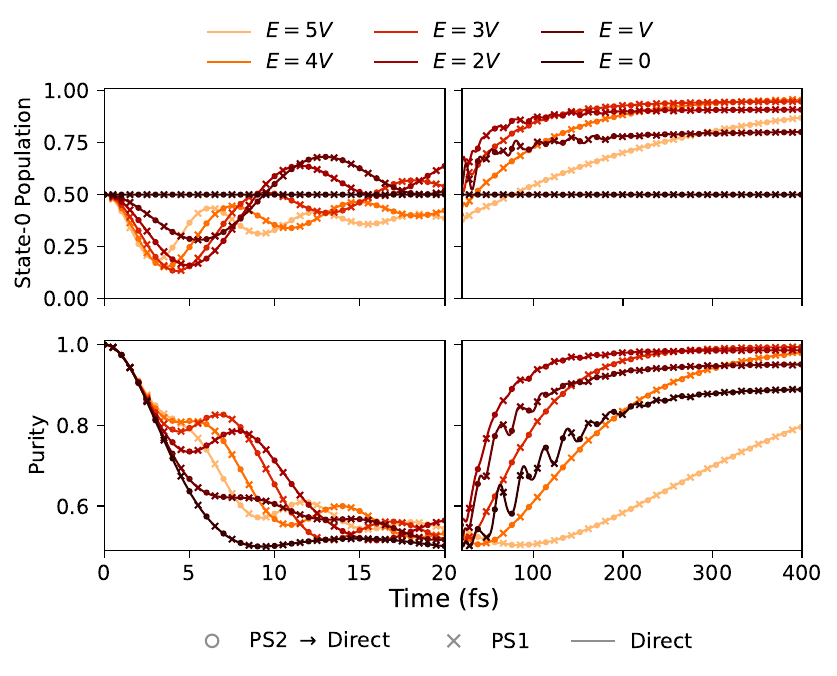}
    \caption{\textbf{TTN-HEOM dynamics captured by different propagation methods} for the two-level system in \eq{eq:qubit}  interacting with a highly structured thermal environment (Fig.~\ref{fig:sd}) using the balanced tree in Fig.~\ref{fig:tn}c. Different colors denote varying model parameters.   
    The direct (solid line) and PS1 (crosses) integration use a rank of $60$ for all tensors. 
    The mixed propagator (PS2 $\to$ direct, circles) starts with PS2 propagator with an initial rank $3$ and switches to direct integration when the adaptive rank grows beyond $60$. Note that the convergence and stability of the dynamics for all model parameters and integrators. 
    }
    \label{fig:ps2}
\end{figure}

Figure~\ref{fig:ps2} shows that using \PyName\ we can obtain identical dynamics with the three implemented propagation strategies. For the PS1 and direct integration, converged results are achieved with a moderate rank $R=60$ for all $R_k$ and a depth $N = 20$ for all $N_k$. 
For the direct integration, the integration of all core tensors is calculated simultaneously with a regularizing parameter $\epsilon=10^{-4}$ using the RK4(5) method, which allows for adaptive time step $h$ during the propagation. 
This method is of $\order{h^4}$ with an error estimator of order $\order{h^5}$ used to determine the integration time step $h$.
For the PS1 and PS2 method, a fixed time step $\Delta$ of $0.1$~fs is applied for splitting the propagation as described in  Algorithms~\ref{alg:ps1-fw}--\ref{alg:ps1-bw} and \ref{alg:ps2-fw}--\ref{alg:ps2-bw}, while the integration of each low-order tensor is calculated by RK4(5). 
Further in PS2 method the truncated SVD is done with an $\epsilon'=10^{-7}$.
Here in all RK4(5) integrator the relative error tolerance is $10^{-5}$ and the absolute error tolerance is $10^{-7}$.

The direct integration strategy offers a practical approach for propagating the bulk of the dynamics.  However, it is numerically challenging in the initial stage ($<2$~fs)  requiring extremely small time steps ($< 0.0001$~fs).  This is because we start from an initially separable system-bath state, which requires regularization to remove the singularity issues in evaluating Eq.~\eqref{eq:su-eom}.  This regularization introduces a small artificial error when this singularity occurs, and affects the stability and accuracy of the numerical integration.  Once this initial stage is overcome, the matrix $D^{(s)}$ becomes numerically invertible as all the eigenvalues $\lambda_i$ in matrix $D^{(s)}$ are greater than zero. That is, if the the regularization constant $\epsilon$ satisfies $\epsilon \le \min_i \sqrt{\lambda_i}$, then the regularization scheme \eq{eq:su-eom-main} becomes a numerical exact method to calculate the inverse of $D^{(s)}$ in \eq{eq:su-eom}.

PS1 is a robust strategy to propagate the TTN-HEOM and a common choice for tensor network methods. The main challenge is that it incurs in Trotterization errors of $\order{\Delta^3}$ in addition to the integration errors within each split-step in $\order{h^4}$ with the actual integration time $h \le \frac{\Delta}{2}$.  As direct integration, PS1 requires a list of initial ranks to capture the entanglement between different core tensors and the convergence with rank requires performing repeated calculations.

In turn, in PS2 the ranks change adaptively during propagation starting from an initial given rank for each contracted bond. The algorithm has the advantage of adapting the ranks as needed to accurately capture the dynamics and, thus, has variable memory requirements. Specifically, the ranks change such that the error introduced in the SVD \eq{eq:truncated-svd} is consistent with the control parameter $\epsilon'$.  These ranks change in a non-uniform fashion as the rank of some bonds can be larger than others. Our PS2 propagation starts with minimal ranks. Thus, these ranks initially grow using PS2 but, eventually, as the dynamics progresses can also decrease.  
Overall, for a TTN with order-$M$ core tensors, the PS2 contains the propagation of $2M-2$ tensors, which is of higher-order computational complexity.

These three methods can be combined on-the-fly to construct strategies that leverages their strengths and overcomes their limitation. There is significant flexibility in combining them as they only require the state of the TTN at the specific propagation time.
For instance, one straightforward PS2$\to$direct strategy is to use the PS2 at initial times followed by direct integration. This mixed strategy has the benefit of determining the proper requirement for the ranks for each bond from the early-time dynamics and avoiding the initial singularity in TTN-HEOM that limits the direct propagation strategy. 
Once the required computational resource requested by PS2 exceed a threshold level, one switches to the direct integration method that has the advantage of allowing integration with higher order adaptive time step methods compared to the order of Trotterization errors in PS1 and PS2. 

\begin{figure}[tb]
    \centering
    \includegraphics[width=\linewidth]{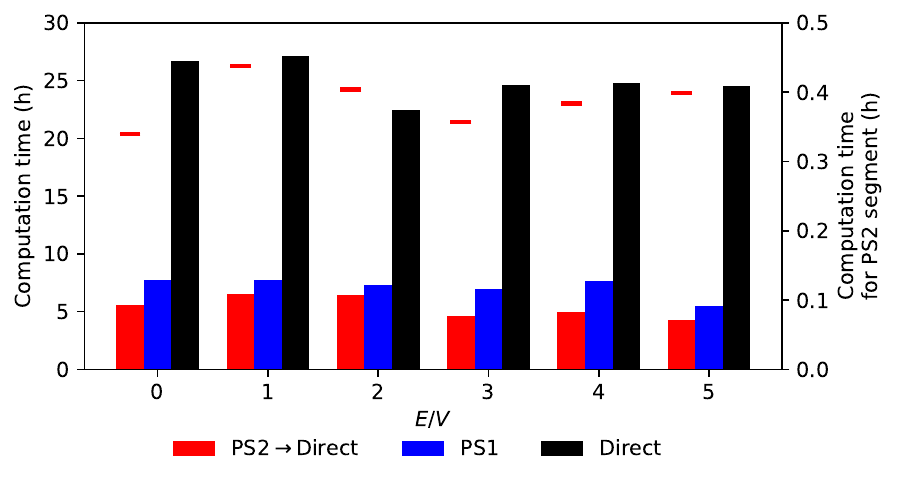}    
    \caption{\textbf{Computation time of each propagation strategy} in Fig.~\ref{fig:ps2} on 8 cores of an Intel Xeon Gold 6330 Processor. 
    Each bar shows the computation time of different propagation strategies for the initial 100~fs of dynamics
    (left axis). 
    The dashes mark the computation time for the PS2 segment ( $\sim 2$~fs) of the PS2 $\to$ direct method (right axis).}
    \label{fig:time}
\end{figure}

Results from this mixed PS2$\to$direct strategy  approach are also shown in Fig.~\ref{fig:ps2}.  In this case, all core tensor in the TTN start with a rank of $3$ and use the PS2 propagator until the maximum rank in the TTN reaches $60$.  After that, the remaining dynamics are propagated using direct integration. 
We find that, compared to the direct integration with regularization and PS1 with all ranks to be the same, the  mixed PS2$\to$direct strategy can achieve the converged results with reduced computational resources. For instance, \fig{fig:time} shows computation time of each of the propagation strategies in \fig{fig:ps2} for the initial 100~fs of dynamics (ran on 8 cores of an Intel Xeon Gold 6330 Processor). By adopting the PS2 for the initial propagation, the mixed strategy with direct integration achieves the best performance, while still providing numerically converged results. The reason for this is that the effective simulation space identified by PS2 where at least one bond has a rank of 60 is smaller than the one used for PS1 and direct where all bonds have a rank of 60.

For the chosen propagation parameters,  PS1 is actually faster than direct integration with the same size of ranks in the TTN.
This is because the main computational effort in the direct integration is in evaluating the $\{f_m^{(s)}\}$ and $\{\bar{D}_m^{(s)}\}$, which requires a series of sequential SVDs that are not parallelized. This is similar to the sequential SVDs required in the PS1. However, the difference is that, while in PS1 the $f_m^{(s)}$ are evaluated only once during Trotterization time $\Delta$, in the direct integrator the $f_m^{(s)}$ and $\bar{D}_m^{(s)}$ are constructed whenever the derivatives of all core tensors are requested by the high-order numerical integrator and this occurs several times per integration time step $h<\Delta$. 
Therefore, the direct integrator is slower than the PS1 for the same size of TTN as it contains more matrix construction steps.
However, if we use the mixed strategy where the most part ($>2$~fs) of the dynamics is obtained by the direct integration, it actually becomes faster than PS1. This is because the direct integrator inherits the non-uniform ranks from the initial PS2 segment that reduces the overall simulation space.

\begin{figure}[tb]
    \centering
    \includegraphics[width=\linewidth]{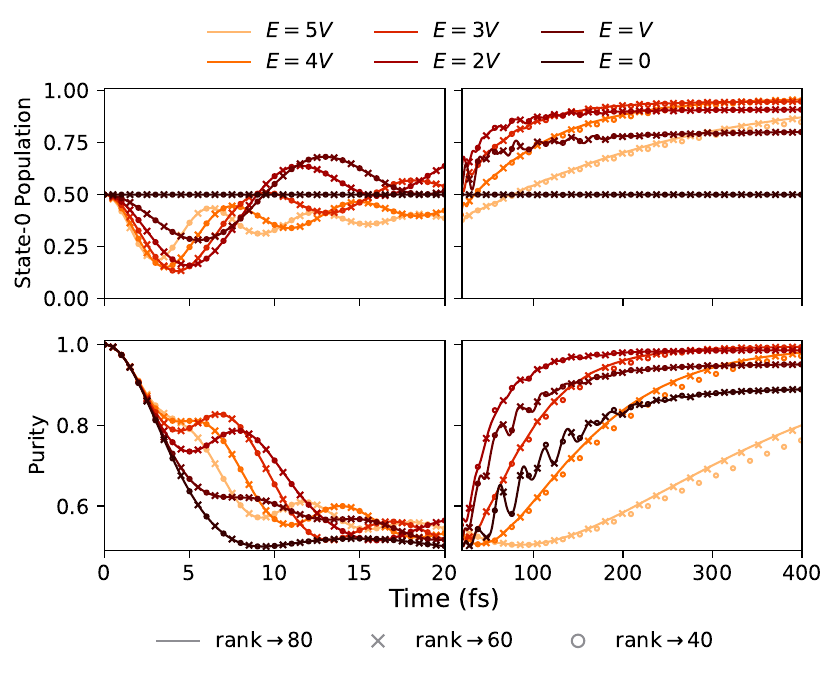}
    \caption{\textbf{TTN-HEOM dynamics for varying maximum rank thresholds} for the system and TTN in Fig.~\ref{fig:ps2} using the mixed propagator (PS2 $\to$ direct). The propagator starts with PS2 with an initial rank $3$ and switches to direct when one of the adaptive ranks grows beyond $80$ (solid lines), $60$ (crosses) and $40$ (circles). }
    \label{fig:ps2-diff-rank}
\end{figure}

\Fig{fig:ps2-diff-rank} shows the effect of adjusting the rank threshold in the mixed propagation strategy.  The other parameters are the same as in Fig.~\ref{fig:ps2}.  The figure shows that for the first $100$~fs convergence is achieved with a rank limit of $40$, while longer dynamics require a rank of $60$.  
As expected, increasing the accuracy of the TTN by increasing the ranks achieves longer converged dynamics.
The maximum rank needed for convergence depends on the system Hamiltonian, revealing that a larger system energy gap demands a higher rank for convergence. We hypothesize that as the gap increases the influence of the bath becomes increasingly Markovian as revealed by the reduction of the partial purity oscillations for long times. However, this Markovian limit is more challenging to capture for this numerically exact non-Markovian TTN-HEOM method.

\subsection{Tree structure choice}

Different TTN structures are expected to influence the computational cost of the TTN-HEOM dynamics and effectiveness of the method to compress the open quantum dynamics. However, given a tree structure, it is challenging to optimize the ranks for efficient computation. The PS2 strategy has the advantage of automatically adapting the compression in different components of the tree to satisfy the criteria \eq{eq:truncated-svd}.

\begin{figure}[tb]
    \centering
    \includegraphics[width=\linewidth]{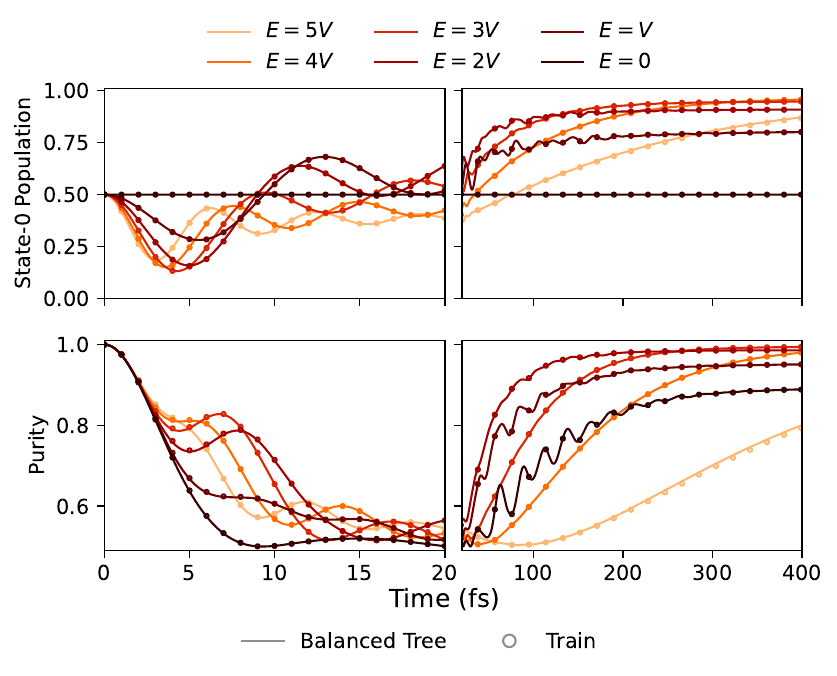}
    \caption{\textbf{Balanced Tensor Tree \emph{vs.}\ Tensor Train TTN-HEOM dynamics} for the system in Fig.~\ref{fig:ps2} using the mixed strategy (PS2 $\to$ direct) with maximum rank threshold $80$. Note that the dynamics is independent of the tree structure.
    }
    \label{fig:ps2-ttn}
\end{figure}

To investigate the influence of the tree structure on the computational resources requested by the PS2$\to$direct propagation, we performed TTN-HEOM simulations for the same model as in \fig{fig:ps2} using this mixed propagation strategy using a tensor tree and tensor train. The tensor train scheme as that shown in \fig{fig:tn}(a) with $K = 20$, and the tree scheme that shown in \fig{fig:tn}(c).
\Fig{fig:ps2-ttn} shows that, as expected, the open quantum dynamics is independent of the TTN employed.

\begin{figure}[tb]
    \centering
    \includegraphics[width=\linewidth]{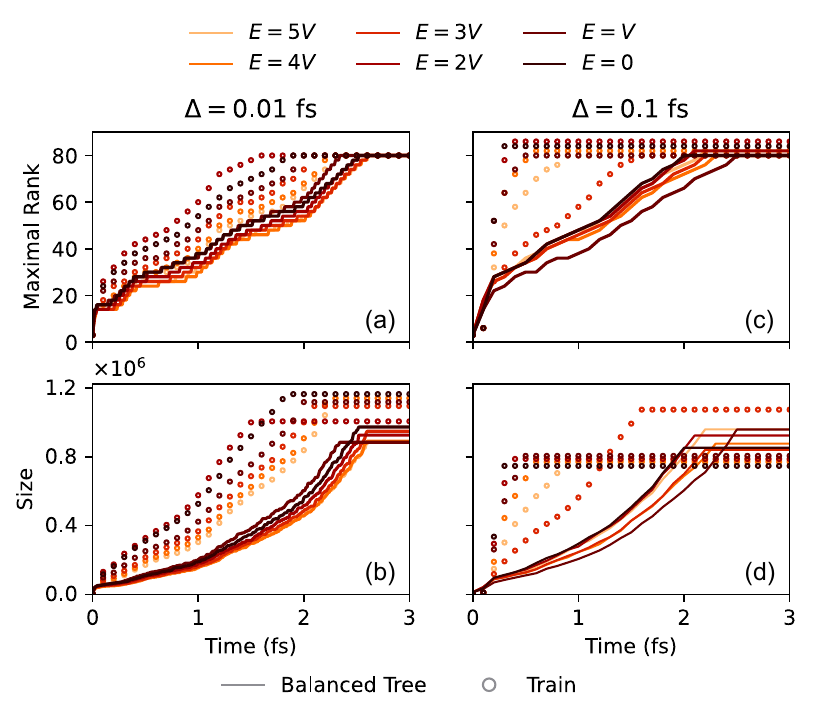}
    \caption{\textbf{Maximal rank and size of the TTN during PS2$\to$direct propagation} with the maximum rank threshold $80$ for a tensor train (circles) and a tensor tree (solid lines) for the dynamics in \fig{fig:tn}(c).  TTN size refers to the overall number of core tensor elements. In (a-b) the splitting time step $\Delta$ is $0.01$~fs while in (c-d) $0.1$~fs.  
    Note that tensor train size grows faster that the tensor tree scheme under the same SVD tolerance ($10^{-7}$) in \eq{eq:truncated-svd}.  Although the rank required from PS2 grows rapidly, switching to direct integration with fixed rank does not introducing appreciable errors.}
    \label{fig:size}
\end{figure}

\Fig{fig:size} shows the growth of the maximal rank and size (overall number of core tensor elements) of the TTN with these two TTN structures for the dynamics in \fig{fig:ps2-ttn}. The maximal rank increases during the PS2 propagation until it satisfies the threshold and changes to the direct method of fixed rank. Overall, the size of the TTN in the tensor train (\fig{fig:tn}(a)) grows faster than that in the tree scheme (\fig{fig:tn}(c)), when applying the same error tolerance in the propagation algorithm.
This is because in the HEOM, the primary entanglements in the tensor occur between the system and each bath feature.  By employing the tensor tree with the system DOFs at the root and the bath features at the leaves, one minimizes the average distance in the TTN between the system and each bath feature to $\order{\log{K}}$.  In turn, for the tensor train the average distance is $\order{K}$. 
{This offers an example where balanced tensor tree are better suited for TTN-HEOM from a computational cost perspective, which may because it keeps strongly correlated parts in the TTN closer to one another \cite{Murg2015}. Our TTN includes all possible tensor tree topological structure, with the balanced TTN and the tensor train being two extreme particular cases.
We expect that the ``optimal'' TTN structure should sit in between these two extreme cases. }

Finally, we point out that simply directly storing the EDO needed in these simulations is just not possible using present-day and foreseeable computational resources. Table~\ref{tab:mem} lists the size of a TTN with fixed rank and compares it to the size of the EDO in conventional HEOM with 20 features and a depth of 20 for each feature. 
Storing a {dense} tensor with that size of $4.2 \times 10^{26}$ would require $6.7$ ronnabytes ($10^{27}$) of memory!
{Here the estimation of the dense HEOM reflects the whole uncompressed Hilbert space dimension.
In reality, from practical experience one can use other numerical techniques such as filtering out  near-zero elements \cite{Shi2009}, or use the standard truncation by total depth instead of the depth of each bexciton to shrink the active dynamical space in HEOM more aggressively.
The depth of 20 is a conservative choice for this case study, and the greatest lower bound of such truncation depths needs to be determined by actual computations or experiences on specific physical model.}
Therefore, either tensor network or other techniques are necessary for practical simulation based on current classical computers.

\begin{table}[tb]
\centering
\begin{tabular}{lccccccc}
\toprule
& \multicolumn{3}{c}{Balanced Tree} & \multicolumn{3}{c}{Train} & Dense HEOM \\
\midrule
Rank & \hspace{1.2ex}40\hspace{1.2ex} & \hspace{1.2ex}60\hspace{1.2ex} & \hspace{1.2ex}80\hspace{1.2ex} & \hspace{1.2ex}40\hspace{1.2ex} & \hspace{1.2ex}60\hspace{1.2ex} & \hspace{1.2ex}80\hspace{1.2ex}  &  ---    \\ 
Size ($\times 10^6$) &  0.7  &  2.2  &  4.9  & 0.6 & 1.3 & 2.3 & 4.2$\times10^{20}$\\
\bottomrule
\end{tabular}
\caption{The size (overall number of core tensor elements) of TTN for a bath described by 20 features and a depth of $20$ for each $N_k$. The memory usage of a dense high-order EDO tensor in HEOM is also showed for comparison. }
\label{tab:mem}
\end{table}

\section{Conclusion}\label{sec:con}

In conclusion, we introduced TTN-HEOM, a {numerically exact} quantum master equation method based on the bexcitonic hierarchical equations of motion (HEOM) and a tree tensor network (TTN) decomposition. TTN-HEOM is designed to capture the dynamics of driven open quantum systems interacting with structured bosonic thermal environments even those of chemical complexity. 

{The specific advances of this paper are as follows:  (1) We introduced a tensor network decomposition of the HEOM based on the bexcitonic generalization. As such, the proposed TTN decomposition applies to all HEOM variants that can be cast into the general bexcitonic form ---including the standard HEOM with and without scaling, and the collective bath coordinate method--- and admits a representation of the bexcitons in number, position or momentum basis.  (2) We showed that the bexcitonic equations of motion can naturally be expressed in sum-of-product form, that the bexcitonic density operator can be decomposed by a tree tensor network, and that a useful set of coupled master equations can be developed for the low-order tensors from Dirac--Frenkel's TDVP. Our developments are analogous to the ML-MCTDH, as the three main design principles (sum-of-product dynamical generator, tree tensor network decomposition, and Dirac--Frenkel's TDVP) are identical.  However, while ML-MCTDH is designed for unitary dynamics, the TTN-HEOM is designed for thermal dissipative dynamics.  (3) We implemented the TTN-HEOM into a general purpose code \PyName\ which stands for Tensor Equations for Non-Markovian Structured Open systems. \PyName\ admits arbitrary tensor tree structure, including tensor trains and balanced tensor trees, and arbitrary orders for the core tensors.  \PyName\ includes three numerically stable propagation strategies (fixed-rank direct integration and PS1, and adaptive rank PS2) for the decomposed master equations based on TDVP.  The direct propagation offers adaptive time steps and the use of integration routines with errors that are high-order in the integration time step.
In turn, the PS1 and PS2 are based on second-order Trotterization.  Specifically, PS1 conserves the size of the TTN by keeping all ranks constant during the dynamics.  In turn, PS2 is an adaptive-rank method that updates the size of the TTN according to the error in a truncated SVD step.  These strategies can be mixed at will in practical simulations.  \PyName\ also includes common decompositions of the bath correlation functions, including the low-temperature corrections, for common environmental spectral densities such as the Drude--Lorentz and Brownian oscillator models.}

{Our TTN-HEOM method can capture both the early times and the asymptotic dissipative dynamics of general quantum systems immersed in thermal environments. This contrasts with some other tensor network techniques such as the TD-DMRG \cite{Schollwoeck2011}, ML-MCTDH \cite{Wang2021} and T-TEDOPA \cite{Tamascelli2019}, which are unitary in nature and can only mimic the dissipative dynamics for a finite amount of time by explicitly capturing the dynamics of a finite discretized version of the bath.}

{Our TTN-HEOM method admits arbitrary tree tensor network structure, tensors with orders that vary across nodes, and variable rank for the core tensors during the propagation. This contrasts with recent advances in combining tensor network techniques into HEOM with specific tensor train decompositions \cite{Shi2018, Borrelli2019, Borrelli2021, Ke2022, Mangaud2023}.}

{With respect to the choice of master equation formalism, the \texttt{TENSO} package implements propagation strategies based on the general sum-of-product master equation generator Eq.~\eqref{eq:sop}.  This contrasts with other approaches where the generator of the dynamics is equivalently expressed as a hierarchical sum-of-product form \cite{Lindoy2025}, or strategies where the generator is decomposed as a matrix product operator (MPO) \cite{Yan2021} or a tree tensor network operator (TTNO) with the same network structure as the extended density operator \cite{Ke2023} (see also Refs.~\onlinecite{Li2024, Cakir2025} for a discussion on how to optimize this strategy). An advantage of the \texttt{TENSO} framework is that it is straightforwardly adaptable to any dynamical master equation method that admits a sum-of-product type of generator.  Using it, we thus avoid repeatedly implementing tensor network techniques for other quantum master equations with sum-of-product generators, such as the Lindblad equation and the time-dependent Schr\"odinger equation.}

We demonstrated the self-consistency and utility of TTN-HEOM and \PyName\ by capturing the open quantum dynamics of two-level molecule interacting with a structured thermal environment with a spectral density composed of one Drude--Lorentz and 8 Brownian Oscillator features. Because of computational cost, such a model is well beyond the applicability of standard versions of the HEOM. We show that the dynamics is independent of the tree structure and propagation method, demonstrating the self-consistency of \PyName. 
Overall, by providing a systematic approach for propagating exact quantum master equations, TTN-HEOM facilitates precise numerical simulations from simulating open quantum dynamics coupled with realistic chemical thermal environments.

We expect that the TTN-HEOM and \PyName\ to be useful to understand and emulate the operation of realistic quantum devices, to
engineer quantum environments that enhance molecular function, to isolate molecular qubits with enhanced
coherence properties as needed for quantum technologies, to understand elementary steps in photosynthesis
and photovoltaics, and to test quantum control strategies in the presence of quantum environments. {Future prospects include investigating other combinations for the mixed propagation strategy, implementing adaptive one-site algorithm \cite{Dunnett2021,Lindoy2025} as an alternative choice of PS2, and the potential use of auto-differentiation techniques \cite{Ansel2024, Liao2019} for propagation.}

\section*{Supplementary Material}
See the Supplementary Material for the generalization of the TTN theory and algorithms to general tree topology{, and a numerical illustration of the coincidence between TTN-HEOM and HEOM for a Drude--Lorentz bath}.

\section*{Data Availability}
The data that support the findings of this study are available from the corresponding author upon reasonable request.

\begin{acknowledgments}
  This material is based on work supported by the U.S.\ Department of Energy, Office of Science, Office of Basic Energy Sciences, Quantum Information Science Research in Chemical Sciences, Geosciences, and Biosciences Program under Award No.~DE-SC0025334.
\end{acknowledgments}

\appendix

\section{Sketch of derivations of Eqs.~\eqref{eq:root-eom} and \eqref{eq:su-eom}}\label{sec:eom}

The derivation of the two master equations \eqref{eq:root-eom} and \eqref{eq:su-eom} for the TTN in \eq{eq:ttn-main} follows from (i) the TDVP (Eq.~\eqref{eq:tdvp}), (ii) the constraint that all $U^{(s)}(t)$ ($0<s<K$) are semi-unitary tensors (Eq.~\eqref{eq:onb}), and (iii) the gauge condition (Eq.~\eqref{eq:gauge}).

We rewrite the TTN decomposition as
\begin{equation}\label{eq:split-1}
    \Omega_{ij n_1 \cdots n_K } (t) 
    = \sum_{a_1}A^{(0)}_{ija_1} \mathcal{U}^{(1)}_{a_1n_1n_2 \cdots n_K}.
\end{equation}
Here $\mathcal{U}^{(1)}$ is the branch of the TTN that contracts with the root $A^{(0)}$, and contains $U^{(1)}$ as
\begin{equation}
    \mathcal{U}^{(1)}_{a_1n_1\cdots n_K} =  [\mathsf{Con} (  U^{(1)} (t), \ldots, U^{(K-1)}(t))]_{a_1 n_1 \cdots n_K }.
\end{equation}
More generally, we define 
\begin{equation}
    \mathcal{U}^{(s)}_{a_s n_{\ell_1}\cdots n_{\ell_{K_s}}} 
    =  [\mathsf{Con} (  U^{(s)} (t), \ldots)]_{a_s n_{\ell_1}\cdots n_{\ell_{K_s}}}
\end{equation}
as the branch of TTN that breaks the bond $a_s$ and includes $U(s)$,
where $\set{{{\ell_1}},\ \ldots,\ {\ell_{K_s}}}$ is a subset of $\set{k}_{k=1}^{K}$, corresponding to the part of indexes ${n_k}$ that show up in the branch $\mathcal{U}^{(s)}$. 
In turn,
$
\mathcal{A}^{(s)}_{i j n_{\ell_{K_s+1}}\cdots n_{\ell_{K}} a_s}
$
is another half of branch of TTN that break the bond $a_s$ and include $A^{(0)}$, and $\set{{{\ell_{K_s+1}}},\ \ldots,\ {\ell_{K}}} = \set{k}_{k=1}^{K} \setminus \set{{{\ell_1}},\ \ldots,\ {\ell_{K_s}}}$, corresponding to the complement part of indexes ${n_k}$ that show up in the branch $\mathcal{A}^{(s)}$. 

In this way, the original EDO $\Omega_{ij n_1 \cdots n_K } (t)$ is
\begin{equation}\label{eq:split-s}
\begin{multlined}
    \Omega_{ij n_1 \cdots n_K } (t) 
    = \sum_{a_s} \mathcal{A}^{(s)}_{i j n_{\ell_{K_s+1}}\cdots n_{\ell_{K}} a_s}
    \mathcal{U}^{(s)}_{a_s n_{\ell_1}\cdots n_{\ell_{K_s}}}
    \\
    \equiv \sum_{a_s} \mathcal{A}^{(s)}_{\mathbf{i} a_s} \mathcal{U}^{(s)}_{a_s \mathbf{j}},
\end{multlined}
\end{equation} 
Here $\mathbf{i}$ and $\mathbf{j}$ are the multi-indexes with $\mathbf{i} = \set{i,\ j,\ n_{\ell_{K_s+1}},\  \ldots,\ n_{\ell_K}}$ and $\mathbf{j} = \set{n_{\ell_1},\  \ldots,\ n_{\ell_{K_s}}}$.
Notice that $\mathcal{A}^{(1)} = A^{(0)}$ as showed in Eq.~\eqref{eq:split-1}.

Because of the TTN can be considered as a series of singular value decomposition, and all $U^{(1)}$ are semi-unitary, 
we have 
\begin{equation}\label{eq:onb-ind}
    \sum_{\mathbf{j}} \mathcal{U}^{(s)\star}_{a'_s \mathbf{j}} \mathcal{U}^{(s)}_{a_s \mathbf{j}}  = \delta_{a'_s a_s}.
\end{equation}
\begin{proof}
This can be proved with an induction on the tensor tree.

(1) Suppose the core tensor $U^{(s)}_{a_s n_{k} n_{l}}$ has the indexes ${n_k}$ and $n_{l}$ on its second and third position in the TTN decomposition Eq.~\eqref{eq:ttn-main}. 
Then $\mathcal{U}^{(s)} = U^{(s)}$ and Eq.~\eqref{eq:onb-ind} is an instance of Eq.~\eqref{eq:onb}.

(2) $U^{(s)}_{a_s a_{u} n_{l}}$ in the TTN decomposition Eq.~\eqref{eq:ttn-main}. 
Suppose $\mathcal{U}^{(u)}$ has indices $\mathcal{U}^{(u)}_{a_u \mathbf{j}_1 }$ where $\mathbf{j}_1 = \mathbf{j} \setminus \set{n_l}$, and
\begin{equation}\label{eq:onb-ind-1}
\begin{aligned}
    \quad \sum_{\mathbf{j}}\mathcal{U}^{(s)\star}_{a'_s \mathbf{j}} \mathcal{U}^{(s)}_{a_s \mathbf{j}} 
    &= \sum_{a'_{u} a_u} \sum_{\mathbf{j}_1  n_l}
    \mathcal{U}^{(u)\star}_{a'_u \mathbf{j}_1}
    U^{(s)\star}_{a'_s a'_{u} n_{l}} 
    U^{(s)}_{a_s a_{u} n_{l}}
    \mathcal{U}^{(u)}_{a_u \mathbf{j}_1}  
    \\
    &= \sum_{ a'_{u} a_u n_{l}} \delta_{a'_{u} a_u}  
    U^{(s)\star}_{a'_s a'_{u} n_{l}}
    U^{(s)}_{a_s a_{u} n_{l}} \\
    &= \sum_{a_u n_{l}}  U^{(s)\star}_{a'_s a_{u} n_{l}}
    U^{(s)}_{a_s a_{u} n_{l}}  = \delta_{a'_s a_s}.
\end{aligned}
\end{equation} 
The second last equal sign is due to the Inductive Hypothesis, and the last one is from Eq.~\eqref{eq:onb}.

(3) $U^{(s)}_{a_s  n_{k} a_{v}}$. This case is similar to (2).

(4) $U^{(s)}_{a_s a_{u} a_v}$. Assume $\mathcal{U}^{(u)}$ has indices $
    \mathcal{U}^{(u)}_{a_s \mathbf{j}_1}$, and
    $\mathcal{U}^{(v)}$ has indices $
    \mathcal{U}^{(v)}_{a_v \mathbf{j}_2}$, 
    with $\mathbf{j}_2 = \mathbf{j} \setminus \mathbf{j}_1$.
    Therefore,
\begin{equation}\label{eq:onb-ind-2}
\begin{aligned}
\quad\sum_{\mathbf{j}}\mathcal{U}^{(s)\star}_{a'_s \mathbf{j}} \mathcal{U}^{(s)}_{a_s \mathbf{j}} 
&=
\sum_{a'_{u} a_u} \sum_{a'_{v} a_v} 
\sum_{\mathbf{j}_1\mathbf{j}_2}  
\mathcal{U}^{(v)\star}_{a'_v \mathbf{j}_2}
\mathcal{U}^{(u)\star}_{a'_u \mathbf{j}_1}
U^{(s)\star}_{a'_s a'_{u} a'_{v}} 
\\&\quad\times
U^{(s)}_{a_s a_{u} a_v}
\mathcal{U}^{(u)}_{a_u \mathbf{j}_1}  
\mathcal{U}^{(v)}_{a_v \mathbf{j}_2} \\
&=\sum_{a'_{u} a_u} \sum_{a'_{v} a_v}  
\delta_{a'_{u} a_u}  
\delta_{a'_{v} a_v}  
U^{(s)\star}_{a'_s a'_{u} a'_v}
U^{(s)}_{a_s a_{u} a_v} \\
&= \sum_{a_u a_{v}}  U^{(s)\star}_{a'_s a_{u} a_v}
U^{(s)}_{a_s a_{u} a_v} = \delta_{a'_s a_s}.
\end{aligned}
\end{equation} 
The second last equal sign is due to the Inductive Hypothesis, and the last one is from Eq.~\eqref{eq:onb}.

From (1)--(4), Eq.~\eqref{eq:onb-ind} is proved for all $s$.
\end{proof}
Similarly, the gauge condition can also been generalized to the branch as
\begin{equation}\label{eq:gauge-ind}
    \sum_{\mathbf{j}} \mathcal{U}^{(s)\star}_{a'_s \mathbf{j}} [\dv{t}\mathcal{U}^{(s)}]_{a_s \mathbf{j}} = {0}.
\end{equation}

To derive Eq.~\eqref{eq:root-eom},  we plug Eq.~\eqref{eq:split-s} into the TDVP Eq.~\eqref{eq:tdvp}:
\begin{equation}
\begin{multlined}
    \delta\left( \mathcal{A}^{(s)} \mathcal{U}^{(s)} \right)^\star \mathcal{L} \mathcal{A}^{(s)}  \mathcal{U}^{(s)} 
    =
    \delta\left( \mathcal{A}^{(s)}  \mathcal{U}^{(s)} \right)^\star \pdv{t} \left(\mathcal{A}^{(s)} \mathcal{U}^{(s)}\right).
\end{multlined}
\end{equation}
That is,
\begin{equation}
\begin{multlined}
    \delta \mathcal{A}^{(s)\star}  \mathcal{U}^{(s)\star}  \mathcal{L}  \mathcal{A}^{(s)}  \mathcal{U}^{(s)} 
    +
    \mathcal{A}^{(s)\star} \delta\mathcal{U}^{(s)\star} \mathcal{L}  \mathcal{A}^{(s)} \mathcal{U}^{(s)} 
    =\\ 
    \delta \mathcal{A}^{(s)\star}  \mathcal{U}^{(s)\star}  \pdv{t}  \mathcal{A}^{(s)}  \mathcal{U}^{(s)} 
    +
    \mathcal{A}^{(s)\star} \delta\mathcal{U}^{(s)\star} \pdv{t} \mathcal{A}^{(s)} \mathcal{U}^{(s)} 
    \\+
    \delta \mathcal{A}^{(s)\star} \mathcal{U}^{(s)\star} \mathcal{A}^{(s)}  \pdv{t}\mathcal{U}^{(s)} 
    + \mathcal{A}^{(s)\star} \delta\mathcal{U}^{(s)\star} \mathcal{A}^{(s)} \pdv{t}\mathcal{U}^{(s)}.
\end{multlined}
\end{equation}
Since the variation of $\delta \mathcal{A}^{(s)\star}$ and $\delta \mathcal{U}^{(s)\star}$ is independent and arbitrary, therefore,
\begin{equation}\label{eq:tdvp-s1}
\begin{multlined}   
\sum_{\mathbf{j}'\mathbf{i}\mathbf{j} a_s}
\mathcal{U}^{(s)\star}_{a'_s \mathbf{j}'}   
\mathcal{L}_{\mathbf{i}'\mathbf{j}',\mathbf{i}\mathbf{j}}  
\mathcal{U}^{(s)}_{a_s \mathbf{j}}   
\mathcal{A}^{(s)}_{\mathbf{i} a_s }   =   \\
\sum_{\mathbf{j} a_s}
\mathcal{U}^{(s)\star}_{a'_s \mathbf{j}}   
\pdv{t}
\mathcal{U}^{(s)}_{a_s \mathbf{j}}   
\mathcal{A}^{(s)}_{\mathbf{i}' a_s }   
+ 
\sum_{\mathbf{j} a_s}
\mathcal{U}^{(s)\star}_{a'_s \mathbf{j}}    
\mathcal{U}^{(s)}_{a_s \mathbf{j}}   
\pdv{t}\mathcal{A}^{(s)}_{\mathbf{i}' a_s }, 
\end{multlined}
\end{equation}
and
\begin{equation}\label{eq:tdvp-s2}
\begin{multlined}   
\sum_{\mathbf{i}' \mathbf{i}\mathbf{j} a_s}
\mathcal{A}^{(s)\star}_{\mathbf{i}'a'_s}   
\mathcal{L}_{\mathbf{i}' \mathbf{j}',\mathbf{i}\mathbf{j}}  
\mathcal{A}^{(s)}_{\mathbf{i} a_s }   
\mathcal{U}^{(s)}_{a_s \mathbf{j}}   
=   \\
\sum_{\mathbf{i} a_s}
\mathcal{A}^{(s)\star}_{\mathbf{i} a'_s}   
\pdv{t}
\mathcal{A}^{(s)}_{\mathbf{i} a_s}   
\mathcal{U}^{(s)}_{a_s \mathbf{j}'}   
+ 
\sum_{\mathbf{i} a_s}
\mathcal{A}^{(s)\star}_{\mathbf{i} a'_s}  
\mathcal{A}^{(s)}_{\mathbf{i} a_s }  
\pdv{t}
\mathcal{U}^{(s)}_{a_s \mathbf{j}'}.
\end{multlined}
\end{equation}
From \eq{eq:tdvp-s1} and the properties of the branch $\mathcal{U}^{(s)}$ in Eqs.~\eqref{eq:onb-ind} and \eqref{eq:gauge-ind} we have
\begin{equation}\label{eq:root-eom-s1}
\begin{multlined}    
\pdv{t}\mathcal{A}^{(s)}_{\mathbf{i}' {a'_s} } =
\sum_{\mathbf{j}'\mathbf{i}\mathbf{j}a_s}
\mathcal{U}^{(s)\star}_{a'_s \mathbf{j}'}   
\mathcal{L}_{\mathbf{i}'\mathbf{j}',\mathbf{i}\mathbf{j}}  
\mathcal{U}^{(s)}_{a_s \mathbf{j}}   
\mathcal{A}^{(s)}_{\mathbf{i} a_s } , 
\end{multlined}
\end{equation}
and when $s=1$, it becomes 
\begin{equation} 
\begin{multlined}    
\pdv{t}{A}^{(0)}_{i'j' {a'_1} } = \\
\sum_{ija_1}\sum_{ \substack{ n'_1\cdots n'_K  \\ n_1\cdots n_K}}
\mathcal{U}^{(1)\star}_{a'_1 n'_1\cdots n'_K}   
\mathcal{L}_{\substack{i'j'n'_1\cdots n'_K\\ijn_1\cdots n_K}} 
\mathcal{U}^{(1)}_{a_1 n_1\cdots n_K}   
{A}^{(0)}_{ij a_1 } , 
\end{multlined}
\end{equation}
Further note that 
\begin{equation}
\begin{multlined}
\sum_{ \substack{ n'_1\cdots n'_K  \\ n_1\cdots n_K}}
\mathcal{U}^{(1)\star}_{a'_1 n'_1\cdots n'_K}  
\mathcal{L}_{\substack{i'j'n'_1\cdots n'_K\\ijn_1\cdots n_K}}
\mathcal{U}^{(1)}_{a_1 n_1\cdots n_K}   = \\
\sum_m [h_m^{>}]_{i'i}[h_m^{<}]_{j'j} [f_{m}^{(1)}]_{a'_1 a_1}.
\end{multlined}
\end{equation} 
Hence, we obtain Eq.~\eqref{eq:root-eom}.

On the other hand, plug \eq{eq:root-eom-s1} into \eq{eq:tdvp-s2} and then we get
\begin{equation} \label{eq:eom2-dev0}
\begin{multlined}   
\sum_{\mathbf{i}'\mathbf{i}\mathbf{j} a_s}
\mathcal{A}^{(s)\star}_{\mathbf{i}'a'_s}   
\mathcal{L}_{\mathbf{i}' \mathbf{j}',\mathbf{i}\mathbf{j}}  
\mathcal{A}^{(s)}_{\mathbf{i} a_s }   
\mathcal{U}^{(s)}_{a_s \mathbf{j}}   
= \\ 
\sum_{\mathbf{i}''\mathbf{j}'' \mathbf{i}\mathbf{j} a''_s a_s }
\mathcal{A}^{(s)\star}_{\mathbf{i} a'_s} 
\mathcal{U}^{(s)\star}_{a_s \mathbf{j}}  
\mathcal{L}_{\mathbf{i}\mathbf{j},\mathbf{i}''\mathbf{j}''}  
\mathcal{U}^{(s)}_{a''_s \mathbf{j}''}   
\mathcal{A}^{(s)}_{\mathbf{i}'' a''_s } 
\mathcal{U}^{(s)}_{a_s \mathbf{j}'}    \\
+ 
\sum_{\mathbf{i} a_s}
\mathcal{A}^{(s)\star}_{\mathbf{i} a'_s}  
\mathcal{A}^{(s)}_{\mathbf{i} a_s }  
\pdv{t}
\mathcal{U}^{(s)}_{a_s \mathbf{j}'} .
\end{multlined}
\end{equation}
To obtain the master equation of $U^{(s)}$, similar to the derivation in Eq.~\eqref{eq:onb-ind}, 
suppose $U^{(s)}$ looks like $U^{(s)}_{a_s a_u a_v}$ in the TTN decomposition \eq{eq:ttn-main} (other cases $U^{(s)}_{a_s n_{k} a_{v}}$, $U^{(s)}_{a_s a_{u} n_{l}} $ and $U^{(s)}_{a_s n_{k} n_{l}}$ are similar), then the master equation of $U^{(s)}$ can be obtained by multiplying $\mathcal{U}^{(u)\star}$ and $\mathcal{U}^{(v)\star}$ on the both sides of Eq.~\eqref{eq:eom2-dev0}.
Suppose $\mathcal{U}^{(u)}_{a_u\mathbf{j}_1}$ and $\mathcal{U}^{(v)}_{a_v\mathbf{j}_2}$ with $\mathbf{j}_2 = \mathbf{j} \setminus \mathbf{j}_1$, then
\begin{equation}\label{eq:eom2-dev1}
\begin{multlined}
\sum_{\mathbf{i}'\mathbf{j}'\mathbf{i}\mathbf{j} a_s}
\mathcal{U}^{(v)\star}_{a'_v\mathbf{j}'_2}
\mathcal{U}^{(u)\star}_{a'_u\mathbf{j}'_1}
\mathcal{A}^{(s)\star}_{\mathbf{i}'a'_s}   
\mathcal{L}_{\mathbf{i}' \mathbf{j}', \mathbf{i}\mathbf{j}}  
\mathcal{A}^{(s)}_{\mathbf{i} a_s }   
\mathcal{U}^{(s)}_{a_s \mathbf{j}}   
=\\
\sum_{m} \sum_{a_{s} a_u a_v} [D^{(s)}_m]_{a_{s}a'_{s}}
     [f_m^{(u)}]_{a'_u a_u} [f_m^{(v)}]_{a'_v a_v}U^{(s)}_{a_{s} a_u a_v},
\end{multlined}
\end{equation} 
\begin{equation}\label{eq:eom2-dev2}
\begin{multlined}
\sum_{\mathbf{i}''\mathbf{j}''\mathbf{i} \mathbf{j}a''_s a_s}
\mathcal{U}^{(v)\star}_{a'_v\mathbf{j}'_2}
\mathcal{U}^{(u)\star}_{a'_u\mathbf{j}'_1}
\mathcal{A}^{(s)\star}_{\mathbf{i} a'_s} 
\mathcal{U}^{(s)\star}_{a_s \mathbf{j}}   
\mathcal{L}_{\mathbf{i}\mathbf{j},\mathbf{i}''\mathbf{j}''}  
\mathcal{U}^{(s)}_{a''_s \mathbf{j}''}   
\mathcal{A}^{(s)}_{\mathbf{i}'' a''_s } 
\mathcal{U}^{(s)}_{a_s \mathbf{j}'} 
\\=
\sum_{m} \sum_{a'_{s} a_{s}} [D^{(s)}_m]_{a_{s}'a_{s}''}
    U^{(s)}_{a_{s}a'_u a'_v}[f_m^{(s)}]_{a_{s}a_{s}'},
\end{multlined}
\end{equation}
and
\begin{equation}\label{eq:eom2-dev3}
\sum_{\mathbf{i} \mathbf{j} a_s}
\mathcal{U}^{(v)\star}_{a_v\mathbf{j}_2}
\mathcal{U}^{(u)\star}_{a_u\mathbf{j}_1}
\mathcal{A}^{(s)\star}_{\mathbf{i} a'_s}  
\mathcal{A}^{(s)}_{\mathbf{i} a_s }  
\pdv{t}
\mathcal{U}^{(s)}_{a_s \mathbf{j}}= 
\sum_{a_{s}} D^{(s)}_{a_{s}a_{s}'}  \dv{t} U^{(s)}_{a_{s} a_u a_v}.
\end{equation}
Here, we have use the definition of $f_m^{(s)}$, $D_{m}^{(s)}$ and $D^{(s)}$ introduced in Eqs.~\eqref{eq:f-recursive}--\eqref{eq:d-recursive-3} to simplify the equation.
Plug Eqs.~\eqref{eq:eom2-dev1}--\eqref{eq:eom2-dev3} into Eq.~\eqref{eq:eom2-dev0} we have
\begin{equation} \label{eq:eom2-1}
\begin{multlined}   
\sum_{a_{s}'} D^{(s)}_{a_{s}'a_{s}''}  \dv{t} U^{(s)}_{a_{s}'a_u' a_v'}
= \\ 
\sum_{m} \sum_{a_{s}' a_u a_v} 
[D^{(s)}_m]_{a_{s}'a_{s}''}
[f_m^{(u)}]_{a'_u a_u} 
[f_m^{(v)}]_{a'_v a_v}
U^{(s)}_{a'_{s} a_u a_v}
\\
- 
\sum_{m} \sum_{a_{s}'a_{s} } 
[D^{(s)}_m]_{a_{s}'a_{s}''}
U^{(s)}_{a_{s} a_u' a_v'}
[f_m^{(s)}]_{a_{s}a_{s}'}.
\end{multlined}
\end{equation}
Similarly, 
for $U^{(s)}_{a_s a_{u} n_{l}} $
\begin{equation} \label{eq:eom2-3}
\begin{multlined}   
\sum_{a_{s}'} D^{(s)}_{a_{s}'a_{s}''}  \dv{t} U^{(s)}_{a_{s}'a_u' n_l'}
= \\ 
\sum_{m} \sum_{a_{s}' a_u n_l} 
[D^{(s)}_m]_{a_{s}'a_{s}''}
[f_m^{(u)}]_{a'_u a_u} 
[h_m^{(l)}]_{n'_l n_l}
U^{(s)}_{a'_{s} a_u n_l}
\\
- 
\sum_{m} \sum_{a_{s}'a_{s} } 
[D^{(s)}_m]_{a_{s}'a_{s}''}
U^{(s)}_{a_{s} a_u' n_l'}
[f_m^{(s)}]_{a_{s}a_{s}'},
\end{multlined}
\end{equation}
for $U^{(s)}_{a_s n_{k} a_{v}}$
\begin{equation} \label{eq:eom2-2}
\begin{multlined}   
\sum_{a_{s}'} D^{(s)}_{a_{s}'a_{s}''}  \dv{t} U^{(s)}_{a_{s}' n_k' a_v'}
= \\ 
\sum_{m} \sum_{a_{s}' n_k a_v} 
[D^{(s)}_m]_{a_{s}'a_{s}''}
[h_m^{(k)}]_{n_k' a_u} 
[f_m^{(v)}]_{a'_v a_v}
U^{(s)}_{a'_{s} n_k a_v}
\\
- 
\sum_{m} \sum_{a_{s}'a_{s} } 
[D^{(s)}_m]_{a_{s}'a_{s}''}
U^{(s)}_{a_{s} n_k' a_v'}
[f_m^{(s)}]_{a_{s}a_{s}'},
\end{multlined}
\end{equation}
and for $U^{(s)}_{a_s n_{k} n_{l}}$
\begin{equation} \label{eq:eom2-4}
\begin{multlined}   
\sum_{a_{s}'} D^{(s)}_{a_{s}'a_{s}''}  \dv{t} U^{(s)}_{a_{s}' n_k' n_l'}
= \\ 
\sum_{m} \sum_{a_{s}'n_kn_l} 
[D^{(s)}_m]_{a_{s}'a_{s}''}
[h_m^{(k)}]_{n_k' n_k} 
[h_m^{(l)}]_{n_l' n_l}
U^{(s)}_{a'_{s} n_k n_l}
\\
- 
\sum_{m} \sum_{a_{s}'a_{s} } 
[D^{(s)}_m]_{a_{s}'a_{s}''}
U^{(s)}_{a_{s} n_k' n_l'}
[f_m^{(s)}]_{a_{s}a_{s}'}.
\end{multlined}
\end{equation}
Hence, Eq.~\eqref{eq:su-eom} is derived.

\section{Example of TTN-HEOM for an open quantum dynamics with 4 bexcitons.}
\label{sec:tree-example}
To better demonstrate the TTN scheme yields a closed set of master equations, here we consider an example of tensor tree decomposition for $\Omega(t)$ with $K=4$ bexcitons
\begin{equation}\label{eq:tree4}
  \begin{multlined}
    \Omega_{ij n_1 n_2 n_3 n_4 } (t) =\\ \sum_{a_1 a_2 a_3}^{R_1 R_2 R_3} A^{(0)}_{ij a_1}(t) U^{(1)}_{a_1a_2 a_3}(t)  U^{(2)}_{a_2 n_1 n_2}(t) U^{(3)}_{a_3 n_3 n_4}(t),
  \end{multlined}
\end{equation} 
where $A^{(0)}$ is the root. 
Inserting Eq.~\eqref{eq:tree4} in Eq.~\eqref{eq:tdvp} and taking into account the gauge condition Eq.~\eqref{eq:gauge}, yields
\begin{equation}\label{eq:rootexample}
    \dv{t} A^{(0)}_{i' j' a'_1} = \sum_{m} \sum_{ij a_1} [h_m^{>}]_{i'i}[h_m^{<}]_{j'j} [f_{m}^{(1)}]_{a'_1 a_1} A^{(0)}_{i j a_1},
\end{equation}
for the root tensor. The $f_m^{(s)}$ are defined as
\begin{align}
    [f_{m}^{(3)}]_{a'_3 a_3} &\equiv \sum_{n'_3 n'_4 n_3 n_4}
    U^{(4)\star}_{a'_3 n'_3 n'_4} 
    [h_m^{(3)}]_{n'_3 n_3} 
    [h_m^{(4)}]_{n'_4 n_4} 
    U^{(4)}_{a_3 n_3 n_4}, \\
    [f_{m}^{(2)}]_{a'_2 a_2} &\equiv \sum_{n'_1 n'_2 n_1 n_2}
    U^{(3)\star}_{a'_2 n'_1 n'_2} 
    [h_m^{(1)}]_{n'_1 n_1} 
    [h_m^{(2)}]_{n'_2 n_2} 
    U^{(3)}_{a_2 n_1 n_2},  \\
    [f_{m}^{(1)}]_{a'_1 a_1} &\equiv \sum_{a'_3 a'_2 a_3 a_2}
    U^{(2)\star}_{a'_1 a'_2 a'_3} 
    [f_{m}^{(3)}]_{a'_3 a_3}
    [f_{m}^{(2)}]_{a'_2 a_2}
    U^{(2)}_{a_1 a_2 a_3},
\end{align}  
and capture the bexciton influence on the system. Note that $f_m^{(1)}$ depends on $f_m^{(2)}$ and $f_m^{(3)}$, while the latter depends on the dissipators $h_m^{(k)}$.  These quantities effectively extract the relevant bath dynamics that influences the system in a compressed fashion.
In turn, the semi-unitary tensors capture the active space of the bexcitons that influences the system's dynamics. The equations of motion for them are given by  
\begin{align}\label{eq:suexample} 
   &\sum_{a'_1} [D^{(1)}]_{a'_1a''_1} \dv{t} U^{(2)}_{a'_1 a'_2 a'_3} =  
   \sum_{m} \sum_{a'_1 a_1 a_2 a_3} 
   [D^{(1)}_m]_{a'_1a''_1} 
   \times \nonumber \\&\quad
   ([f_m^{(2)}]_{a'_2a_2} [f_m^{(3)}]_{a'_3a_3}U^{(2)}_{a'_1 a_2 a_3} -  U^{(2)}_{a_1 a'_2 a'_3}[f_m^{(1)}]_{a_1a'_1}),  \\
   &\sum_{a'_2} [D^{(2)}]_{a'_2 a''_2} \dv{t} U^{(3)}_{a'_2 n'_1 n'_2} = 
   \sum_{m} \sum_{a'_2 a_2 n_1 n_2} 
   [D^{(2)}_m]_{a'_2 a''_2} 
   \times \nonumber \\&\quad
   ([h_m^{(1)}]_{n'_1n_1} [h_m^{(2)}]_{n'_2n_2}   U^{(3)}_{a'_2 n_1 n_2}
   -   U^{(3)}_{a_2 n'_1 n'_2}[f^{(2)}_m]_{a_2a'_2}),   \\
   &\sum_{a'_3} [D^{(3)}]_{a'_3a''_3} \dv{t} U^{(4)}_{a'_3 n'_3 n'_4} =  
   \sum_{m} \sum_{a'_3 a_3 n_3 n_4}[D^{(3)}_m]_{a'_3a''_3}  
   \times \nonumber 
   \\&\quad
   ([h_m^{(3)}]_{n'_3 n_3}  [h_m^{(4)}]_{n'_4 n_4} U^{(4)}_{a'_3 n_3 n_4}
   - U^{(4)}_{a_3 n'_3 n'_4}[f_m^{(3)}]_{a_3a'_3} ) .
\end{align} 
Here, the quantities $D^{(s)}_m$ and $D^{(s)}$ are defined by 
\begin{align}
    [D^{(1)}]_{a'_1 a_1} &\equiv \sum_{ij}
    A^{(0)}_{i j a'_1} 
    A^{(1)\star}_{i j a_1}, \\
    [D^{(2)}]_{a'_2 a_2} &\equiv \sum_{a'_1 a_1 a_3}
    U^{(2)}_{a'_1 a'_2 a_3} 
    [D^{(1)}]_{a'_1 a_1}  
    U^{(2)\star}_{a_1 a_2 a_3}, \\
    [D^{(3)}]_{a'_3 a_3} &\equiv \sum_{a'_1 a_1 a_2}
    U^{(2)}_{a'_1 a_2 a'_3} 
    [D^{(1)}]_{a'_1 a_1}  
    U^{(2)\star}_{a_1 a_2 a_3}.
\end{align} 
and  
\begin{align}
  [D^{(1)}_m]_{a'_1 a_1} &\equiv \sum_{i' j' i j} 
  [h_m^{>}]_{ii'}[h_m^{<}]_{jj'} 
  A^{(0)}_{i' j' a'_1}
  A^{(1)\star}_{i j a_1}, \\
  [D^{(2)}_m]_{a'_2 a_2} &\equiv \sum_{a'_1 a'_3 a_1 a_3}
  [f_{m}^{(3)}]_{a_3 a'_3}  
  U^{(2)}_{a'_1 a'_2 a'_3} 
  [D^{(1)}_m]_{a'_1 a_1}
  U^{(2)\star}_{a_1 a_2 a_3}, \\
  [D^{(3)}_m]_{a'_3 a_3} &\equiv \sum_{a'_1 a'_2 a_1 a_2}
  [f_{m}^{(2)}]_{a_2 a'_2}  
  U^{(2)}_{a'_1 a'_2 a'_3} 
  [D^{(1)}_m]_{a'_1 a_1}
  U^{(2)\star}_{a_1 a_2 a_3}.
\end{align}

\bibliography{references}

\clearpage %
\pagenumbering{gobble}
\thispagestyle{empty}
\begin{figure*}
    \centering
    \includegraphics[page=1,width=\linewidth]{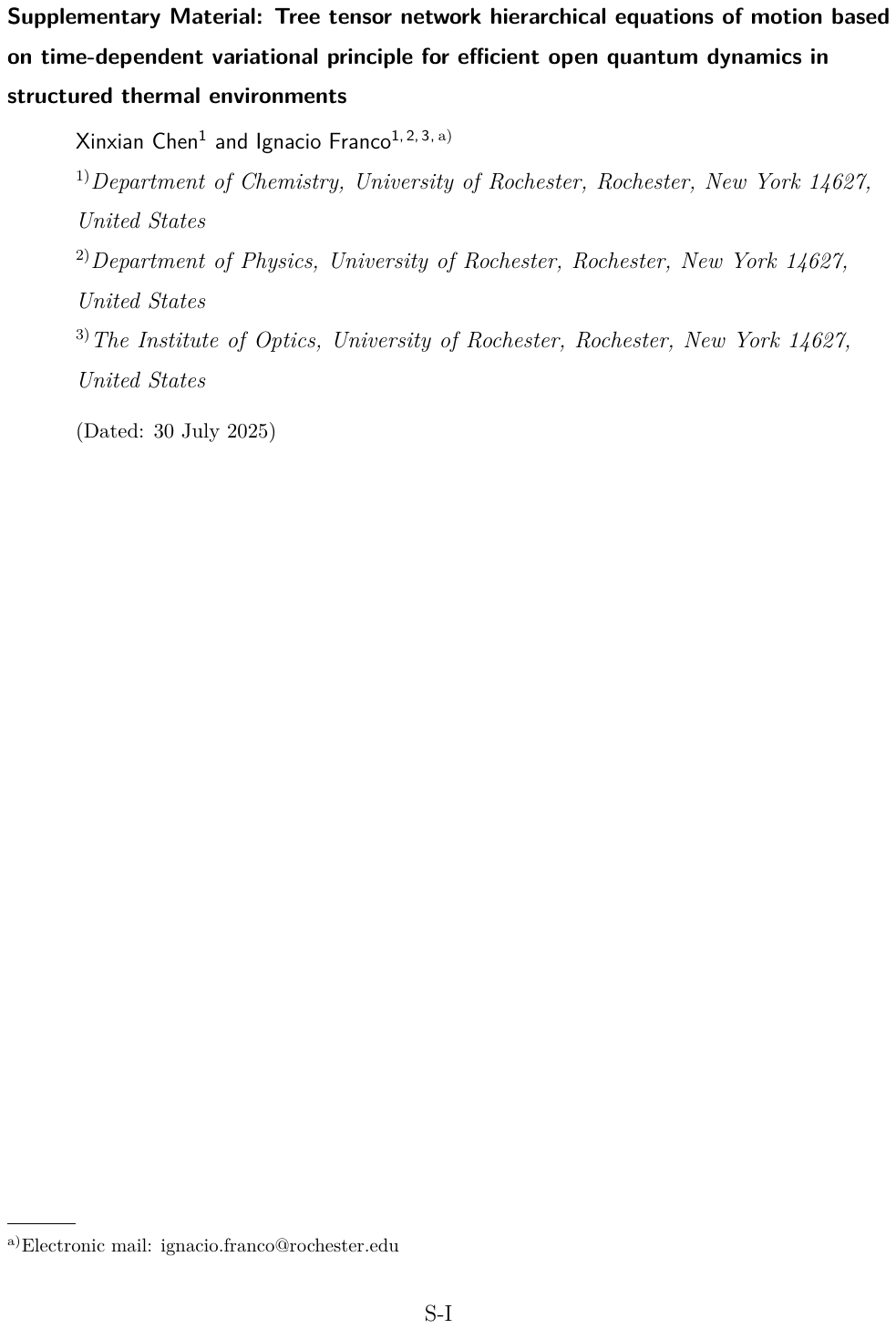} %
\end{figure*}
\thispagestyle{empty}
\begin{figure*}
    \centering
    \includegraphics[page=2,width=\linewidth]{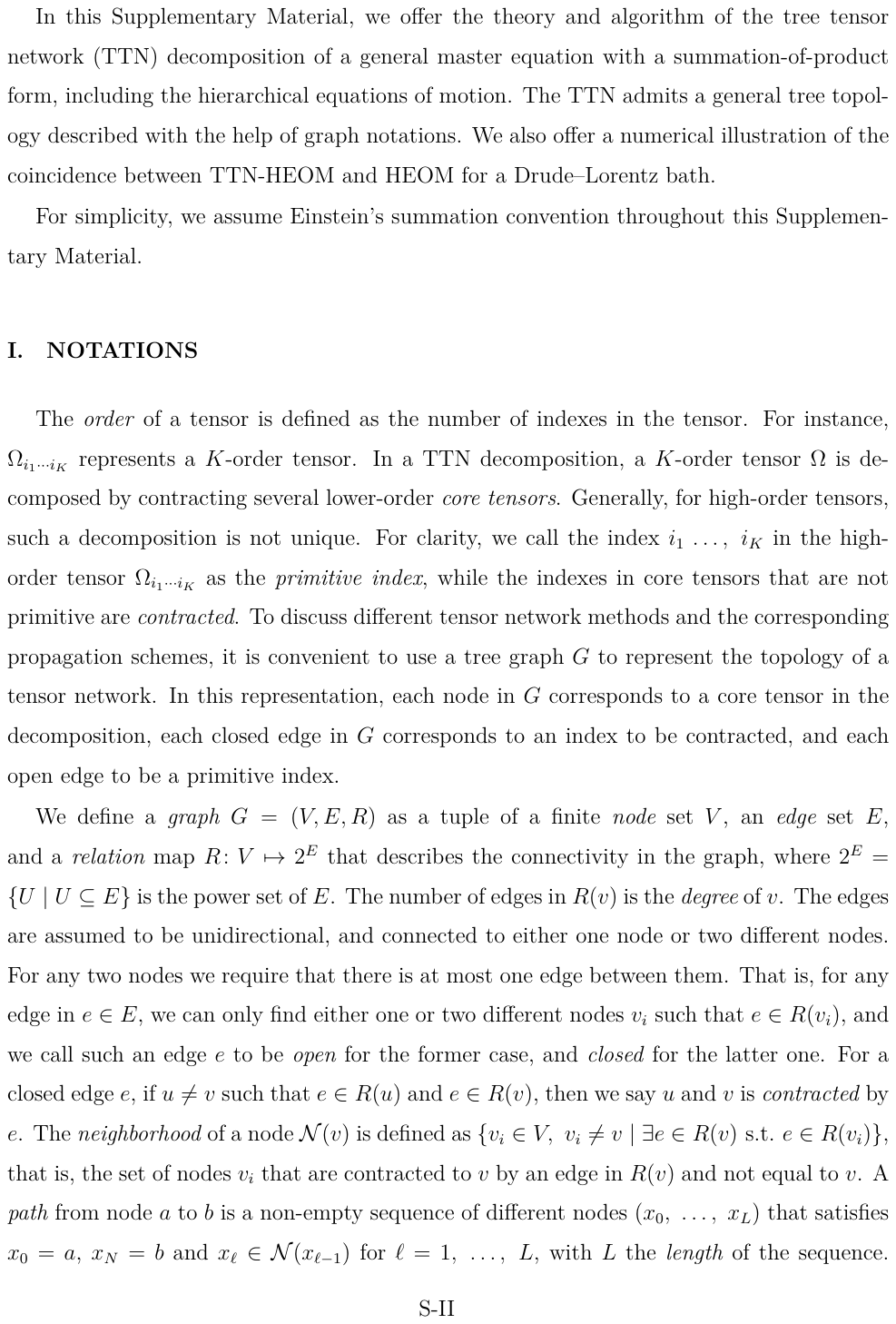} %
\end{figure*}
\thispagestyle{empty}
\begin{figure*}
    \centering
    \includegraphics[page=3,width=\linewidth]{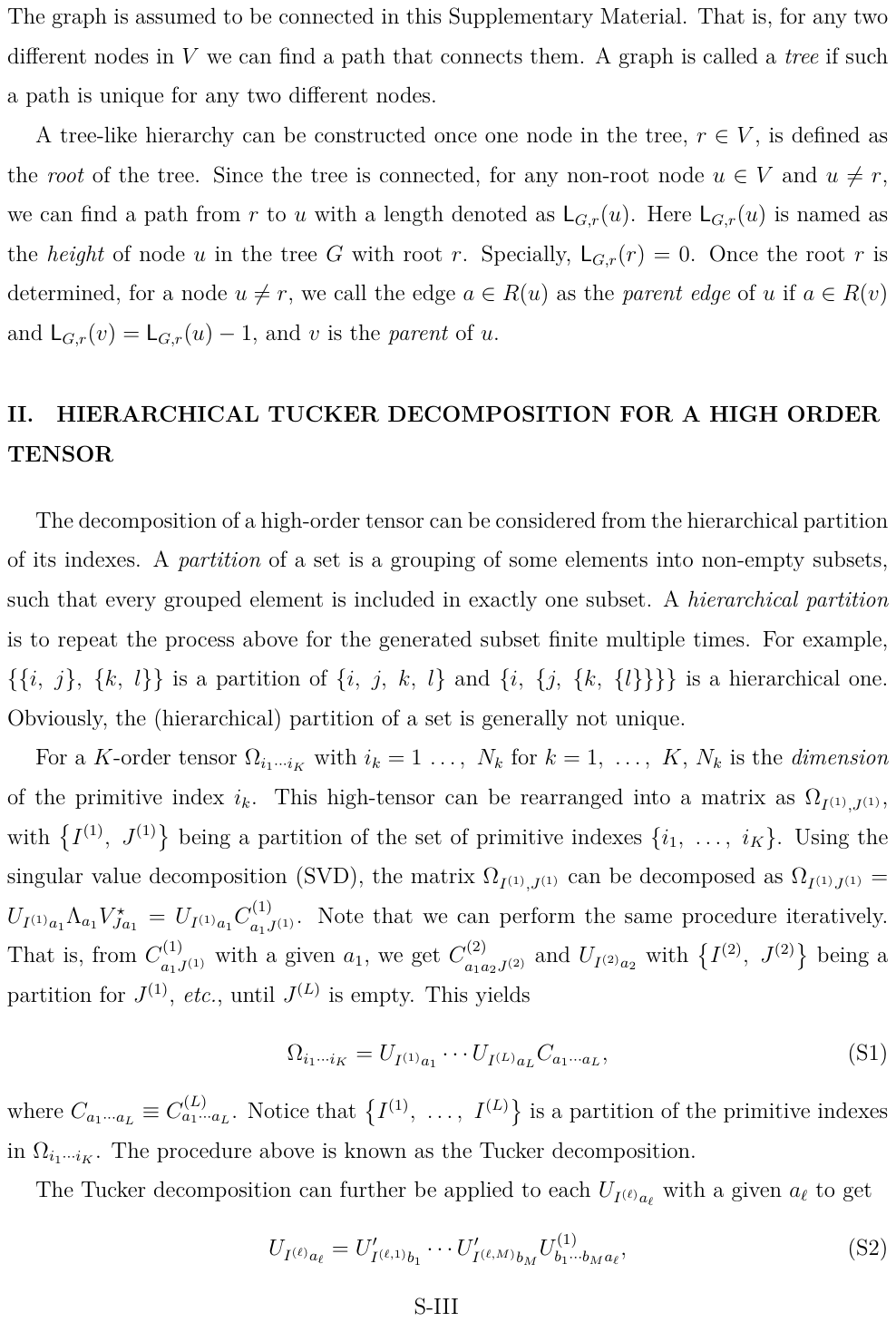} %
\end{figure*}
\thispagestyle{empty}
\begin{figure*}
    \centering
    \includegraphics[page=4,width=\linewidth]{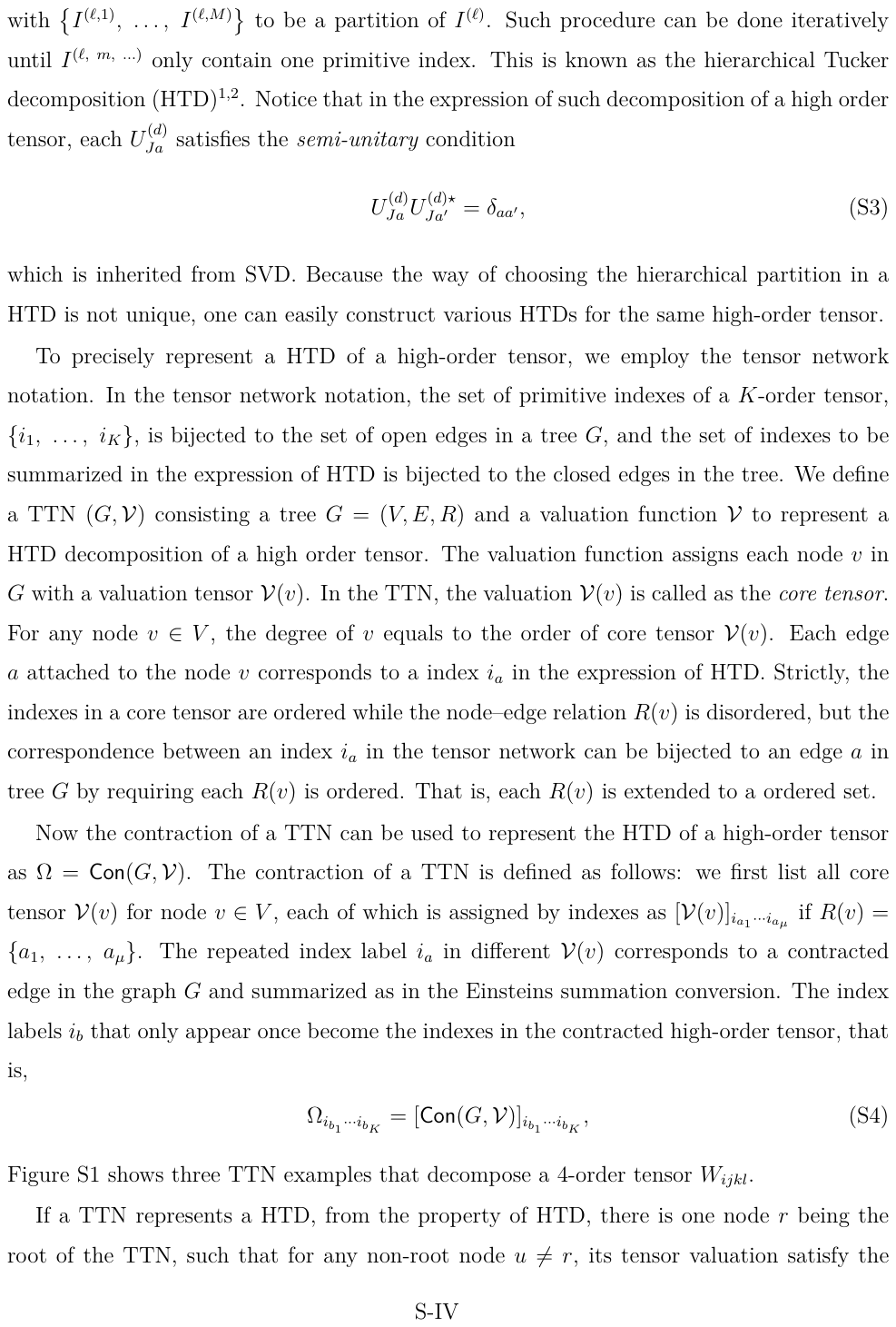} %
\end{figure*}
\thispagestyle{empty}
\begin{figure*}
    \centering
    \includegraphics[page=5,width=\linewidth]{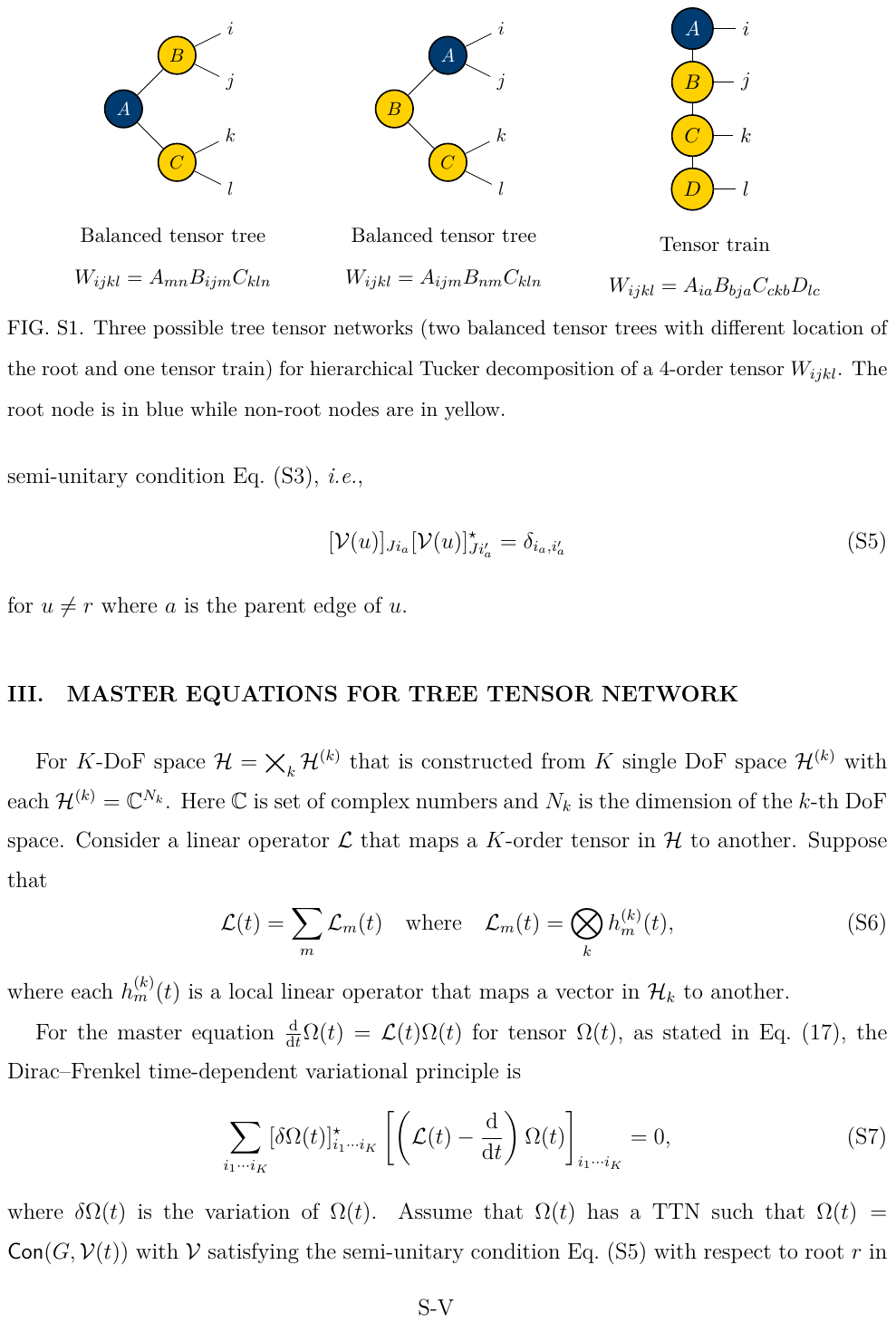} %
\end{figure*}
\thispagestyle{empty}
\begin{figure*}
    \centering
    \includegraphics[page=6,width=\linewidth]{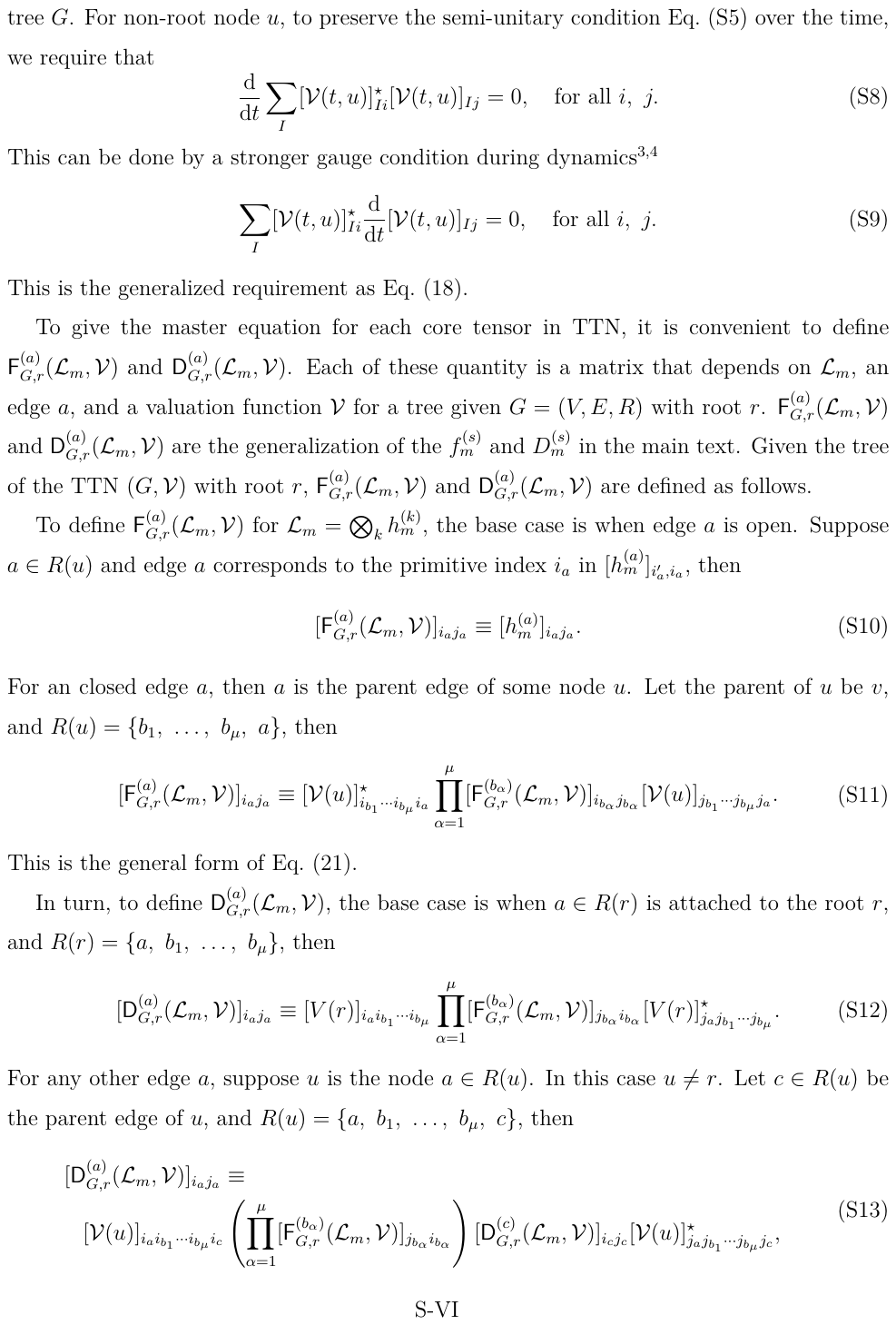} %
\end{figure*}
\thispagestyle{empty}
\begin{figure*}
    \centering
    \includegraphics[page=7,width=\linewidth]{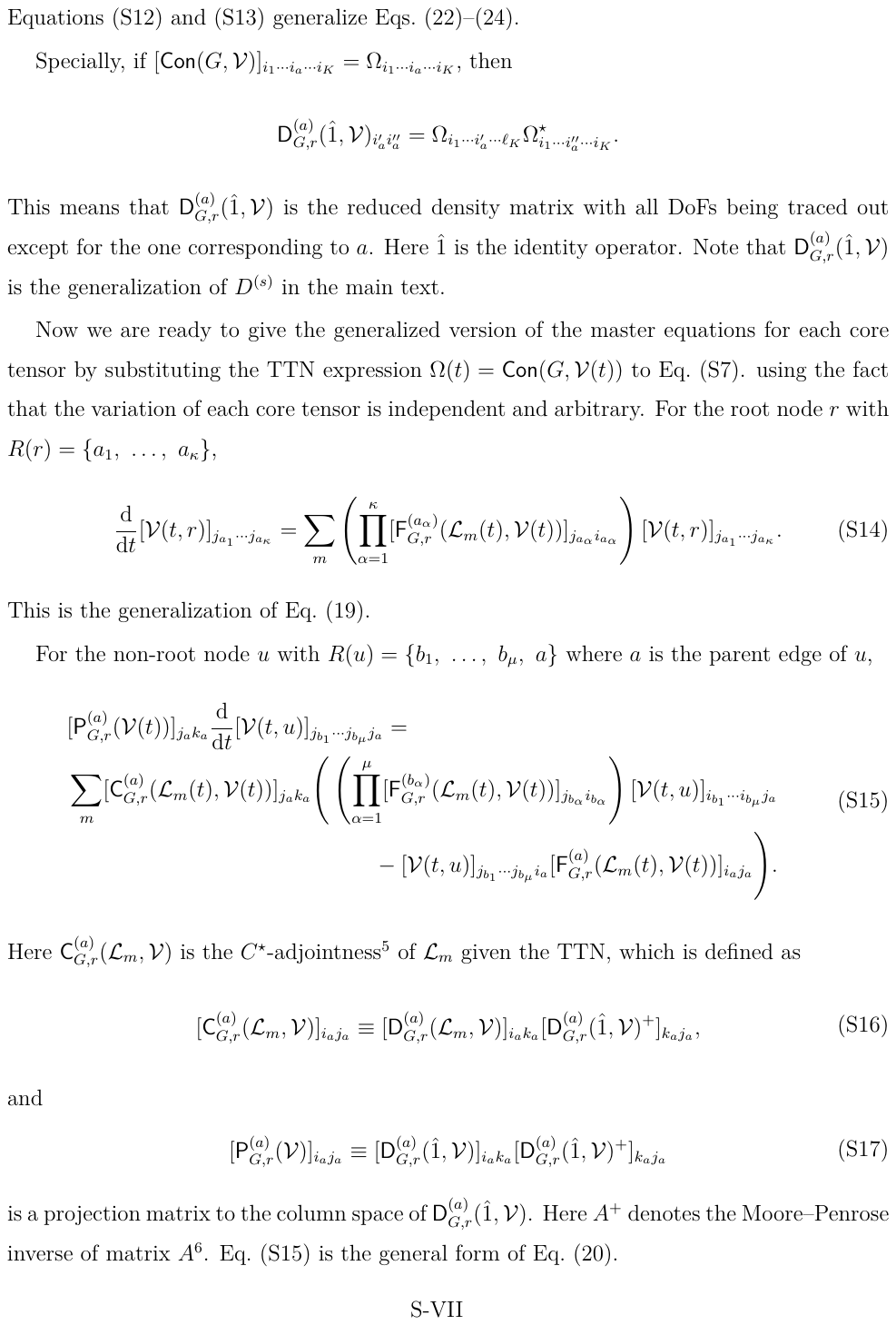} %
\end{figure*}
\thispagestyle{empty}
\begin{figure*}
    \centering
    \includegraphics[page=8,width=\linewidth]{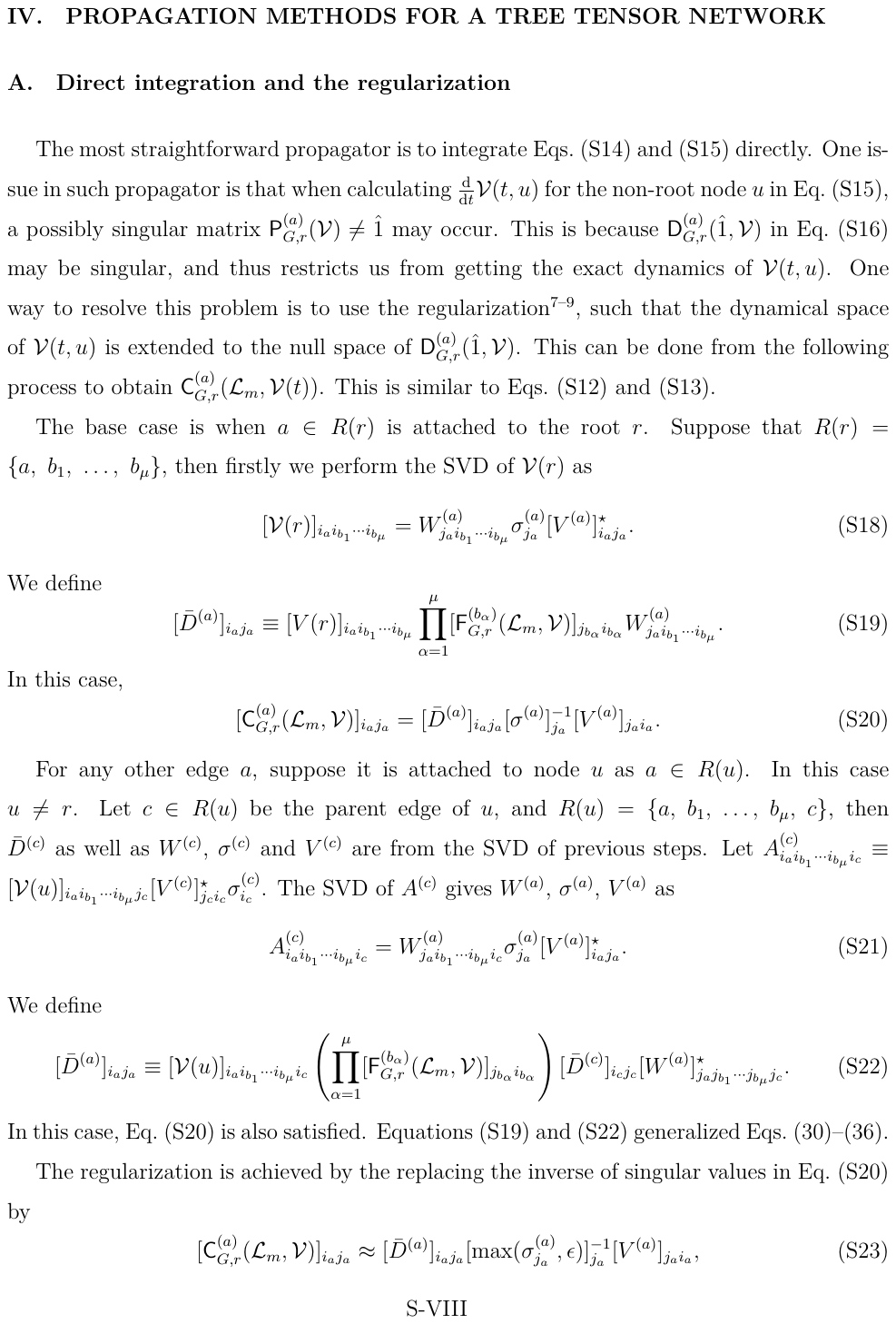} %
\end{figure*}
\thispagestyle{empty}
\begin{figure*}
    \centering
    \includegraphics[page=9,width=\linewidth]{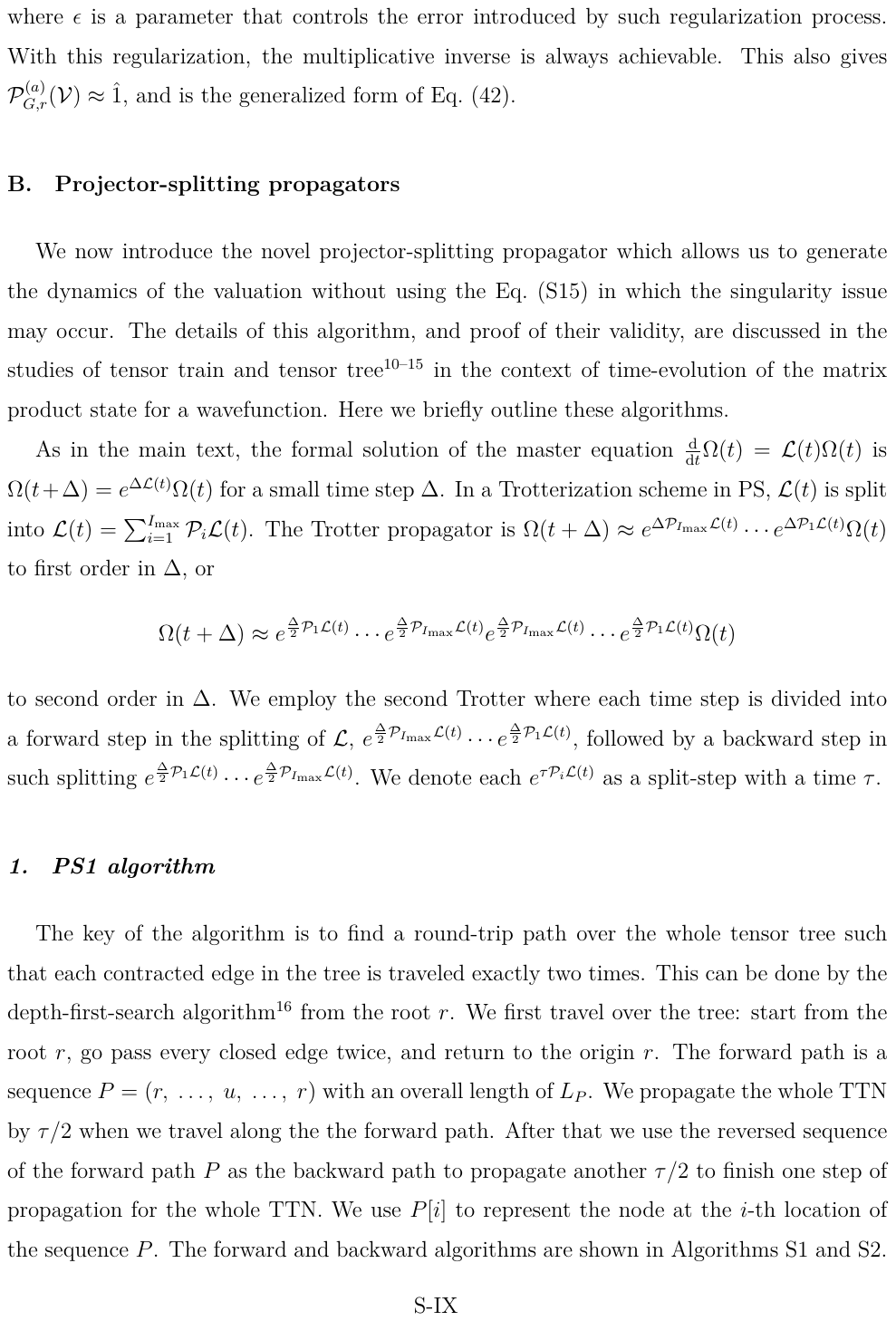} %
\end{figure*}
\thispagestyle{empty}
\begin{figure*}
    \centering
    \includegraphics[page=10,width=\linewidth]{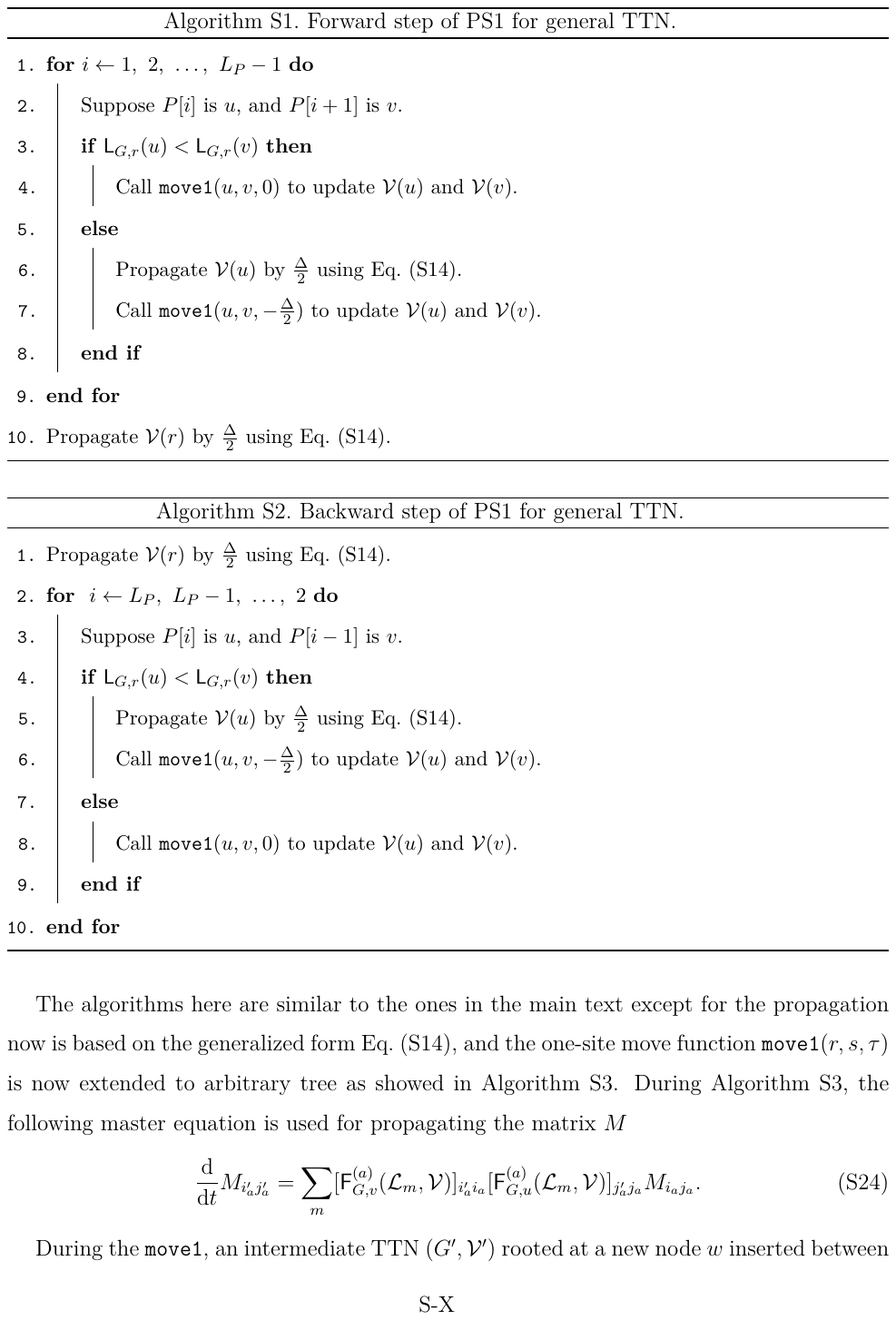} %
\end{figure*}
\thispagestyle{empty}
\begin{figure*}
    \centering
    \includegraphics[page=11,width=\linewidth]{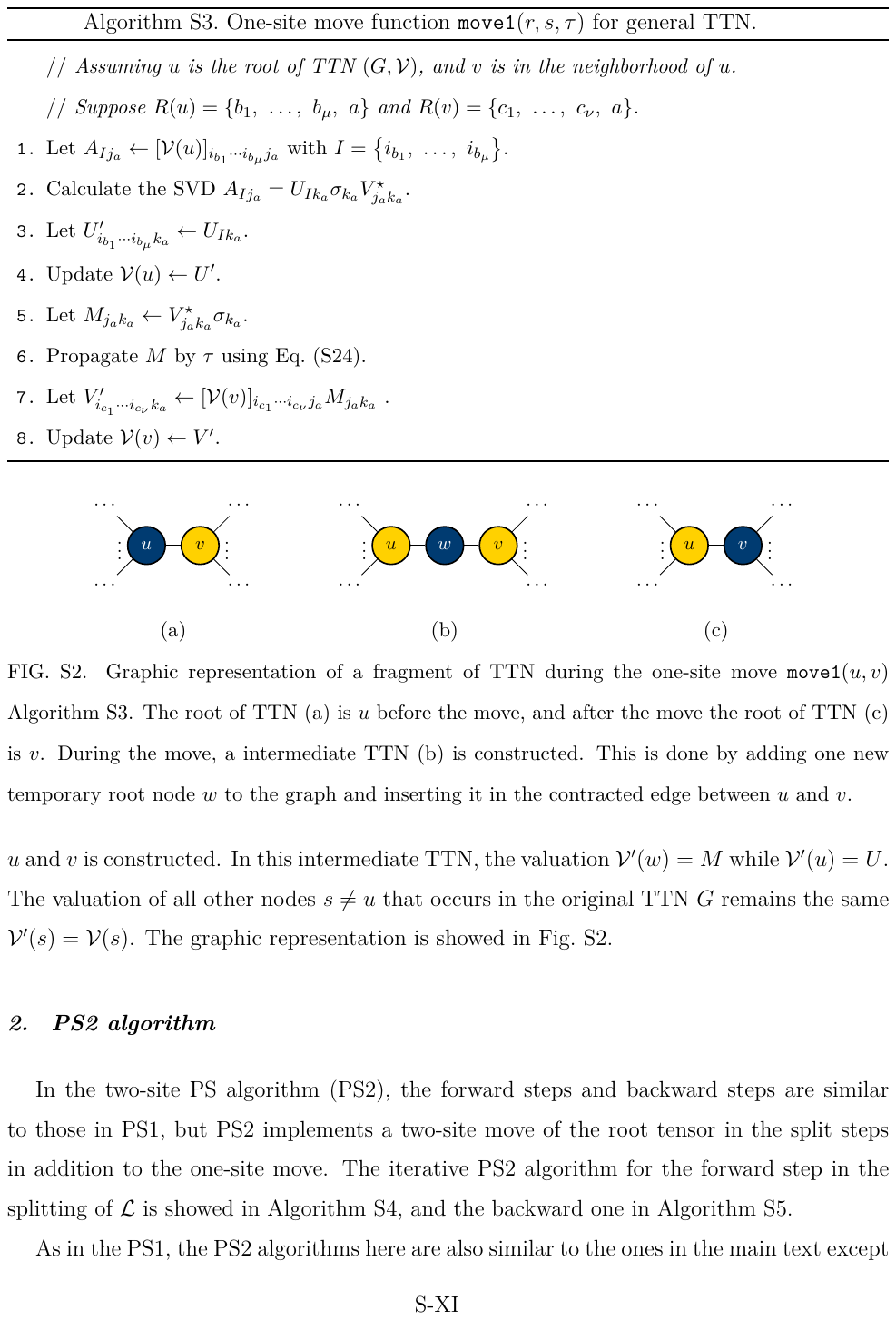} %
\end{figure*}
\thispagestyle{empty}
\begin{figure*}
    \centering
    \includegraphics[page=12,width=\linewidth]{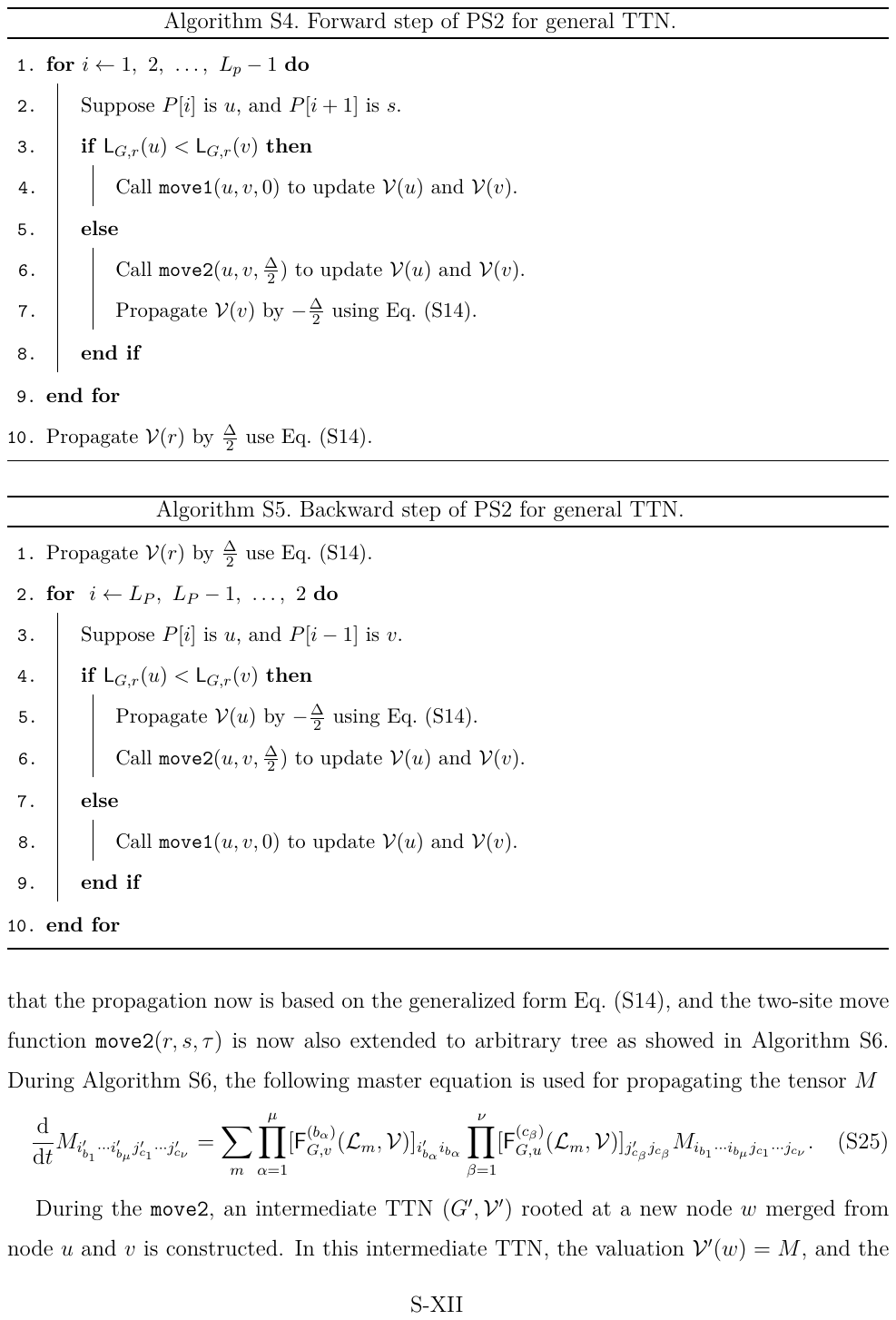} %
\end{figure*}
\thispagestyle{empty}
\begin{figure*}
    \centering
    \includegraphics[page=13,width=\linewidth]{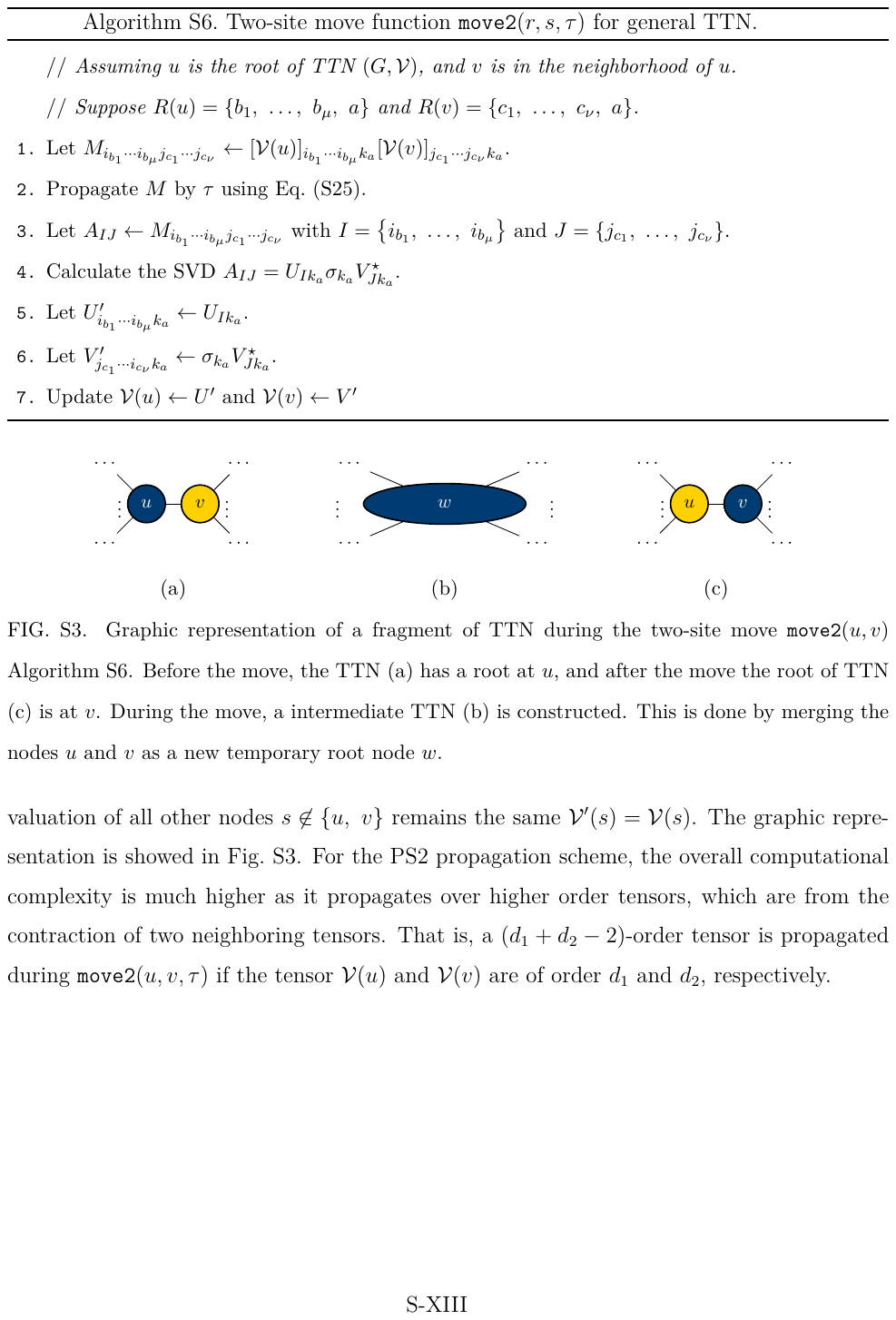} %
\end{figure*}
\thispagestyle{empty}
\begin{figure*}
    \centering
    \includegraphics[page=14,width=\linewidth]{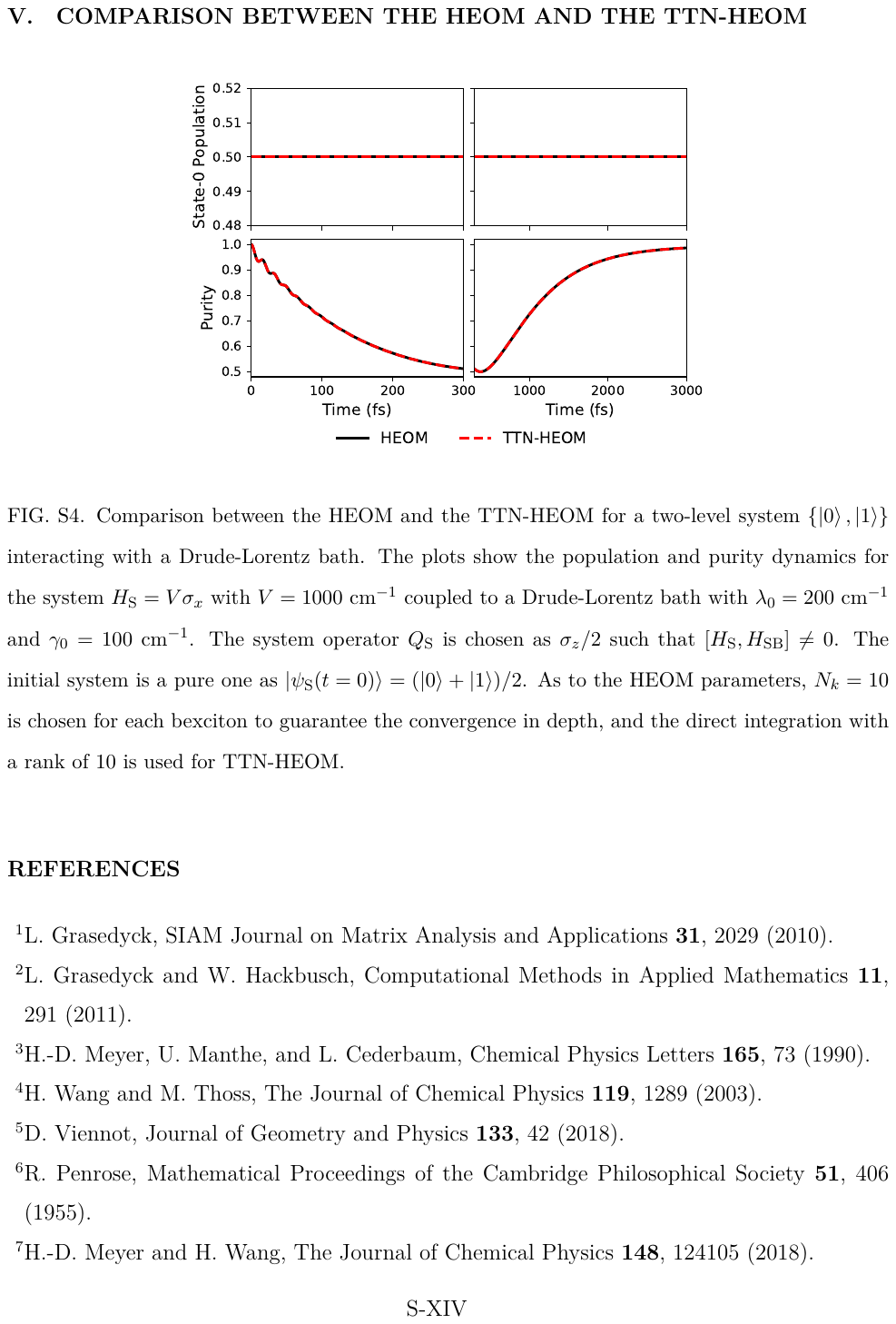} %
\end{figure*}
\thispagestyle{empty}
\begin{figure*}
    \centering
    \includegraphics[page=15,width=\linewidth]{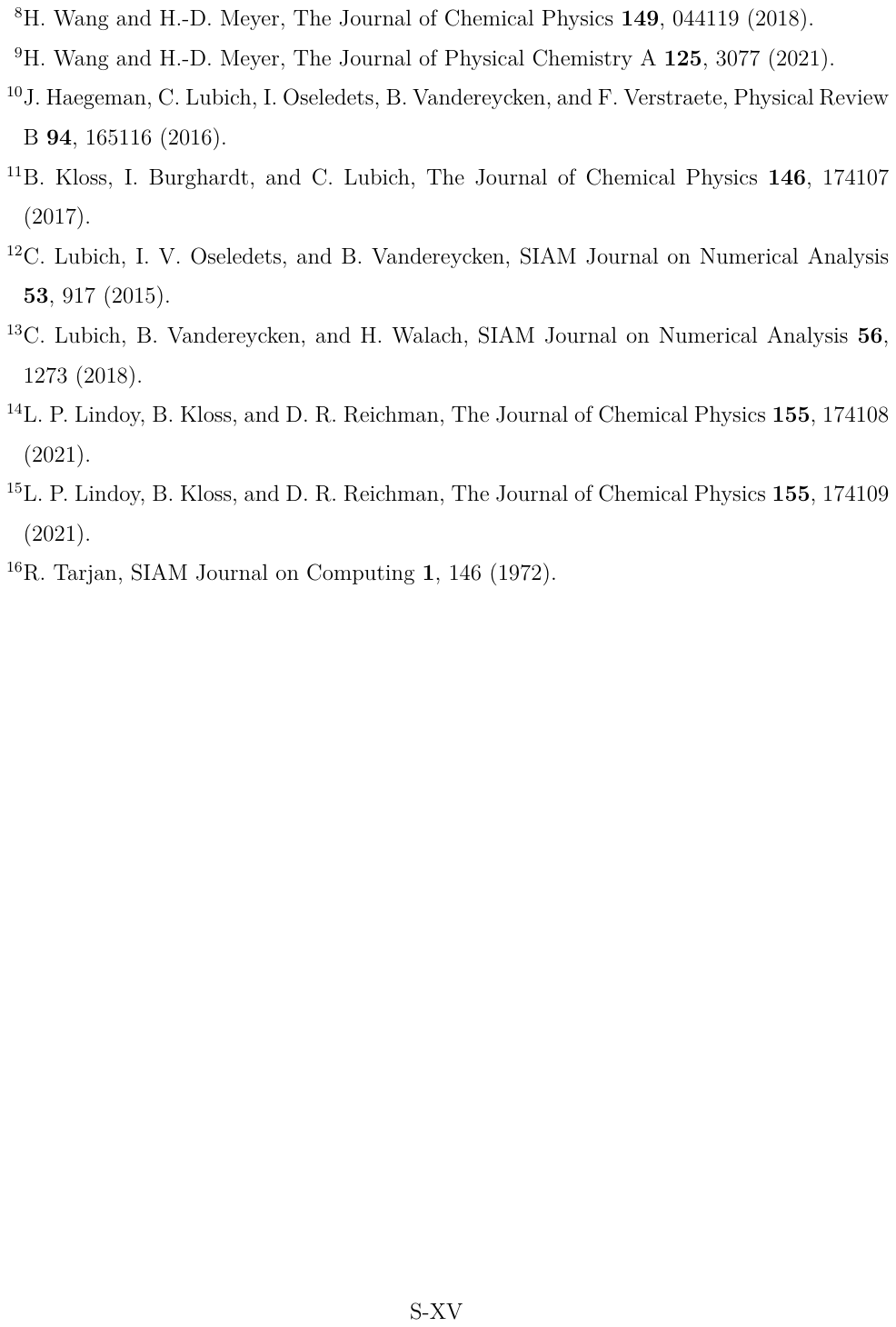} %
\end{figure*}

\end{document}